\documentclass[nofootinbib,superscriptaddress,a4paper, twocolumn, floatfix]{revtex4-2}
\usepackage[english]{babel}
\usepackage[utf8]{inputenc}
\usepackage[colorinlistoftodos, color=green!40, prependcaption]{todonotes}
\usepackage{amsthm}
\usepackage{mathtools}
\usepackage{physics}
\usepackage{braket}
\usepackage{listings}
\usepackage{xcolor}
\usepackage{graphicx}
\usepackage[left=23mm,right=13mm,top=35mm,columnsep=15pt]{geometry} 
\usepackage{adjustbox}
\usepackage{placeins}
\usepackage[T1]{fontenc}
\usepackage{lipsum}
\usepackage{csquotes}
\usepackage[pdftex, pdftitle={Article}, pdfauthor={Author}]{hyperref} 
\usepackage{booktabs}
\usepackage{amssymb}
\usepackage{bm}
\usepackage[ruled,vlined]{algorithm2e}
\usepackage[caption=false]{subfig}

\usepackage{dsfont}

\begin{document}

\title{\textbf{Information flow in parameterized quantum circuits}}

\author{Abhinav Anand}
\thanks{These authors contributed equally to this work}
\email[E-mail:]{abhinav.anand@mail.utoronto.ca}
\affiliation{Chemical Physics Theory Group, Department of Chemistry, University of Toronto, Canada.}

\author{Lasse Bjørn Kristensen~}
\thanks{These authors contributed equally to this work}
\email[E-mail:]{l.kristensen@mail.utoronto.ca}
\affiliation{Chemical Physics Theory Group, Department of Chemistry, University of Toronto, Canada.}
\affiliation{Department of Computer Science, University of Toronto, Canada.}

\author{Felix Frohnert}
\affiliation{Niels Bohr Institute, University of Copenhagen, Denmark}

\author{Sukin Sim}
\affiliation{Zapata Computing Inc., 100 Federal Street, Boston, MA 02110, USA.}

\author{Al\'{a}n Aspuru-Guzik}
\email[E-mail:]{aspuru@utoronto.ca}
\affiliation{Chemical Physics Theory Group, Department of Chemistry, University of Toronto, Canada.}
\affiliation{Department of Computer Science, University of Toronto, Canada.}
\affiliation{Department of Chemical Engineering and Applied Chemistry,  University of Toronto, Canada.}
\affiliation{Department of Materials Science and Engineering, University of Toronto, Canada.}
\affiliation{Vector Institute for Artificial Intelligence, Toronto, Canada.}
\affiliation{Canadian  Institute  for  Advanced  Research  (CIFAR)  Lebovic  Fellow,  Toronto,  Canada.}
\date{\today}

\begin{abstract}
    In this work, we introduce a new way to quantify information flow in quantum systems, especially for parameterized quantum circuits.
    We use a graph representation of the circuits and propose a new distance metric using the mutual information between gate nodes.
    We then present an optimization procedure for variational algorithms using paths based on the distance measure.
    We explore the features of the algorithm by means of the variational quantum eigensolver, in which we compute the ground state energies of the Heisenberg model. 
    In addition, we employ the method to solve a binary classification problem using variational quantum classification. 
    From numerical simulations, we show that our method can be successfully used for optimizing the parameterized quantum circuits primarily used in near-term algorithms.
    We further note that information-flow based paths can be used to improve convergence of existing stochastic gradient based methods.
\end{abstract}

\maketitle

\section{Introduction}
Parameterized quantum circuits (PQCs) are a central component of many variational quantum algorithms (VQAs) with applications in quantum chemistry and combinatorial optimization~\cite{bharti2022noisy, cerezo2021variational, anand2022quantum}, such as the variational quantum eigensolver \cite{vqe_cite} and the quantum approximate optimization algorithm \cite{qoao_cite}. 
In addition, variational algorithms have more recently been applied to a number of machine learning tasks, including data classification \cite{mlc_1,vqc_ref_1,vqc_ref_2} and generative modeling \cite{mlg_1,mlg_2,mlg_3,mlg_4,mlg_5,mlg_6,anand2021noise}.
The general concept behind VQAs is to employ a PQC to generate a trial wavefunction on a quantum device. 
The resulting state is then repeatedly measured to estimate expectation values of some Hermitian operator(s) using the current trial wavefunction. 
These expectation values are used to evaluate an objective function that a classical optimizer maximizes or minimizes by varying the parameter values of the PQC.

In recent years, significant progress has been made to better understand PQCs, their design, and their relationship to algorithm performance. 
For example, expressibility and entanglement capability have been proposed as meaningful metrics to compare different PQCs and exclude circuits with limited capabilities \cite{vqc_ansatz}. 
Moreover, expressibility has been correlated with other performance metrics of certain variational quantum algorithms \cite{exp_metric, anand2022exploring}, and high expressibility has been associated with the presence of barren plateaus in cost function landscapes \cite{exp_barren}.
However, there is still a general lack of understanding of how the parameterization of quantum circuits impacts algorithm performance. 

In this work, we propose a novel method to characterize the information flow in PQCs to incorporate correlations between parameters in a quantum circuit.
To quantify the information flow, we first consider a graph representation of quantum circuits to define paths between the two-qubit unitaries present in the quantum circuit.
We then introduce a measure of distance between two points in a circuit by mapping local unitaries into a multi-qubit state and using mutual information between the resulting single-qubit states.
Finally, using the distance metric, we propose a new method for stochastic optimization of variational algorithms using a subset of gate parameters from a selected (random or shortest) path. \color{black} The proposed method fits into a larger family of approaches based on parameter-subset optimization, including sequential single-parameter optimization~\cite{nakanishi2020sequential}, \color{black}layer-wise strategies~\cite{skolik2021layerwise, xiao2022reconstructing, jattana2023improved,campos2021training}, and adaptive machine-learning approaches~\cite{halder2023machine,liu2023training}.\color{black} However, the explored path-based subsets differ from the subsets explored in existing works. Furthermore, the proposed path-based subsets have the advantage that they directly allow for the use of distance-based metrics as physics-inspired heuristics for adaptively guiding the choice of parameters throughout the optimization---as illustrated using the metric proposed in this work. \color{black}

We perform numerical simulations to analyze the performance of the proposed method for two different tasks; ground state energy estimation and binary classification.
Applying our method for optimization of VQAs, we observe that using (shortest) paths consistently outperforms the stochastic gradient-based method. 

The rest of the paper is organized as follows: we present some preliminary information required to understand the article in section~\ref{sec:prelimns}, and details of the method used in the article are presented in section~\ref{sec:method}. 
Next, the results from numerical simulations is presented in section~\ref{sec:simres} and we finally present some concluding remarks in section~\ref{sec:conclusion}. 

\section{Preliminaries}\label{sec:prelimns}
\subsection{Mutual Information}
For a bipartite quantum system on two subsystems $A$ and $B$, the state space can be represented as the tensor product of the individual sub-spaces as: $\mathcal{H}^{AB} = \mathcal{H}^A \otimes \mathcal{H}^B$.
The quantum mutual information $I(A:B)$ is a measure of the amount of correlation between the two sub-systems and is defined as:
\begin{equation}\label{eq:MI}
    I(A:B) = S(\rho^A) + S(\rho^B) - S(\rho^{AB}),
\end{equation}
where $\rho^{AB}$ represents the density matrix of the full system on $ \mathcal{H}^{AB}$, $\rho^{A} = \Tr_B (\rho^{AB})$ and  $\rho^{B} = \Tr_A (\rho^{AB})$ are the reduced state of $\rho^{AB}$ on systems $A$ and $B$ respectively, and $S(\rho) = - \Tr (\rho \log \rho)$ represents the von Neumann entropy of a density matrix $\rho$.


\subsection{Quantum Circuits as Graphs}\label{sec:qcandp}
A quantum circuit is a sequence of operations that can be used to prepare a state of interest on a quantum device.
It typically comprises of simple unitary operations called quantum gates that depend on some parameters. 
The action of the circuit $U(\theta)$ on an initial state $\ket{\psi_0}$ \color{black} is to prepare \color{black} the state $\ket{\psi(\theta)} = U(\theta) \ket{\psi_0}$.

\begin{figure}[htbp]
\centering
\subfloat[An illustration of the graph representation of single and two qubit gates.]{\includegraphics[width=0.99\columnwidth]{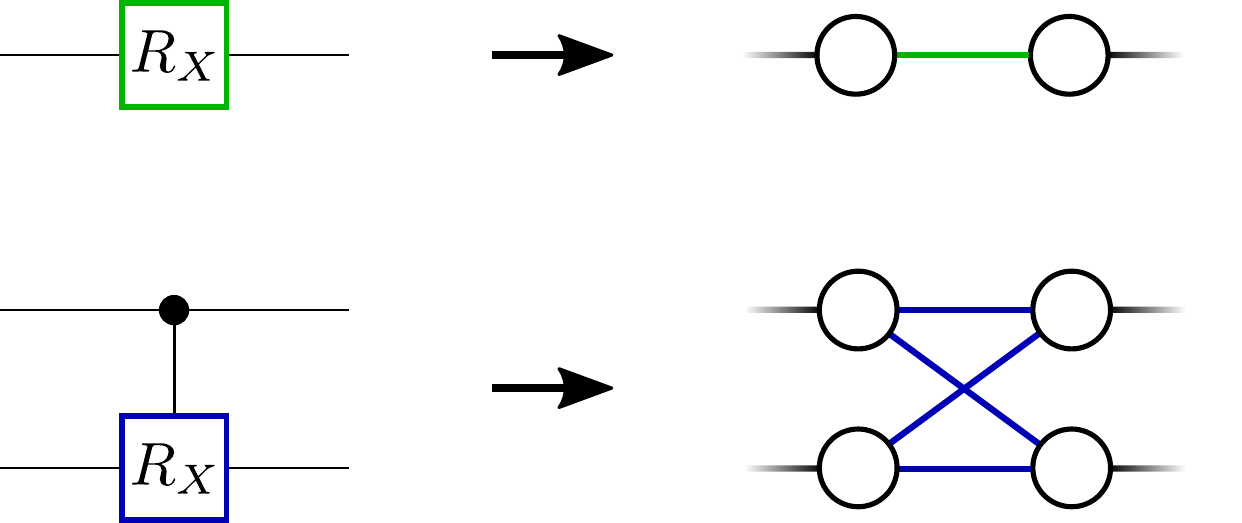}} \\
\subfloat[An illustration of the graph representation of the full circuit.]{\includegraphics[width=0.99\columnwidth]{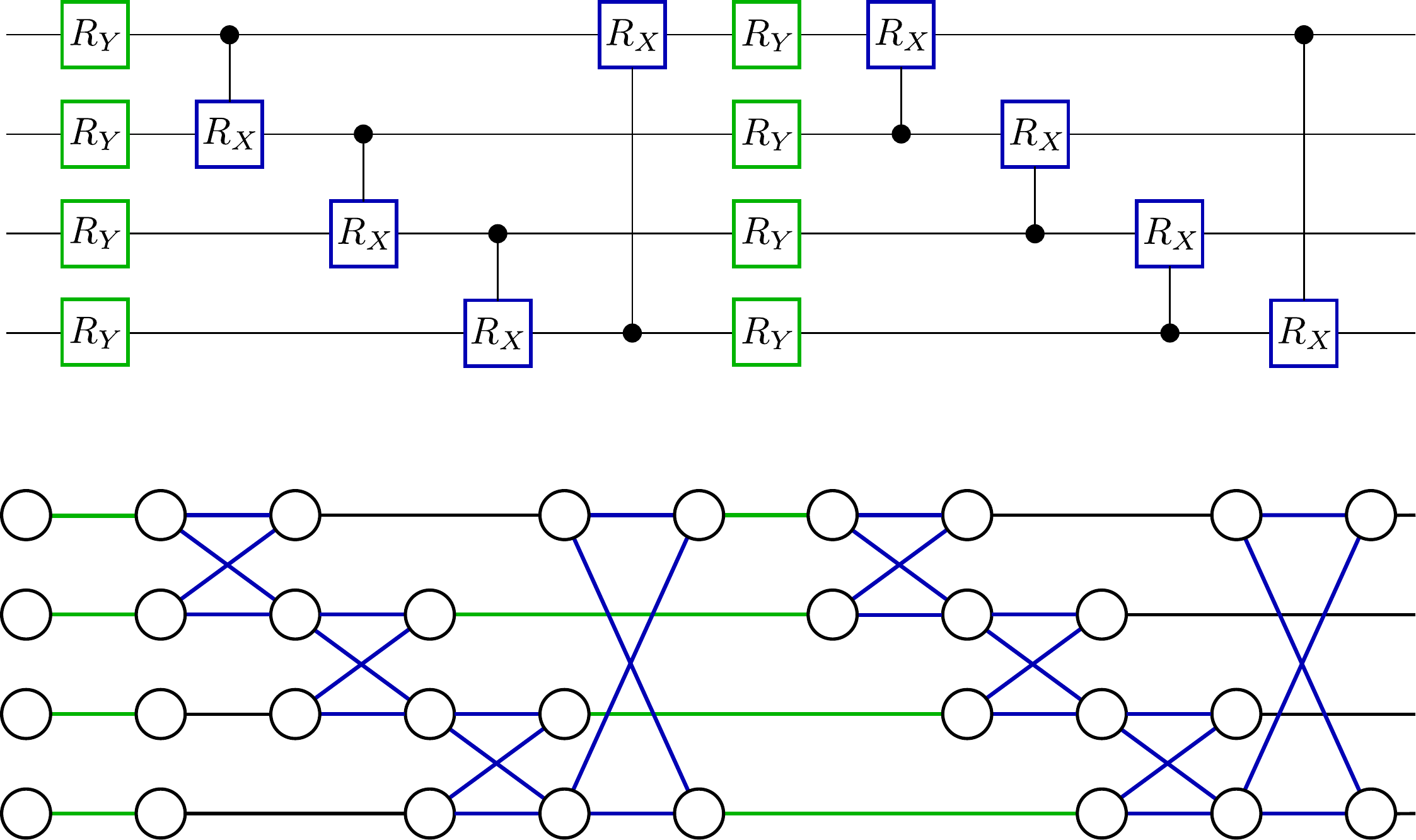}}
\caption{A figure showing the graph representation of a given circuit.}
\label{fig:cir_graph_conversion}
\end{figure}

We develop a simple strategy for converting a quantum circuit into a directed graph by representing every quantum gate as edges in the graph and the states at different moments in the circuits as nodes.
Using this picture, a single qubit gate is represented as an edge that connects two nodes, while a two-qubit gate is represented as four edges between four nodes.
The nodes are then arranged as per the time steps in the quantum circuit.
The open edges represent the initial state and the \color{black} final \color{black} state of the qubits.
A simple illustration of this conversion is shown in Fig.~\ref{fig:cir_graph_conversion}.
\newline 

\subsection{Information Flow}
\begin{figure}[htbp]
  \centering
   \includegraphics[width=0.99\columnwidth]{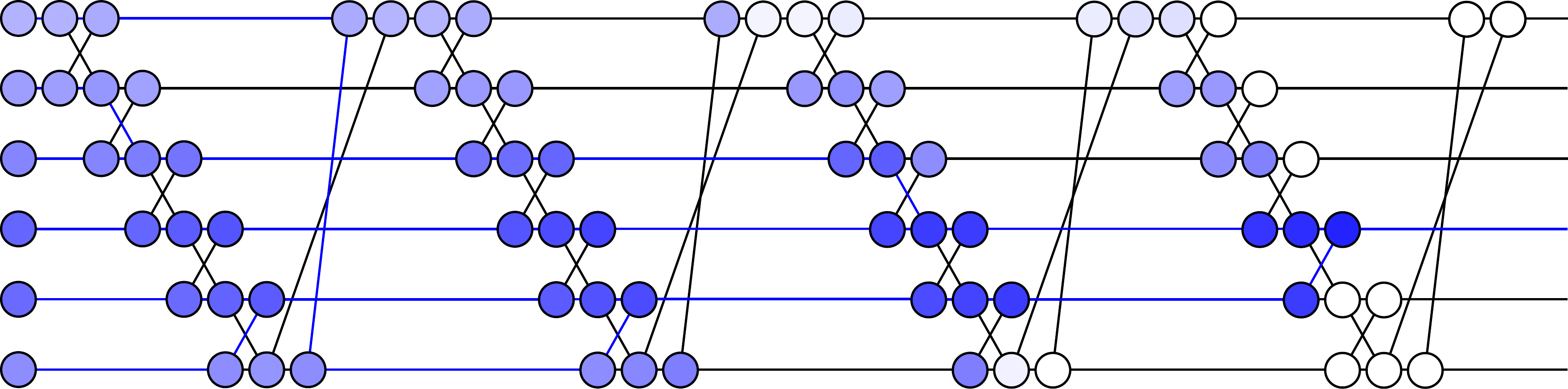}
   \caption{A figure showing the causal cone and the various paths in a graph representation of a circuit. The nodes present in the causal cone of the observable on qubit 4 are colored blue, and the color gradient represent the distance (darker implies smaller).  The blue lines denote \color{black} a subset of the \color{black} different paths within the causal cone.}
   \label{fig:cir_cones_paths}
\end{figure}

Information flow is defined as the transfer of information from one variable to another in a process.
In the context of quantum circuits, this can be regarded as the spread of correlation between qubits as a result of multi-qubit gates. 
Analogously one can think of it as the spread of correlation between gates in their corresponding causal cones.
The spread of correlation has been considered in previous studies~\cite{zhang2021mutual, zhang2021variational} to design better ansatz for variational algorithms.
In this work, we use the notion of information flow (spread of correlation) to define paths through the circuit, using which we can control the spread of correlation between qubits.
We combine the concept of causal cones with paths and utilize the graph representation of a parameterized quantum circuit to identify and control the flow of information (correlation) between gates. 
An illustration of these concepts is depicted in Fig.~\ref{fig:cir_cones_paths}.
\newline 

\subsection{Optimization of variational quantum algorithms}
In variational quantum algorithms we usually optimize the parameters of states prepared by a PQC, $U(\theta)$, in order to minimize an objective function of the form:
\begin{equation}
   f(\theta) = \langle H \rangle_{\theta} =  \sum_{j=1}^{M} \langle H_j \rangle_{\theta},
\end{equation}
where $H$ represent a problem Hamiltonian, which is a linear combination on $M$ Pauli terms $H_j$.
The optimization of the objective function is carried out by iteratively updating the parameters of the quantum circuit as:
\begin{equation}\label{eq:update_rul}
    \mathbf{\theta}^{(t+1)}_i = \mathbf{\theta}^{(t)}_i - \alpha \frac{\partial f(\mathbf{\theta})}{\partial \mathbf{\theta}_i},
\end{equation}
where $\alpha$ is the learning rate and $\frac{\partial f(\mathbf{\theta})}{\partial \mathbf{\theta}_i}$ denotes the partial derivative of the objective function with respect to the variable $\mathbf{\theta}_i$.
The analytical gradient can be calculated using a K-term parameter-shift rule~\cite{schuld2019evaluating, mari2021estimating, kottmann2021feasible, wierichs2022general} as:
\begin{equation}\label{eq:shiftrule}
    \frac{\partial f(\mathbf{\theta})}{\partial \mathbf{\theta}_i}  =  \sum_{j=1}^{M} \frac{\partial \langle H_j \rangle_{\theta}}{\partial \mathbf{\theta}_i} = \sum_{j=1}^{M}  \sum_{k=1}^{K} \gamma_{k,i} \langle H_j \rangle_{\theta_{k,i}}, 
\end{equation}
where, $\gamma_{k,i}$ is the $K-$term coefficient in the shift rule.
The calculation of the analytical gradient requires a large number of measurements for a single parameter update, as every gradient calculation requires $K$ objective evaluations, which in turn requires a large number of measurements (shots) for accurate estimation of each objective.
To overcome this, $n$-shot stochastic gradient descent was proposed in Ref.~\cite{harrow2021low, sweke2020stochastic} where one uses $n$-shot estimators of the gradient instead of the exact ones,
\begin{equation}
    \frac{\partial f(\mathbf{\theta})}{\partial \mathbf{\theta}_i}  = \sum_{j=1}^{M}  \sum_{k=1}^{K} \gamma_{k,i} \tilde{h}_j^{(n)}({\theta_{k,i}}), 
\end{equation}
where, $\tilde{h}_j^{(n)}({\theta_{k,i}})$ is the $n$-sample mean estimator of $\langle H_j \rangle_{\theta_{k,i}}$.
The parameters are then updated similarly to Eq.~\ref{eq:update_rul}.

\subsection{Variational Quantum Classification}\label{vqc_intro}
Variational quantum classifiers (VQC) are quantum circuits that are trained for supervised learning tasks \cite{vqc_ref_1,vqc_ref_2}. There exist several strategies to design a quantum classifier, including \color{black}generalizations of \color{black} well-known
classical machine learning techniques such as artificial neural networks \cite{nn_1, nn_2} or kernel methods~\cite{kernel_1,kernel_2}. 

\begin{figure}[htbp]
  \centering
   \includegraphics[width=0.8\columnwidth]{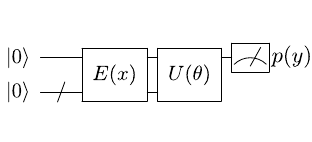}
   \caption{Structure of an $n$-qubit variational quantum binary
classifier: a state preparation circuit $E(\boldsymbol x)$ encoding the input $\boldsymbol x$ into
the amplitudes of a quantum system, a model circuit $U(\theta)$, and
a single qubit measurement. The measurement retrieves the
probability $p(y)$ of the model predicting 0 or 1, from which the binary prediction can be inferred. The classification
circuit parameters $\theta$ are trained by a
variational scheme.}
   \label{fig:vqc_structure}
\end{figure}
In this study, the circuit-centric architecture in Fig.~\ref{fig:vqc_structure} is used to study a binary classification problem \cite{vqc_ref_2}.
The general objective is to train the VQC on a data set $\{\boldsymbol x_i, y_i\}_{train}$ to find a mapping between input $\boldsymbol x_i$ and label $y_i$. The trained parameterized quantum circuit can then be used as a black box to predict labels $\hat{y}_i$ for a given set of test inputs $\{\boldsymbol x_i, y_i\}_{test}$.

The circuit used for this classification task consists of three distinct parts: The state preparation circuit, the model circuit, and a measurement scheme. Preparing a quantum state that embeds some classical data is achieved by applying a static quantum routine on the initial ground state, $\ket{\Phi(x)} = E(x)\ket{0}$. \color{black} Here, we use \color{black} the so-called basis encoding method
\begin{align}
    \ket{\Phi(x)} = \bigotimes_{i=0}^n   \ket{x_i} & & \color{black} x\in \mathbb{B}^n \color{black}
\end{align}

\begin{figure}[htbp]
  \centering
   \includegraphics[width=0.99\columnwidth]{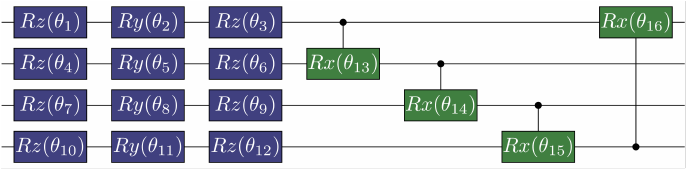}
   \caption{Gate composition of a single layer of the ansatz used in the model circuit $U(\theta)$ for the classification task.}
   \label{fig:vqc_1l_ansatz}
\end{figure}

\color{black} to encode \color{black} the input data $\boldsymbol x$ as computational basis states. 
The prepared state $\ket{\Phi(x)}$ is then further processed with a given parametrized quantum circuit $U(\theta)$, resulting in a state $\ket{\psi_{vqc}(\boldsymbol x,\theta)}= U(\theta)E(\boldsymbol x)\ket{0}$. The ansatz structure used in this study contains parameterized
single and two-qubit gates with trainable parameters. 
The circuit is shown in Fig.~\ref{fig:vqc_1l_ansatz}. 
The output state $\ket{\psi_{vqc}(\boldsymbol x,\theta)}$ is finally measured using the Pauli-Z operator on a qubit (we chose the first qubit) to obtain the expectation value $\mathbb{E}(\sigma_z)$. This yields the predicted label  $(\hat{y}_i)$.

The parameters of the variational block are trained by minimizing the square loss cost function

\begin{align}
f(\theta) &= \frac{1}{n'}\sum_{i=0}^{n'} \left( y_i - \hat{y}_i(\theta) \right)^2
\end{align}
with $n'$ as the training set size and the predicted label output of the VQC as
\begin{align}
\hat{y}_i(\theta) &= \bra{\psi(x_i,\theta)} \hat{Z} \otimes \mathbb{I}^{\otimes (n-1)}\ket{\psi(x_i,\theta)}.
\end{align}

The binary classification problem studied in this paper is the $n$-bit parity problem. The corresponding dataset contains $2^n$ distinct binary vectors, where each label indicates whether the sum of the $n$ components of the binary vector is odd or even. 
The Boolean $n$-bit parity function to be modeled is
\begin{align}
    f : \{0,1\}^{\otimes n} \xrightarrow{} \{0,1\},
\end{align}
with the property that $f(x)=1$ if the number of ones in the vector $x\in \{0,1\}^{n}$ is odd and zero otherwise.

In what follows, we describe the details of how to quantify information flow for designing algorithm to optimize variatonal algorithms.

\section{Method}\label{sec:method}
\subsection{Measure of distance}\label{sec:mod}
A given unitary $\hat{U}$ on two qubits can be conveniently described using a $4 \times 4$ matrix $U_{(a,b),(c,d)}$ with entries:
\begin{align}
U_{(a,b),(c,d)} &= \left< c,d \right| \hat{U} \left| a,b \right> & & a,b,c,d \in \{0,1\} \; ,
\end{align}
where the indices $(a,b)$ are thought of as a single unified row-index, and $(c,d)$ similarly plays the role of a unified column-index. 
Using this description, a four qubit state corresponding to the unitary transformation can be defined:
\begin{align}
\left| \psi_{U} \right> &= \frac{1}{\mathcal{N}} \sum_{a,b,c,d \, \in \, \{0,1\}} U_{(a,b),(c,d)} \left|a,b,c,d\right>  ,
\label{eq:Mapping_to_States}
\end{align}
where $\mathcal{N}$ is a normalization-constant. 

Using this state, we can now define reduced density matrices of different subsystems of qubits. For instance, we can define the single- and two-qubit reduced density-matrices:
\begin{align*}
\rho^a &= \text{Tr}_{\{b,c,d\}} \left( \left| \psi_U \right> \left< \psi_U \right| \right)\\
\rho^{ab} &= \text{Tr}_{\{c,d\}} \left( \left| \psi_U \right> \left< \psi_U \right| \right)  ,
\end{align*}
where the subscripts on the trace indicates the degrees of freedom that are traced over.

\begin{figure}[htbp]
  \centering
   \includegraphics[width=0.8\columnwidth]{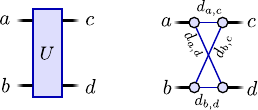}
   \caption{An illustration of the definition of labels on the unitary operator, as well as the corresponding labeling of the metric distances. Image adapted from Ref.~\cite{hyatt2017extracting}.}
   \label{fig:Label_Definition}
\end{figure}

Using this description of the unitary, Ref.~\cite{hyatt2017extracting} proposed the following distance metric across the legs of the unitary (see Fig. \ref{fig:Label_Definition}):
\begin{align}
d_{i,j} &= - \log \left( \frac{I(i:j)}{2 \log\left(2\right)} \right) ,
\end{align}
where $I(i:j)$ is the mutual information between the sites $i$ and $j$ as defined in Eq.~\ref{eq:MI}.

\begin{figure}[htbp]
  \centering
   \includegraphics[width=0.95\columnwidth]{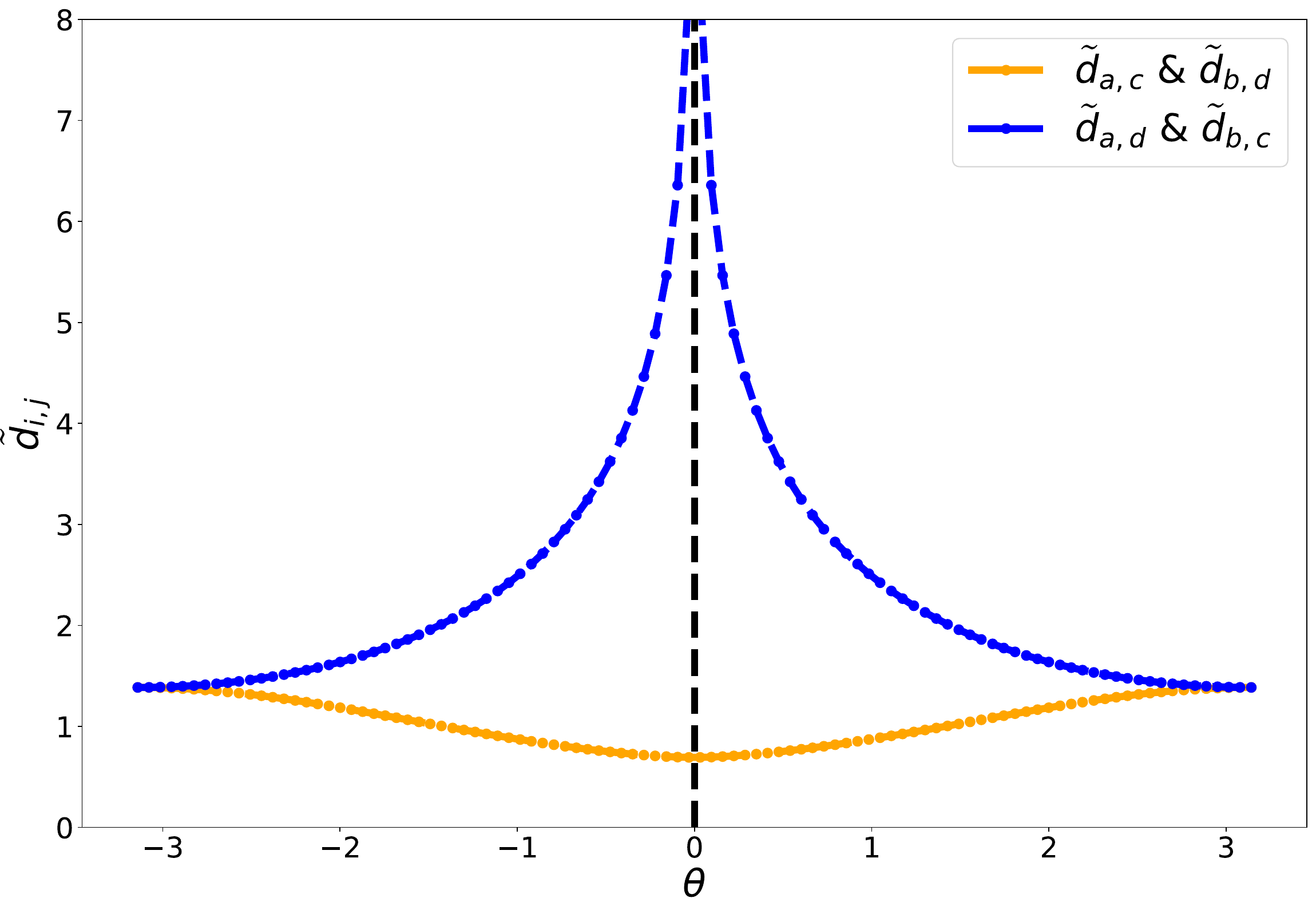}
   \caption{A plot of the distance vs. the parameter value for a C-Ry($\theta$) gate.}
   \label{fig:distance}
\end{figure}

In this article we introduce a further modification to the distance metric as:
\begin{align}
    \tilde{d}_{i,j}=\begin{cases}
         - \log \left( \frac{I(i:j)}{4 \log\left(2\right)} \right) \quad \text{if } (i,j) \in \{(a,c), (b,d)\} \\
        \\
    - \log \left( \frac{I(ac:bd)}{4 \log\left( 2 \right)} \right) \quad \text{else}
    \end{cases}
    \label{eq:metric}
\end{align}
where, $I(ij:kl)$ is defined as:
\begin{align}
    I(ij:kl) &= S\left(\rho^{ij}\right) + S\left(\rho^{kl}\right) - S\left(\rho^{ijkl}\right) \nonumber \\ 
            &= S\left(\rho^{ij}\right) + S\left(\rho^{kl}\right) .
\end{align}
\color{black}
This modification was motivated mainly by a need to more accurately treat the small class of gates relevant in a PQC context---see Appendix~\ref{ap:Motivation} for details. \color{black} To develop some intuition about this metric, we present an illustration of the distance between the legs of a two qubit gate, C-Ry($\theta$), in Fig.~\ref{fig:distance}.
As expected, we observe that the distance (weight of the diagonal edges) approaches infinity when theta equals zero because the two qubit gate at this value acts as identity.

\begin{figure*}[htbp]
\centering
\includegraphics[width=0.95\linewidth]{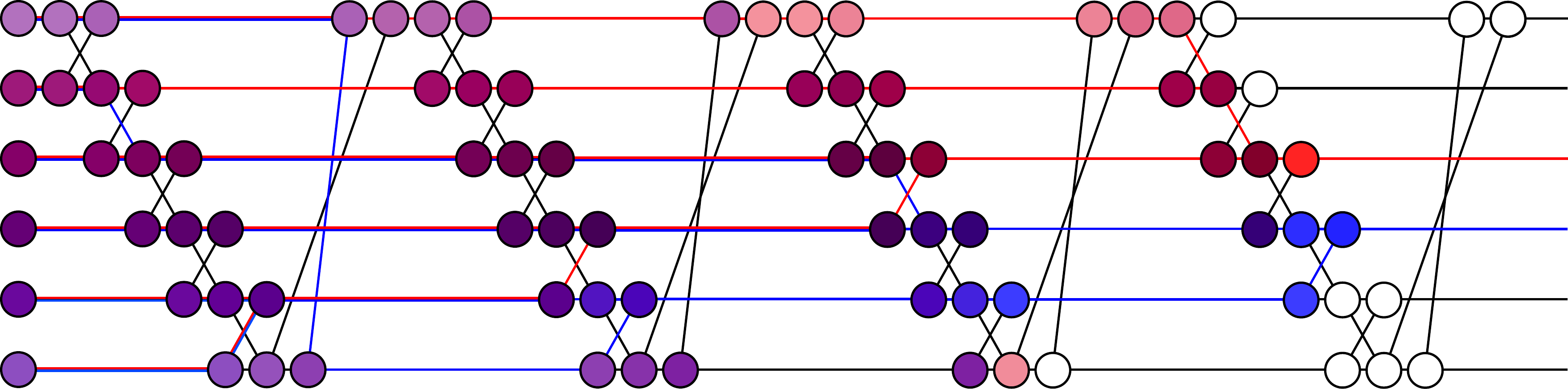}
\caption{A figure showing the different path options in the causal cone of an observable. The intensity of the colors (darker implies smaller) denote the qualitative distance from the last point on the qubit to measured.}
\label{fig:cir_diff_paths}
\end{figure*}

\subsection{Paths}\label{sec:path}
The objective functions $f(\theta)$ for commonly used quantum algorithms are constructed as a linear combination of expectation values as:
\begin{equation}
    f(\theta) = \sum_{j=1}^{M} \langle H_j \rangle_{\theta},
\end{equation}
where $H_j$ represent individual Pauli-strings in the Hamiltonian defining the problem.
These individual Pauli-strings usually act on the state of a subset of the full qubits, and their expectation value thus depends on the gate parameter in the causal cone of these qubits~\cite{benedetti2021hardware}.
We define different paths in the causal cone of the qubits to be measured by converting the quantum circuits into graphs as described in section~\ref{sec:qcandp}.
An illustration of the choices of paths in the causal cone is shown in Fig.~\ref{fig:cir_diff_paths}.
 
We further introduce the notion of shortest paths by adding weights to the edges in the graph representation of the circuit according to the distance measure proposed in section~\ref{sec:mod}.
This is inspired from Ref.~\cite{hyatt2017extracting} where the authors use the length of geodesics for disentangling a quantum state.
We use the \textsc{networkx} package~\cite{hagberg2008exploring} to create the graph representation of the quantum circuit and for finding different paths between nodes in the graph representation.

\subsection{Optimization with paths}
In this section, we present a strategy for optimizing variational algorithms based on paths in the causal cone of the individual Pauli-strings used to define the objective function.
We use two different strategies of choosing sets of parameters by randomly sampling paths from either the set of all possible paths or the set of shortest paths. 
The overall algorithm is presented in Algorithm~\ref{al:stoc_op}.

\begin{algorithm}[htbp!]\label{al:stoc_op}
\SetAlgoLined
	\textbf{Input}: path selection strategy, problem Hamiltonian and ansatz, learning rate\\
	\textbf{Output}: estimate of the ground state energy of the Hamiltonian\\
	\textbf{Initialize}: Generate the circuit graph corresponding to the ansatz \\
     \While{not converged}
     {  
        0. Initialize an empty list, $\mathcal{L}$ \\
        \For{term in Hamiltonian}
        {   
            1. Calculate all paths for that term \\
            2. Select a path based on the strategy (random or shortest) \\
            3. Add  all the parameter ids in the path which are not already in the list $\mathcal{L}$ to it\\
        }
        4. Update all parameters corresponding to ids in  $\mathcal{L}$ using the matching gradients and learning rate by applying Eq.~\ref{eq:update_rul}
     }
     \Return Estimate of the energy of the Hamiltonian using the ansatz and optimized parameters
 \caption{An outline of the path optimization algorithm.}
\end{algorithm}

One can also modify the presented algorithm to add more stochasticity by randomly sampling the Hamiltonian terms to optimize in each step~\cite{kivlichan2019phase, sweke2020stochastic} or by updating parameters before calculating paths for terms. 
We leave these variations for future studies.


\section{Simulation and Results}\label{sec:simres}
In this section, we numerically demonstrate the applications of the proposed algorithm for training variational quantum algorithms.
The training is implemented in \textsc{Tequila}~\cite{kottmann2020tequila}, an open-source python package which uses \textsc{Qulacs}~\cite{suzuki2021qulacs} as the backend for the execution of all the numerical simulations.
We also used the \textsc{Pennylane}~\cite{bergholm2018pennylane} package for running some of the numerical simulations.
We first present the details of the experiments for finding ground state energies.

\subsection{VQE - XXZ-Heisenberg model}
We use the VQE framework to find the ground state energy of the XXZ-Heisenberg model.
The Hamiltonian of such a system can be written as follows:
\begin{equation}
    \hat{H} = \sum_{\langle i,j\rangle} \left(  - J_X \hat{X}_{i} \hat{X}_{j} - J_Y \hat{Y}_{i} \hat{Y}_{j} - J_Z \hat{Z}_{i} \hat{Z}_{j} \right) - h \sum_{k} \hat{Z}_k ,
\end{equation}
where $\hat{X}, \hat{Y} , \hat{Z}$ are the Pauli matrices, $\langle i,j\rangle$ denotes all the pairs of adjacent lattice sites, and $J_X, J_Y, J_Z$ are the coupling constant and $h$ on the right represents the external magnetic field. 
The coupling constants for the XXZ-model follow the relation:  $ J = J_X = J_Y \ne J_Z \equiv \Delta$.
For all the experiments, we fix the values of the different constants to $h=0$ (no external magnetic field), $\Delta = -20.0$ and $J = 1.0$.
This corresponds to the model having a ferromagnetic ground state. \color{black} As illustrated in Appendix~\ref{ap:more_heisenberg}, qualitative behavior is similar in other parameter regimes. \color{black}
We use \color{black}the XXZ-Heisenberg \color{black} model due to the fact that all the terms in the Hamiltonian depend on only the state of two qubits, e.g. $\hat{X}_i\otimes\hat{X}_j$, and thus our method can \color{black} potentially \color{black} be very useful in reducing the cost of parameter updates in every iteration of the optimization\color{black}, as investigated in more detail below.\color{black}

\begin{figure*}[htbp!]
    \centering
    \begin{tabular}{c c c c}
    \toprule
    \multicolumn{4}{c}{\textbf{1. 6 Qubit Ansatz}}\\
    \midrule
    a) 1 layer & b) 2 layers & c) 3 layers & d) 4 layers\\
    \midrule
    \includegraphics[width=0.24\textwidth]{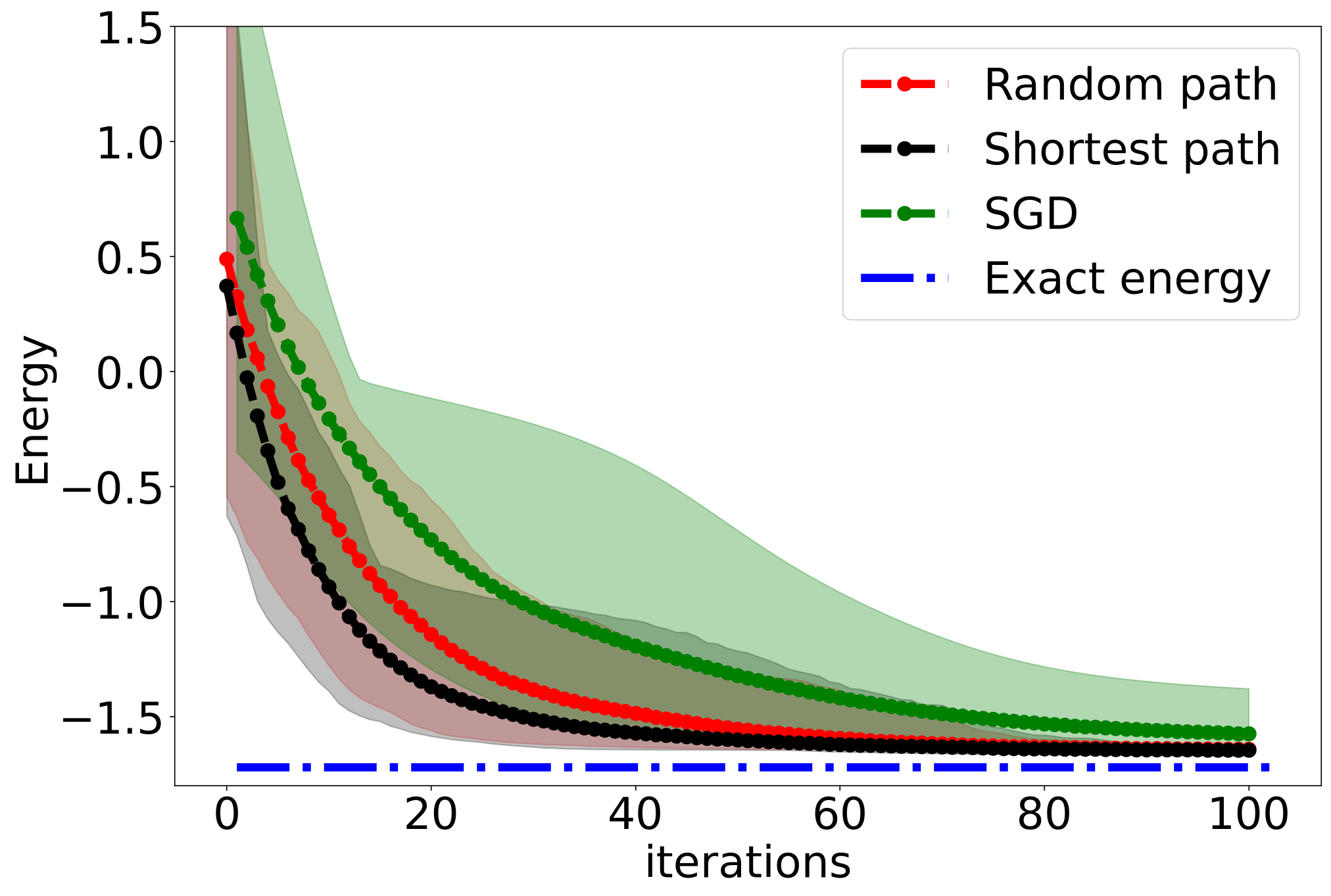} & 
    \includegraphics[width=0.24\textwidth]{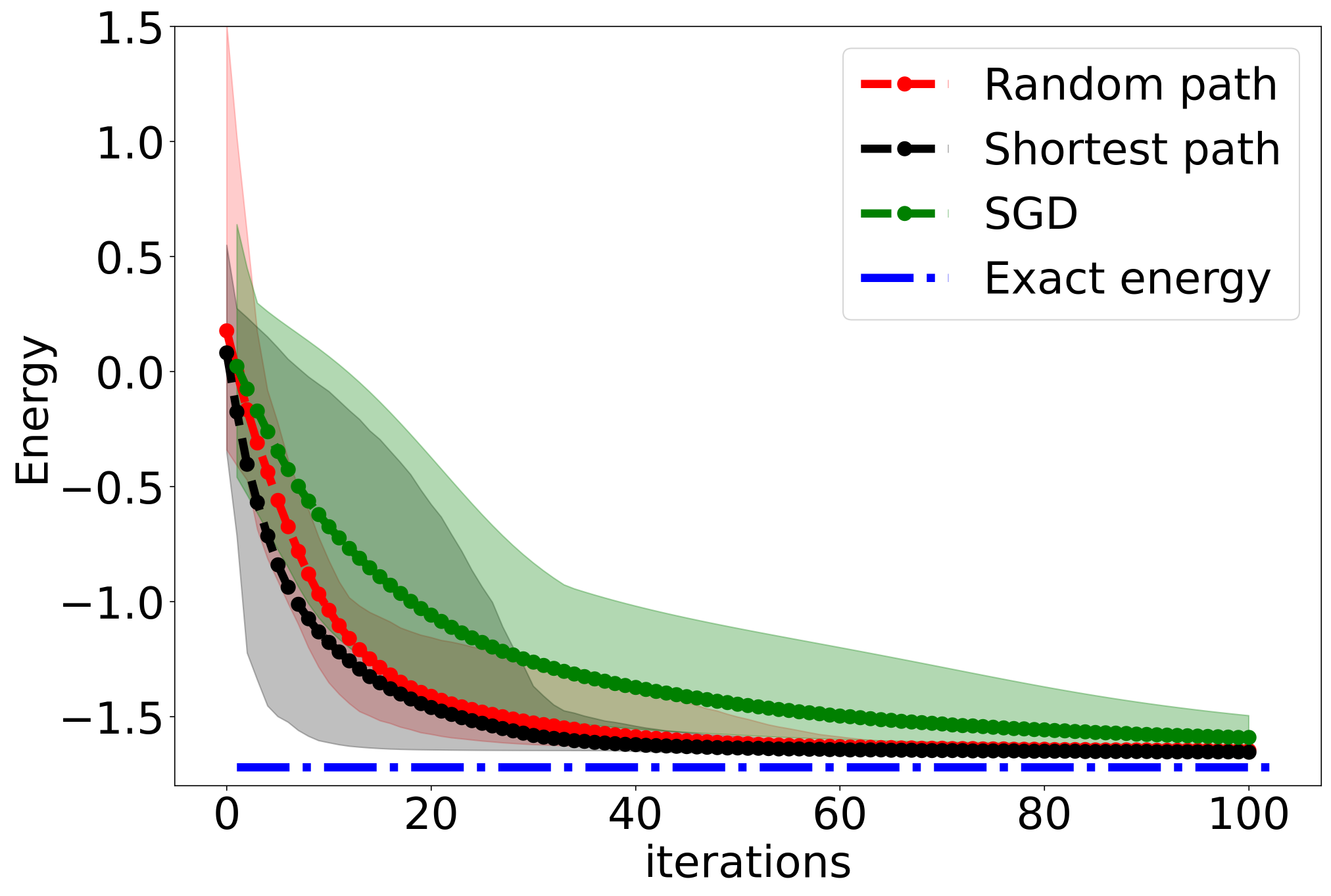} &
    \includegraphics[width=0.24\textwidth]{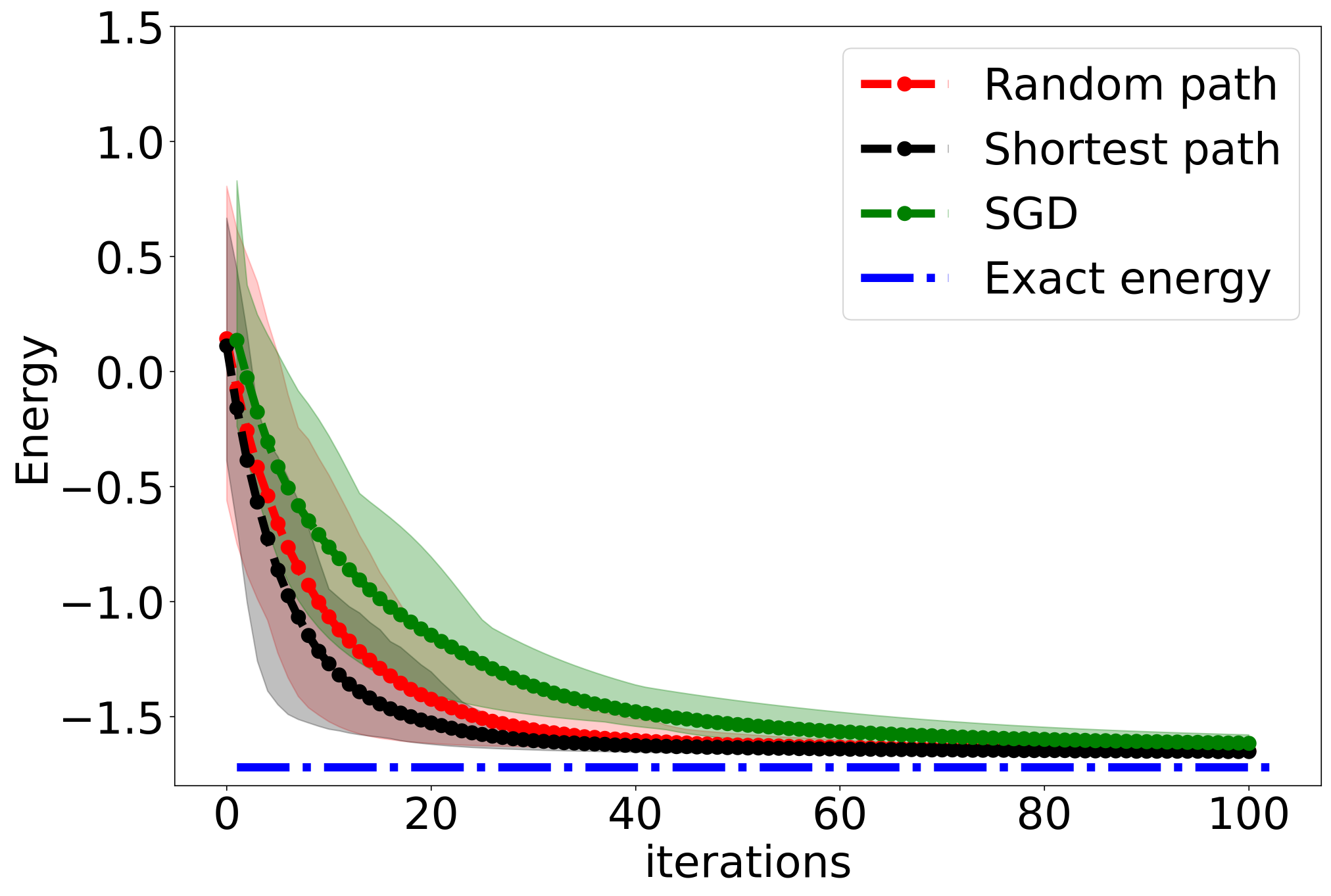} &
    \includegraphics[width=0.24\textwidth]{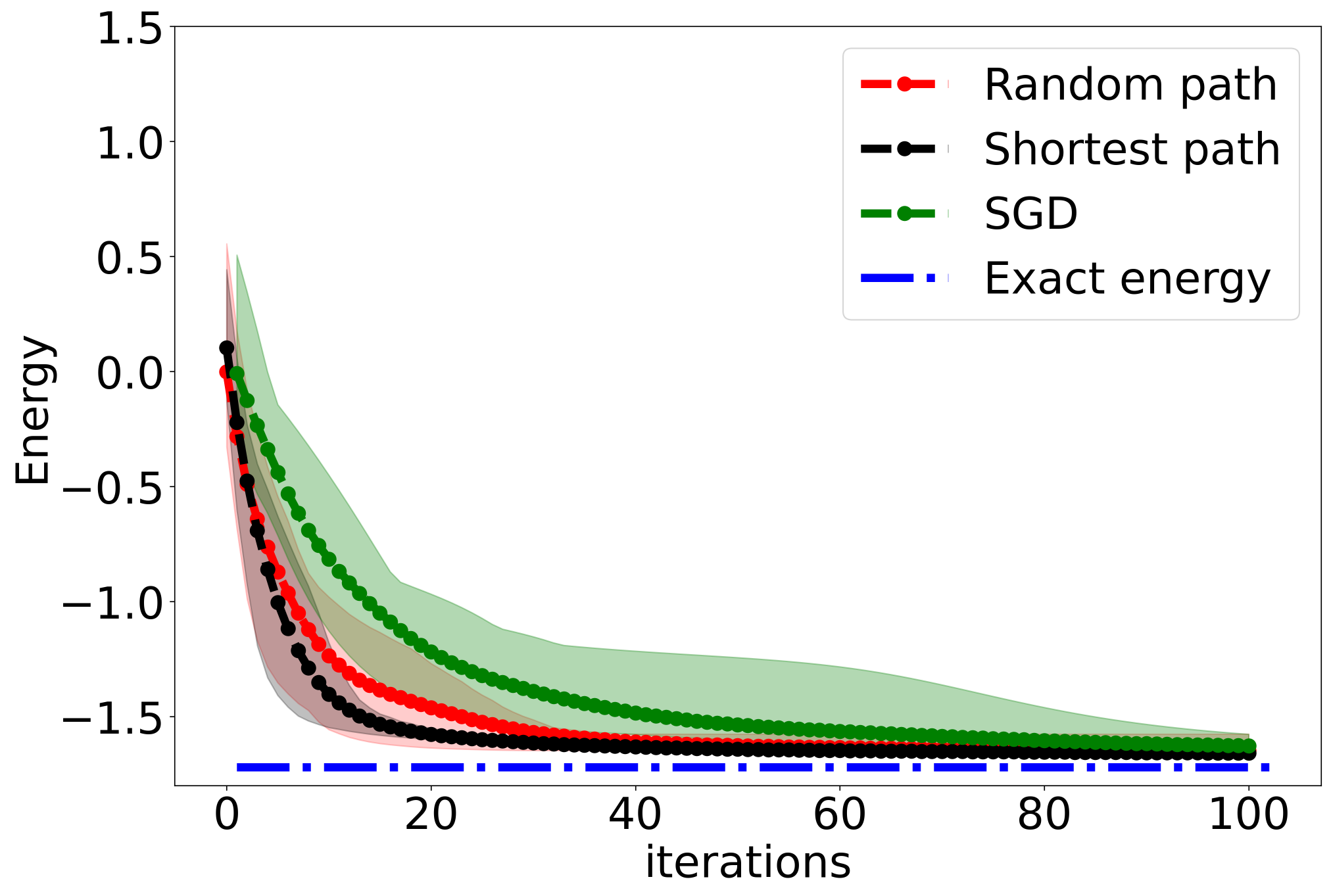}\\
    \midrule
    \multicolumn{4}{c}{\textbf{2. 8 Qubit Ansatz}}\\
    \midrule
    a) 1 layer & b) 2 layers & c) 3 layers & d) 4 layers\\
    \midrule
    \includegraphics[width=0.24\textwidth]{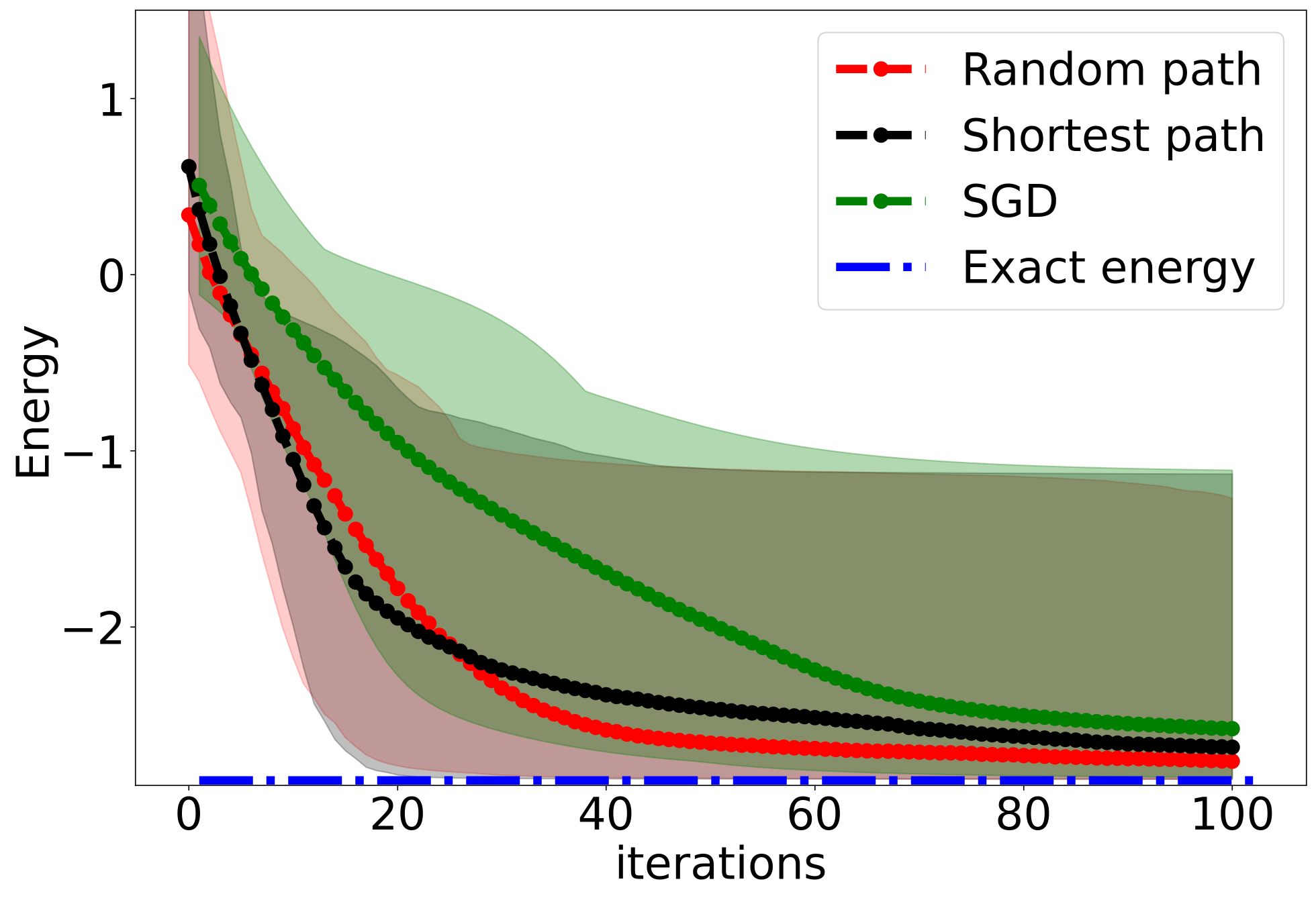} & 
    \includegraphics[width=0.24\textwidth]{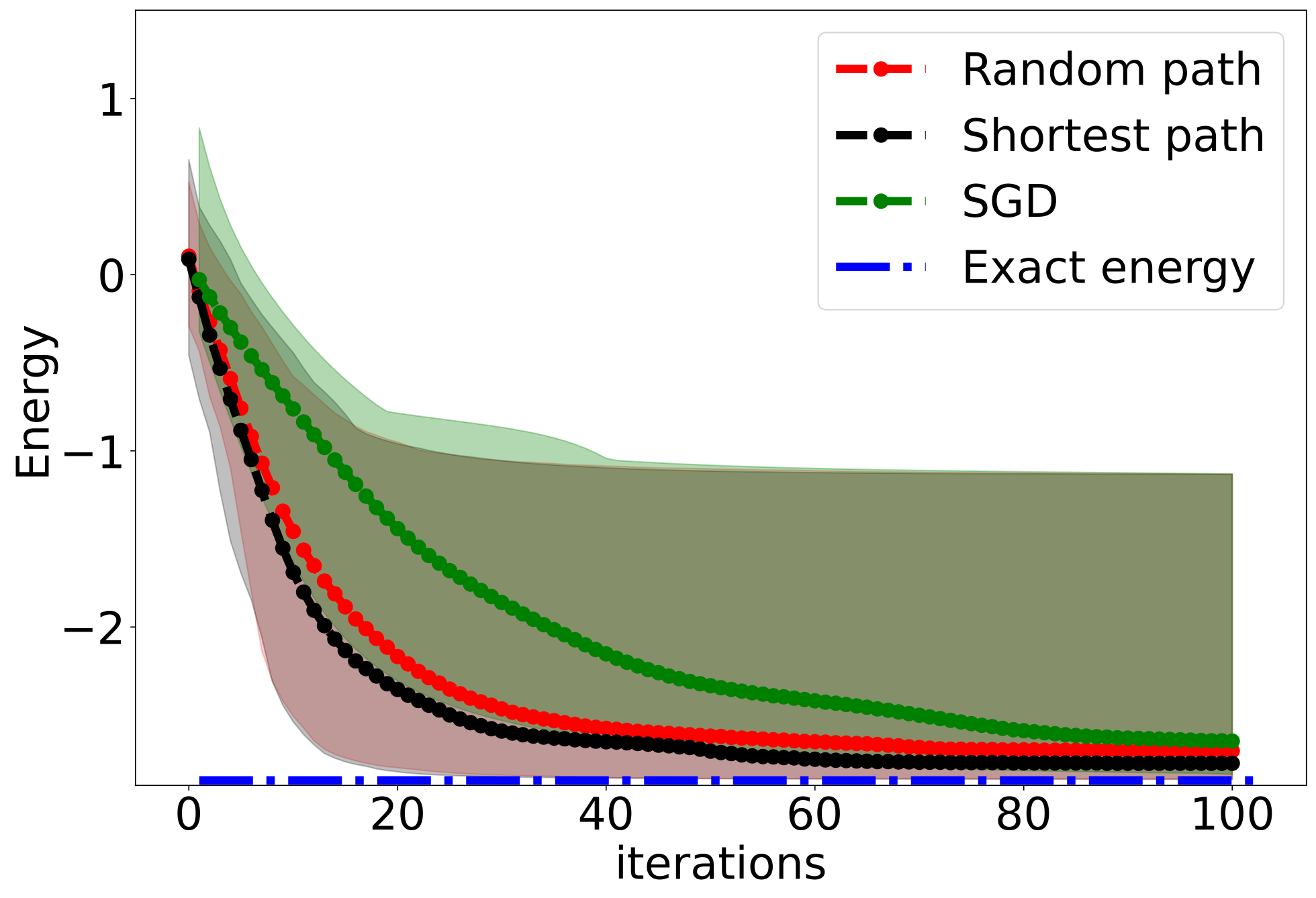} &
    \includegraphics[width=0.24\textwidth]{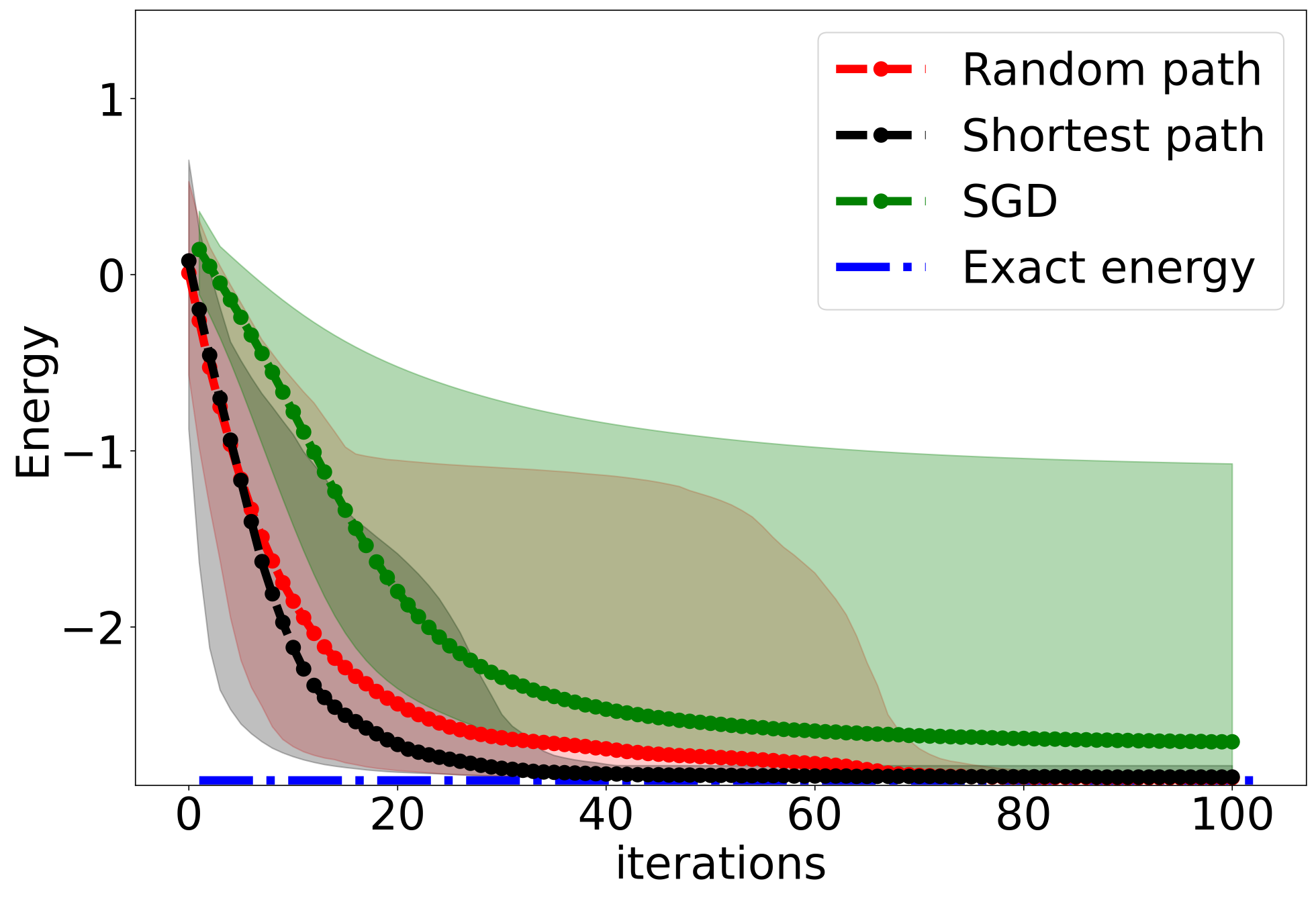} &
    \includegraphics[width=0.24\textwidth]{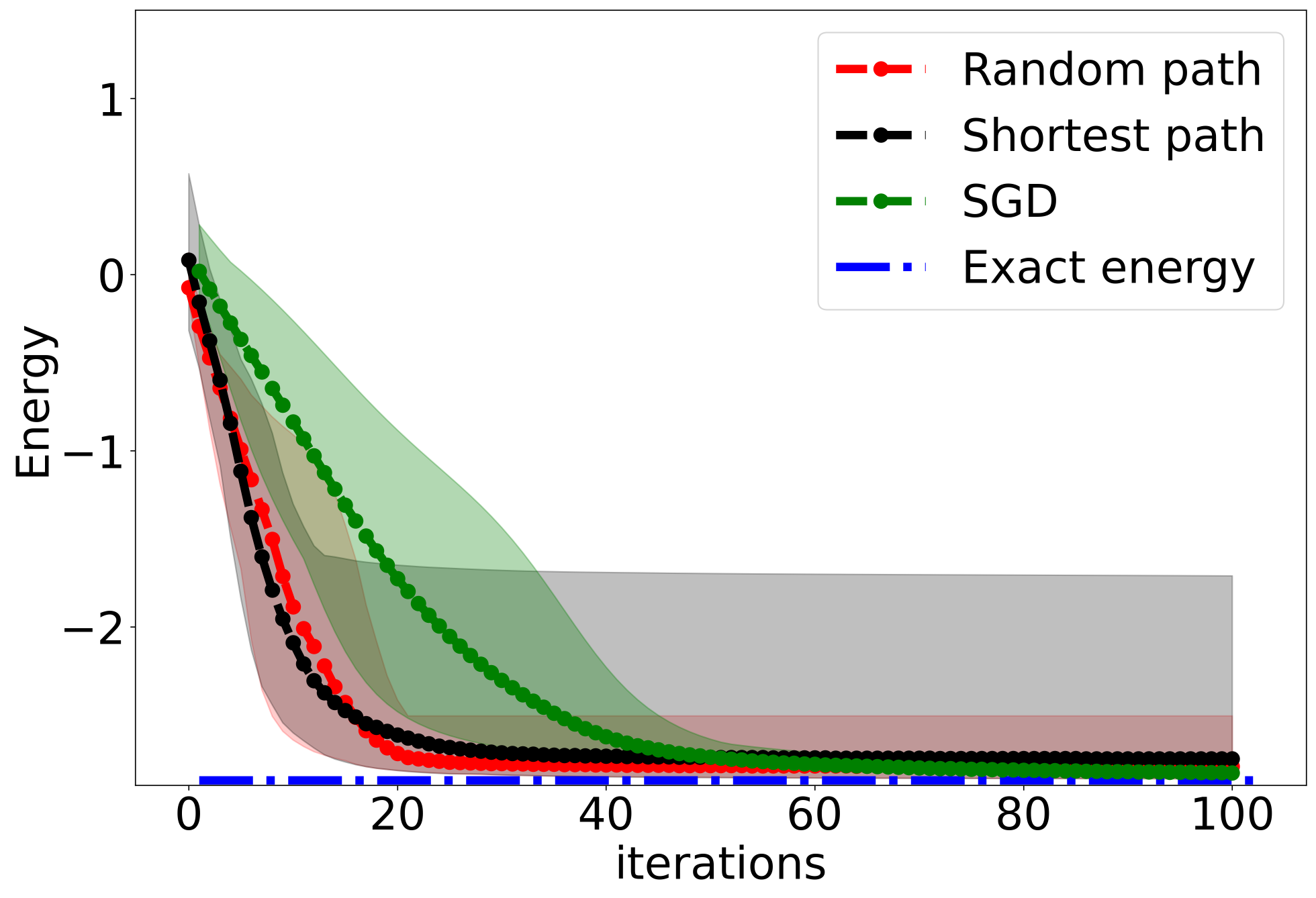}\\
    \end{tabular}
    \begin{tabular}{c c c }
    \midrule
    \multicolumn{3}{c}{\textbf{3. 10 Qubit Ansatz}}\\
    \midrule
    a) 1 layer & b) 2 layers & c) 3 layers \\
    \midrule
    \includegraphics[width=0.275\textwidth]{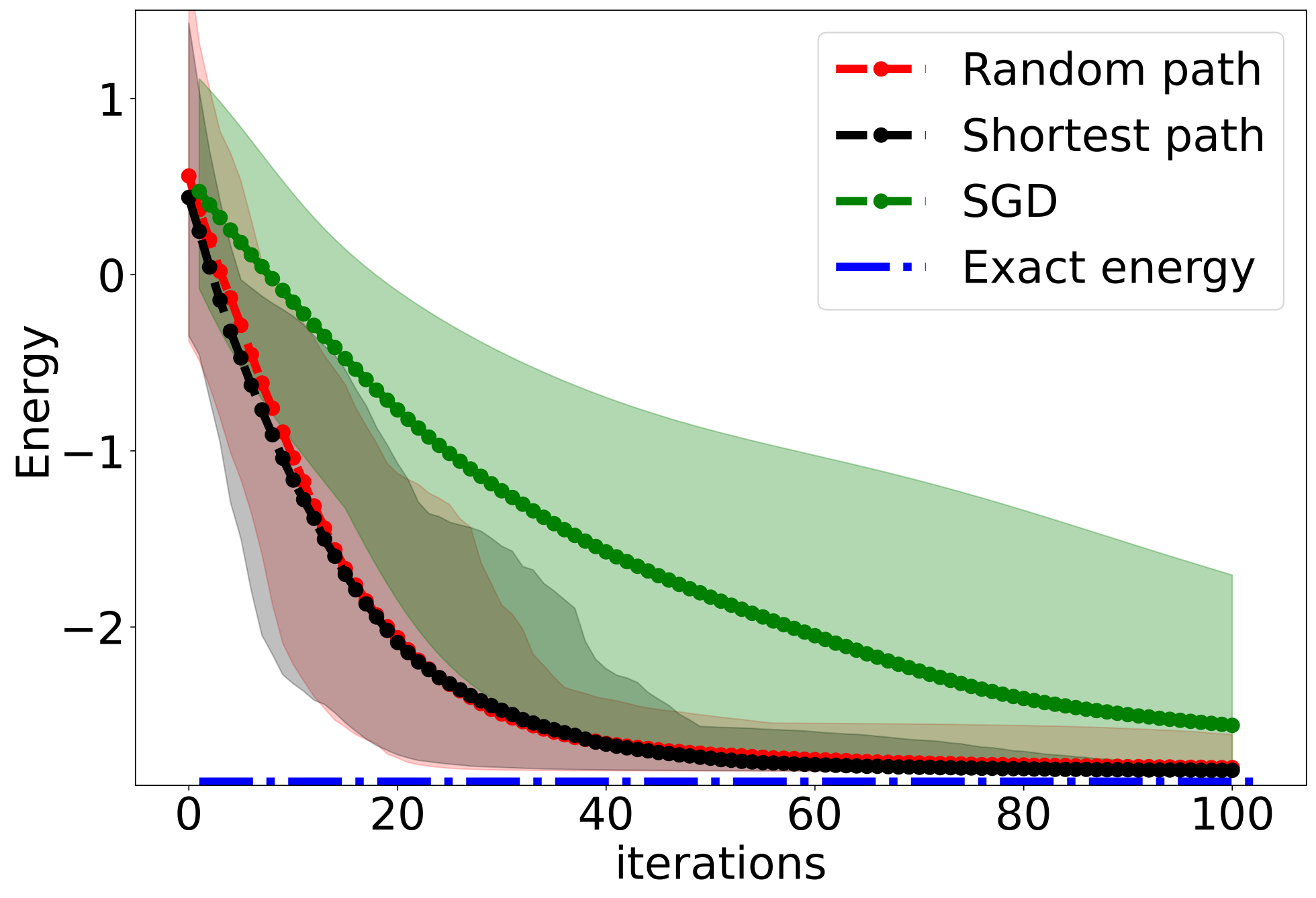} & 
    \includegraphics[width=0.275\textwidth]{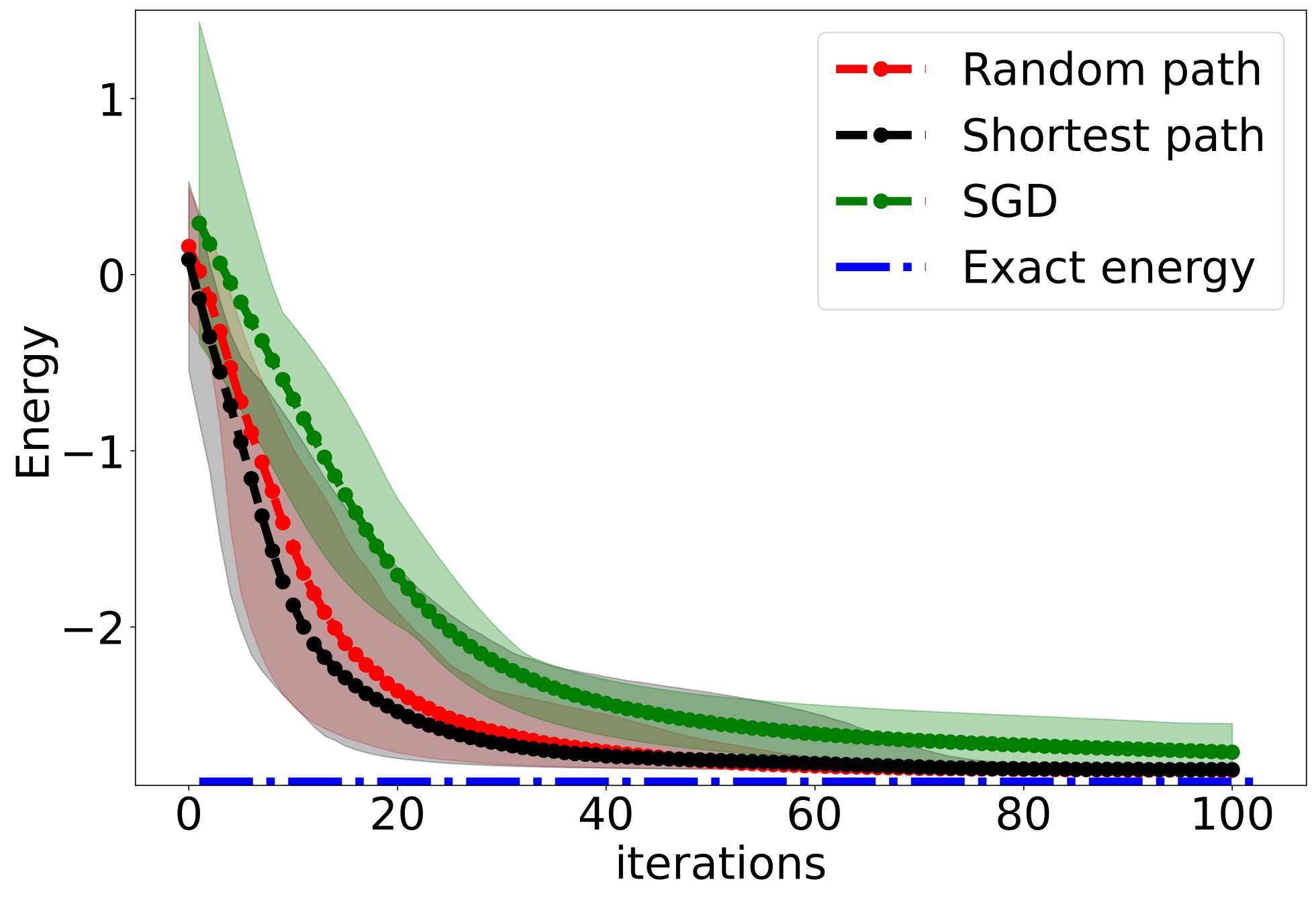} &
    \includegraphics[width=0.275\textwidth]{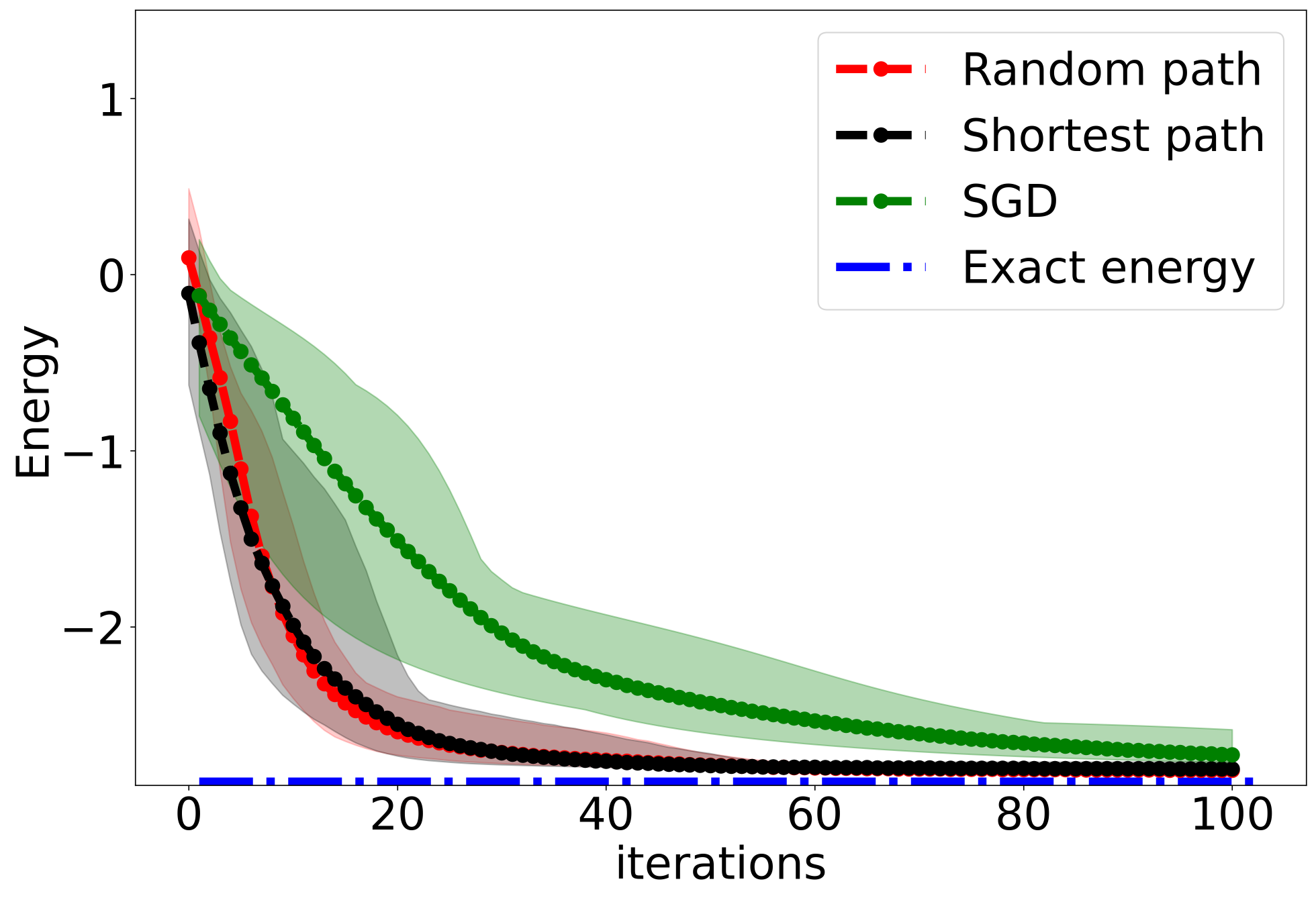} \\
    \end{tabular}
    \begin{tabular}{c c}
    \midrule
    \multicolumn{2}{c}{\textbf{4. 12 Qubit Ansatz}} \\
    \midrule
    a) 1 layer & b) 2 layers \\
    \midrule
    \includegraphics[width=0.27\textwidth]{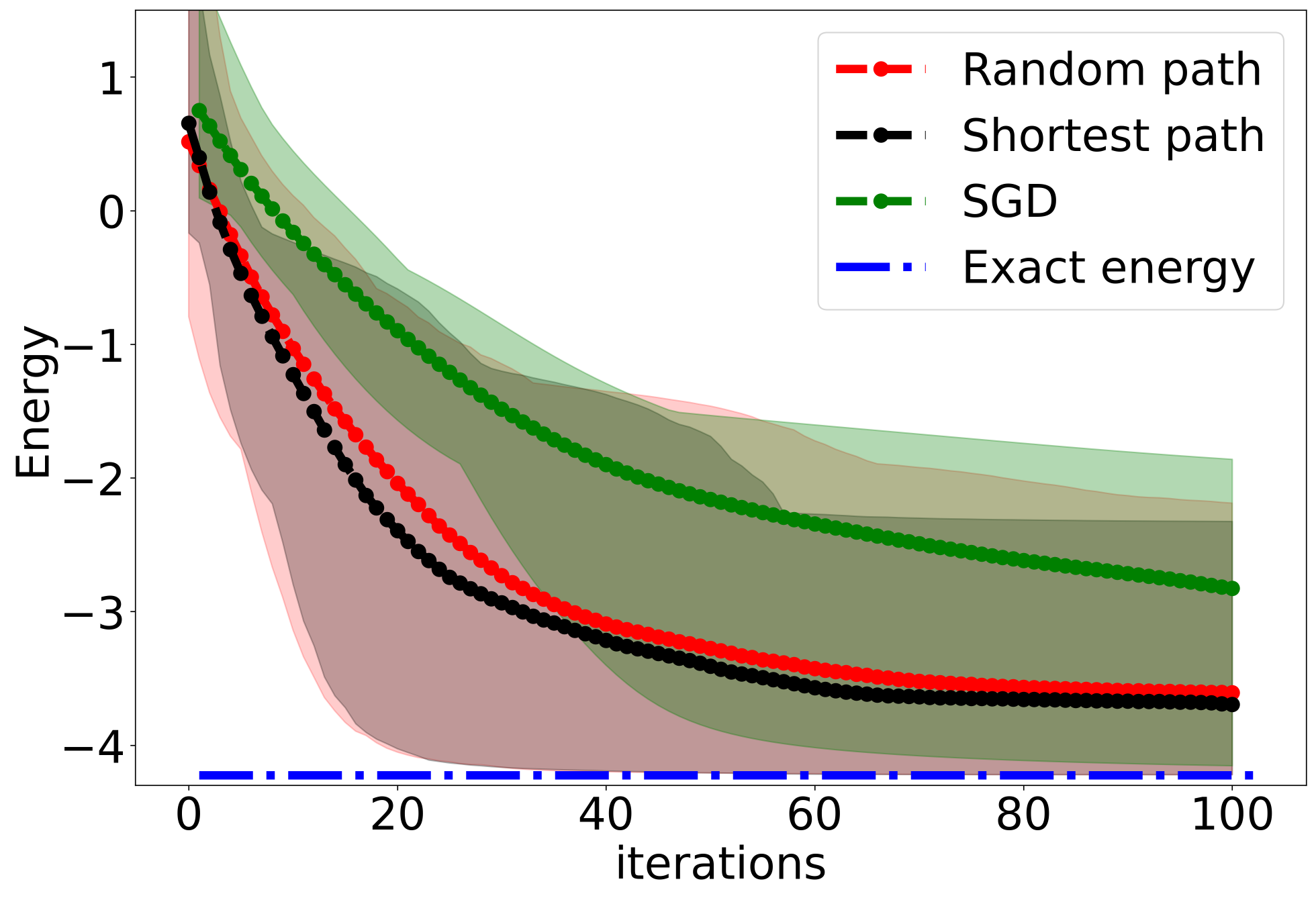} & 
    \includegraphics[width=0.27\textwidth]{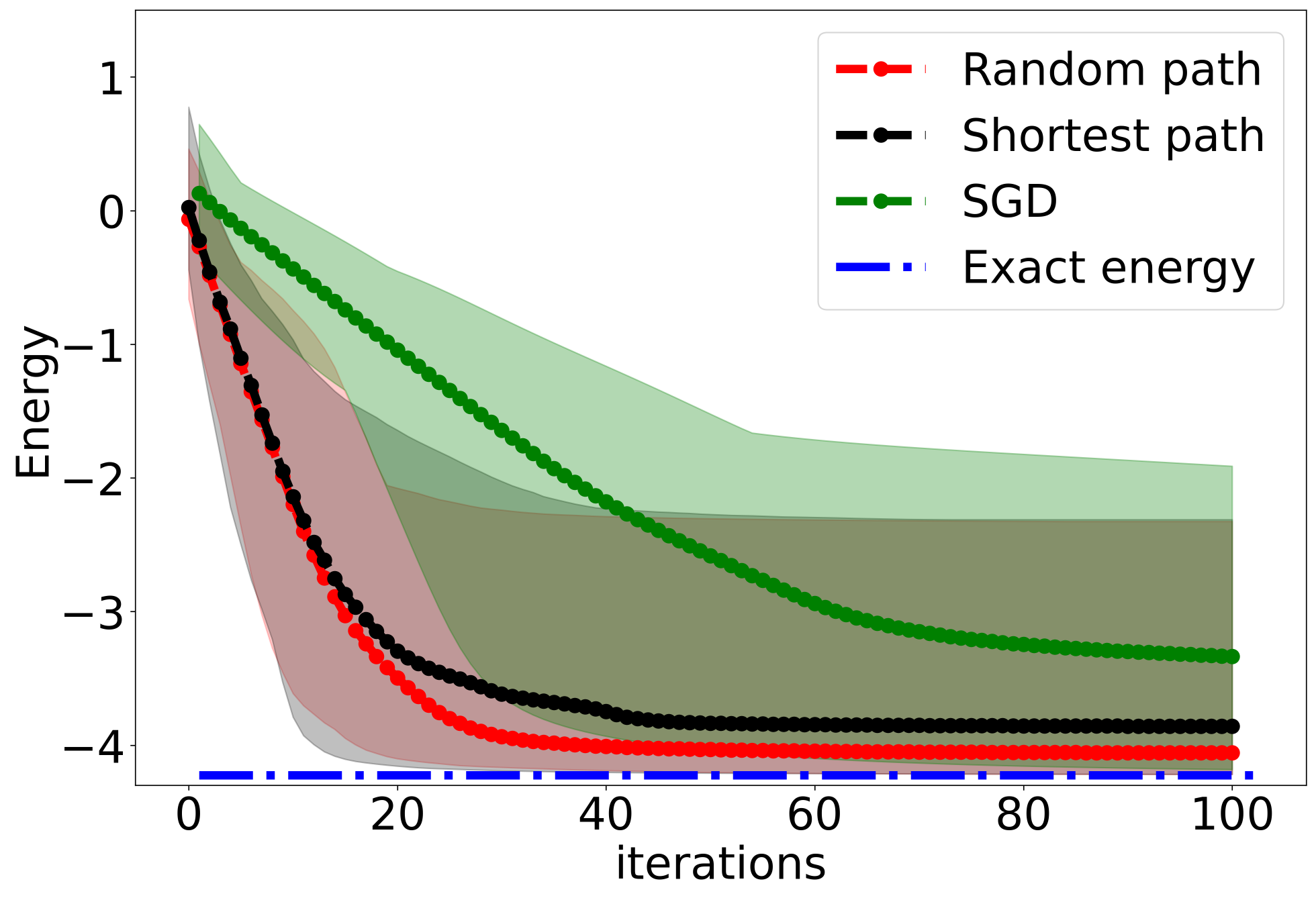} \\
    \midrule
    \multicolumn{2}{c}{\textbf{5. 14 Qubit Ansatz}} \\
    \midrule
    a) 1 layer & b) 2 layers \\
    \midrule
    \includegraphics[width=0.27\textwidth]{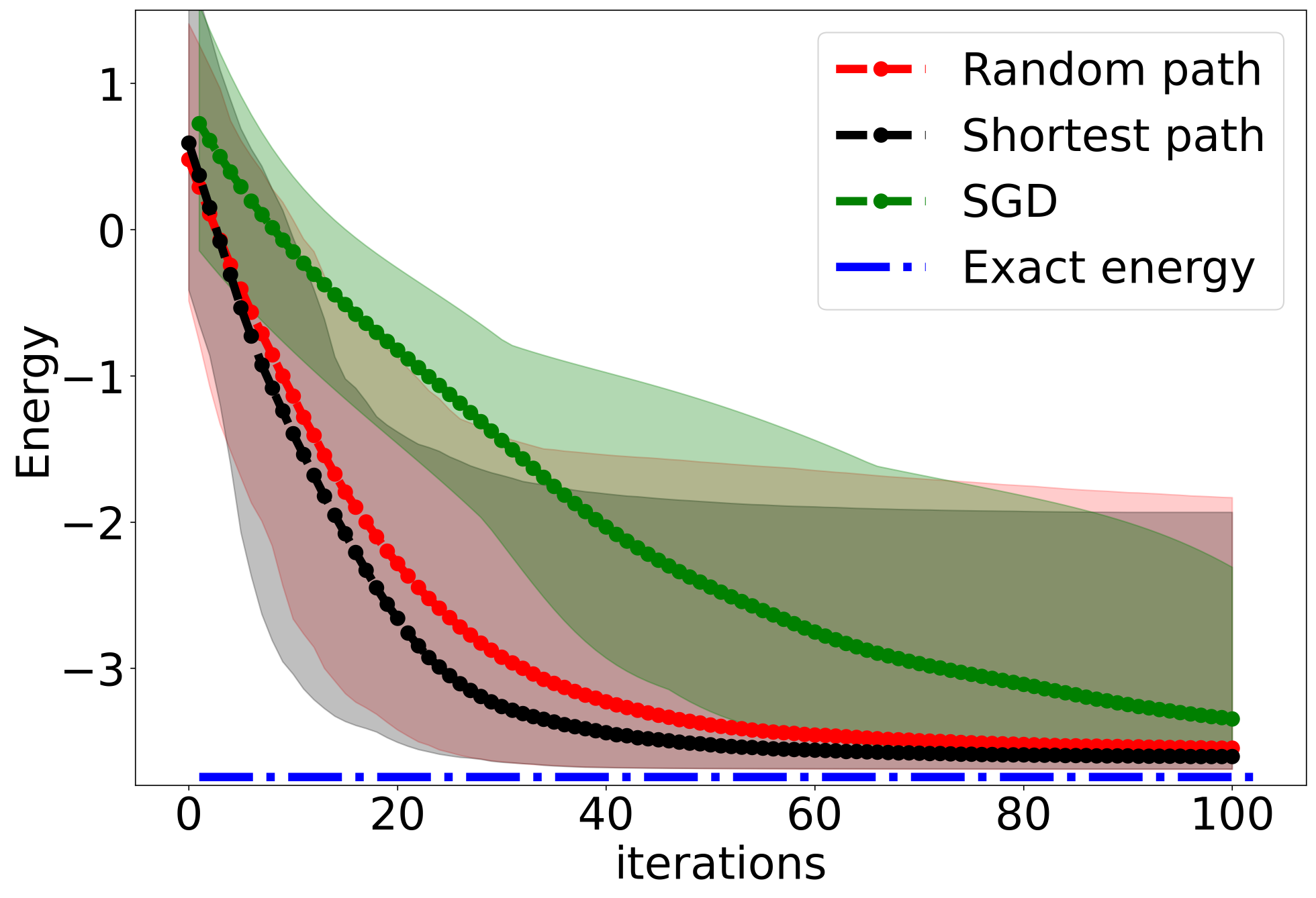} &
    \includegraphics[width=0.27\textwidth]{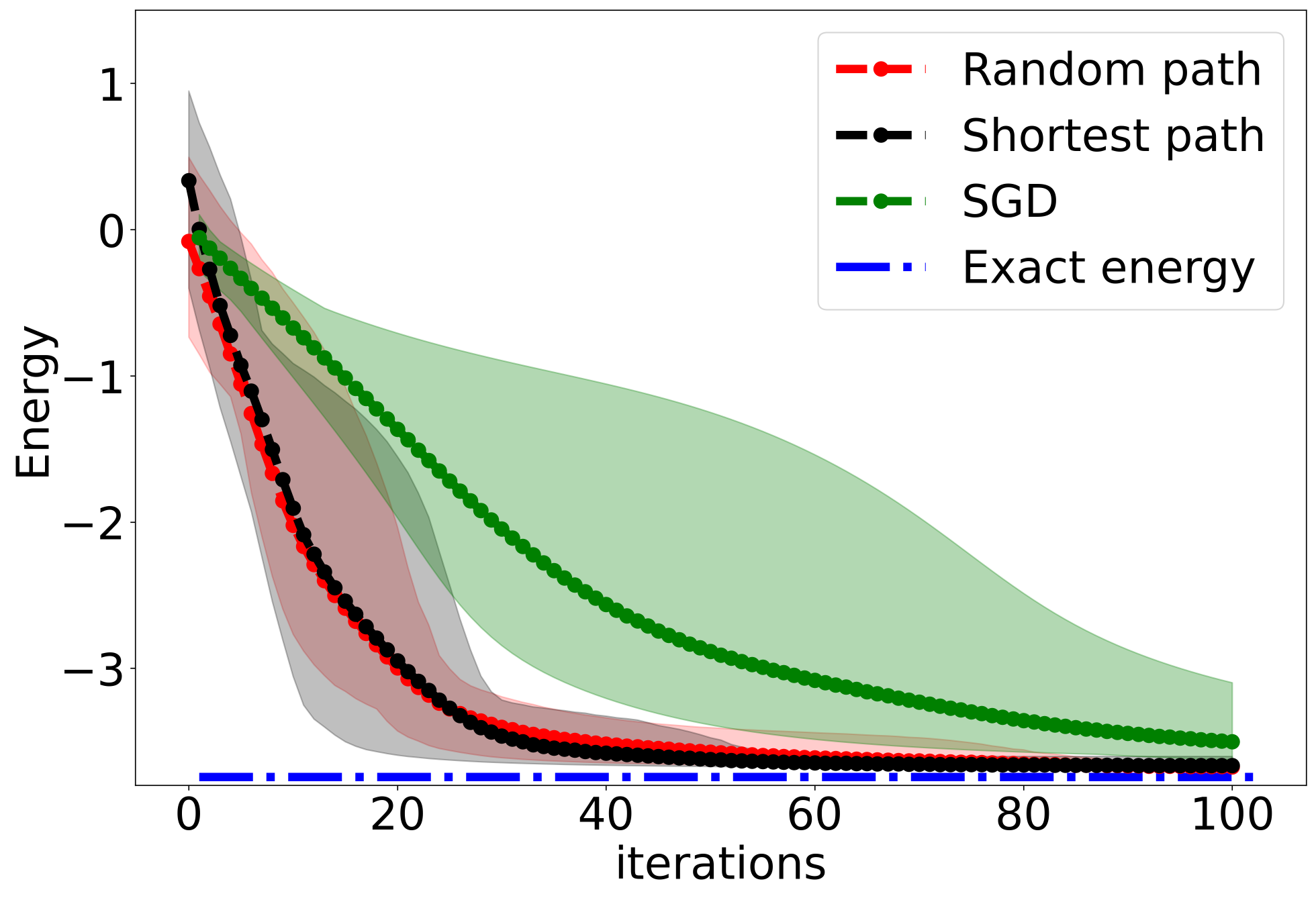} \\
    \end{tabular}
    \caption{\label{fig:vqe_results} Optimization trajectories from different VQE simulations of the different XXZ-Hamiltonians. The lines correspond to the mean of the trajectories from different runs, and the shadow represent the area between the best and worst values from the simulations.}
\end{figure*}

We carried out simulations for five different lattice sizes, starting with $3\times 2$ qubits, $4\times 2$ qubits, $5\times 2$  qubits, $6\times 2$ qubits and $7\times 2$ qubits (6, 8, 10, 12, and 14 qubits, respectively).
For all the simulations, we use the ansatz shown in Fig.~\ref{fig:cir_graph_conversion}(b) with varying number of layers.
The optimization is carried out using the algorithm presented in Algorithm~\ref{al:stoc_op} and the stochastic gradient descent algorithm, with a fixed learning rate of $0.1$.
The results from all the simulations are plotted in Figure~\ref{fig:vqe_results}.
We only plot the first 100 iterations of the trajectories for all the cases, as it is sufficient for comparing the different methods.
All the simulations were repeated at least 5 times with random initialization of the gate parameters to collect the statistics for comparison.

We look at the optimization trajectories from the different simulations using a single layer of the ansatz plotted in Fig.~\ref{fig:vqe_results}.
First, we observe that optimization using Algorithm~\ref{al:stoc_op} with either a random or shortest path always performs better when compared with stochastic gradient descent.
Second, we observe that the optimization trajectories using the shortest path on average have steeper initial convergence, however, the final energies achieved by all optimization methods were very close.
This indicates that choosing a set of parameters based on information transfer between qubits via gates can help accelerate the overall convergence of an algorithm.
Also, it should be noted that forcing the flow of information along a particular path can be useful in cases where the spread of information can lead to convergence issues~\cite{marrero2021entanglement}. 
Finally, we point out that the spread in the trajectories corresponding to the runs with the shortest path tends to be smaller, however, all the methods have some runs (particularly in the case of the 12 qubit model) where they converge to a local minimum.
This is a common occurrence in stochastic optimization methods and can be mitigated using different methods.~\cite{swenson2020distributed}

As we increase the number of layers of the ansatz used in the numerical simulations, we observe that the rate of convergence for all the methods increases.
However, we point out that optimization trajectories with the shortest path are still the fastest converging among all the methods.
This implies that using (shortest) paths for optimization of algorithms with objectives depending on only a subset of the qubits might be useful. \\

\color{black}
\subsubsection{Resource analysis}
As mentioned above, an important property of the path-based algorithm is that it selects only a subset of the gate parameters to be updated at each step in the optimization. Since the gradient related to each circuit parameter generally has to be estimated separately using parameter-shift rules, and there would be no need to calculate gradients for parameters that are not going to be updated, fewer parameters to update generally leads to fewer measurements required. Furthermore, since our parameter selection is based on a classical algorithm requiring no additional measurements, there is no additional measurement overhead to consider. Combining these facts, it seems likely that the path-based approach should require fewer measurements than stochastic gradient descent, even if the same number of iterations were required. To investigate this question further, we take the number of updated parameters as a proxy for the required number of measurements, and track it across the 6- and 8-qubit XXZ-Heisenberg runs of Appendix~\ref{ap:more_heisenberg}. In Fig.~\ref{fig:param_update_dist}, the number of parameters updated at each step for the two path-selection strategies is compared to each other and to stochastic gradient descent. As can be seen from this figure and the statistics in \color{black}Table~\ref{tab:param_savings}, \color{black} using the strategy based on shortest paths allows for a reduction in the number of updated parameters of about 8\%, while using random paths allow for an even larger reduction of about 12\%. A priori, this would seem to indicate an advantage for the random-path sampling over shortest-path sampling when it comes to minimizing the amount of measurements required. However, as observed in Fig.~\ref{fig:vqe_results}, the shortest-path strategy sometimes converges in fewer iterations. Thus, the question of the best method with respect to the number of measurements to some extent boils down to the question of using fewer but more expensive shortest-path iterations or more but cheaper random-path iterations. To showcase how these competing effects look in practice, Fig.~\ref{fig:vqc_results_updated_params} shows the XXZ-Heisenberg results of Appendix~\ref{ap:more_heisenberg} plotted against the total number of parameter updates. As shown in these figures, the path-based approaches in general tend to outperform stochastic gradient descent, while the best-performing strategy varies from problem to problem and run to run, and depends on whether you value rapid early convergence or better convergence late in the process. This highlights how the metric of updates per iteration in  \color{black}Table~\ref{tab:param_savings} \color{black} isn't necessarily sufficient to judge relative performances of the methods. 

\begin{figure}[htbp!]
\centering
\subfloat[6 qubits]{\includegraphics[width=0.495\textwidth]{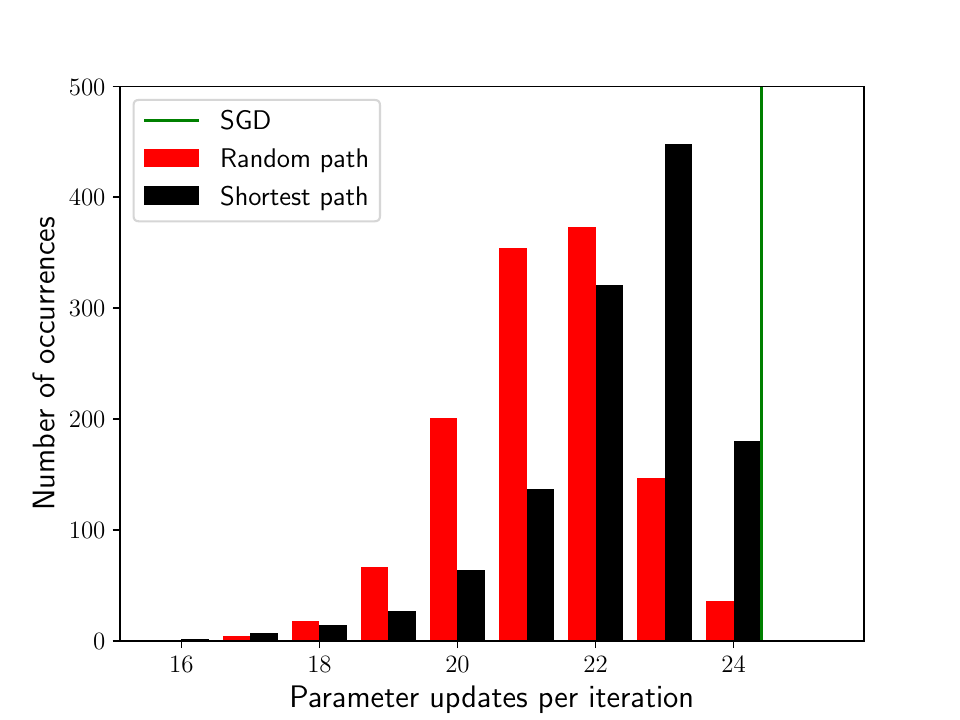}} \\
\subfloat[8 qubits]{\includegraphics[width=0.495\textwidth]{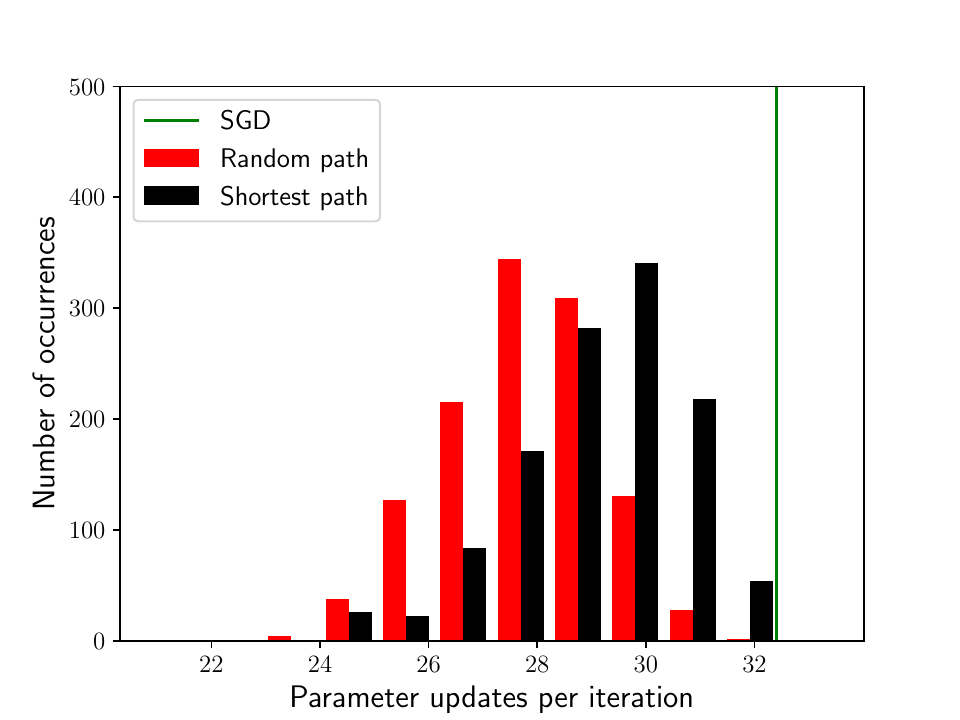}} \\
\caption{\label{fig:param_update_dist} \color{black} Histograms showing the number of updated parameters per iteration for the random-path and shortest-path strategies aggregated over the complete sets of 6 and 8-qubit runs of XXZ-Heisenberg benchmarks described in Appendix~\ref{ap:more_heisenberg}. For reference, the total number of parameters in the circuit, i.e., the number updated at each iteration using the SGD approach, is marked with a green vertical line. \color{black}}
\end{figure}

\begin{table}[]
    \centering
    \renewcommand{\arraystretch}{2.0}
    \begin{tabular}{c|c|c|c}
         & \; SGD \; & \; Shortest path \; & \; Random path \;  \\ 
        \hline
         6 Qubits \; & \; 24 \; & 22.3 (8.3\%) & 21.3 (12.4\%) \\
         8 Qubits \; & \; 32 \; & 29.3 (7.1\%) & 28.0 (11.2\%) 
    \end{tabular}
    \caption{\color{black}Average number of parameters updated at each step in the optimization for different methods compared in this paper, and their reductions compared to baseline stochastic gradient descent. For a more detailed look at the corresponding distributions, see Fig.~\ref{fig:param_update_dist}.\color{black}}
    \label{tab:param_savings}
\end{table}

\begin{figure*}[htbp!]
    \centering
    \begin{tabular}{c c c}
    \toprule
    \multicolumn{3}{c}{\textbf{6 Qubit Ansatz}}\\
    \midrule
    a) $\Delta = 0.5$ & b) $\Delta = 1.0$ & c) $\Delta = 1.5$ \\
    \midrule
    \includegraphics[width=0.32\textwidth]{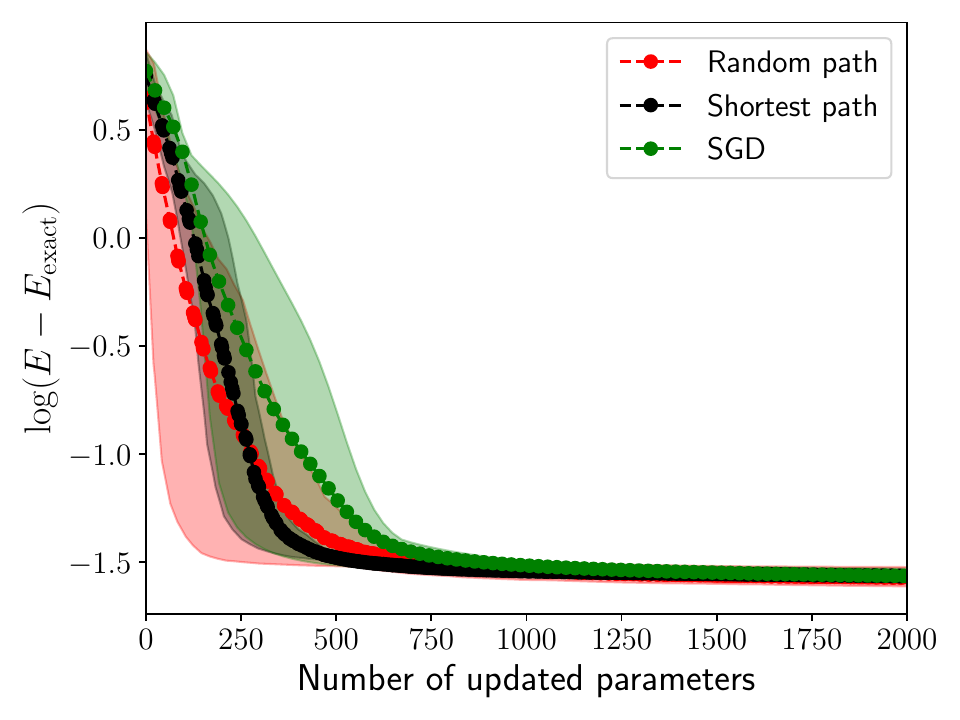} & 
    \includegraphics[width=0.32\textwidth]{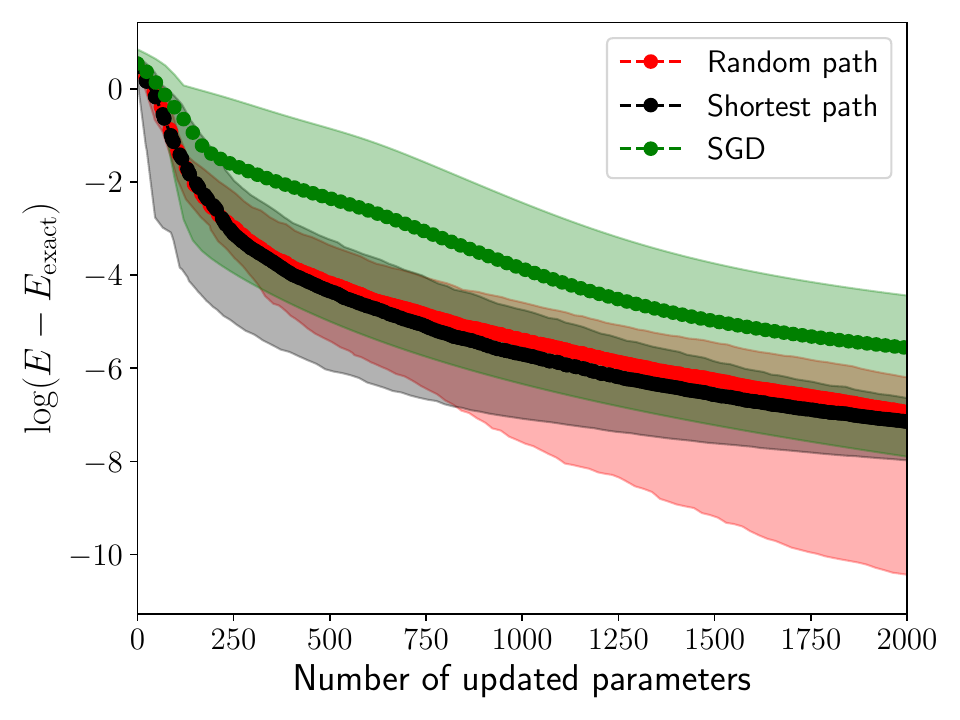} &
    \includegraphics[width=0.32\textwidth]{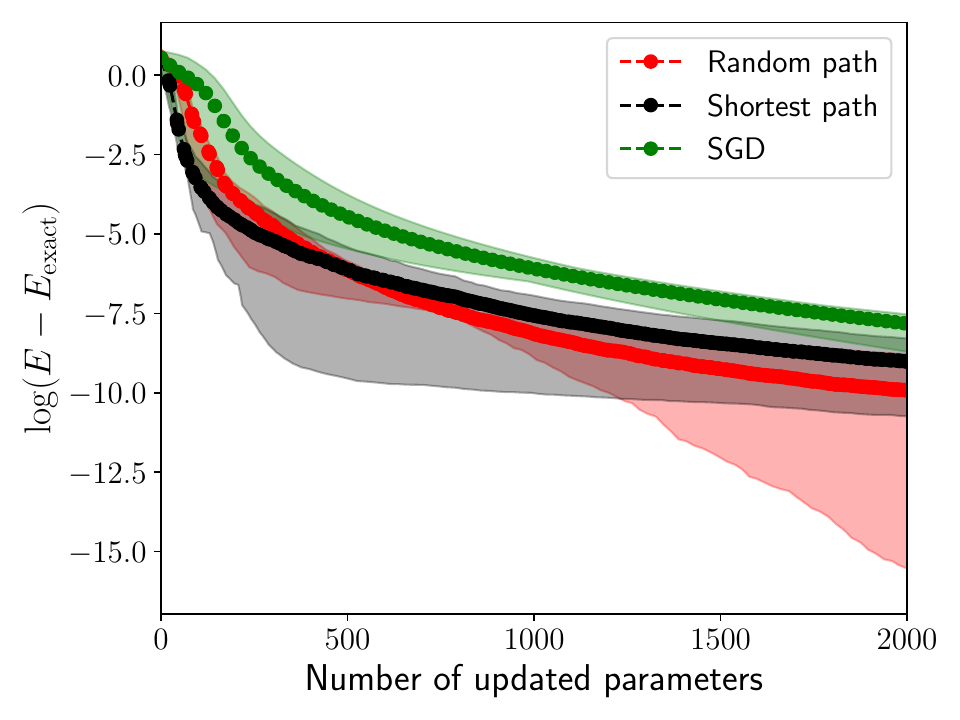}\\
    \midrule
    \multicolumn{3}{c}{\textbf{8 Qubit Ansatz}}\\
    \midrule
    a) $\Delta = 0.5$ & b) $\Delta = 1.0$ & c) $\Delta = 1.5$ \\
    \midrule
    \includegraphics[width=0.329\textwidth]{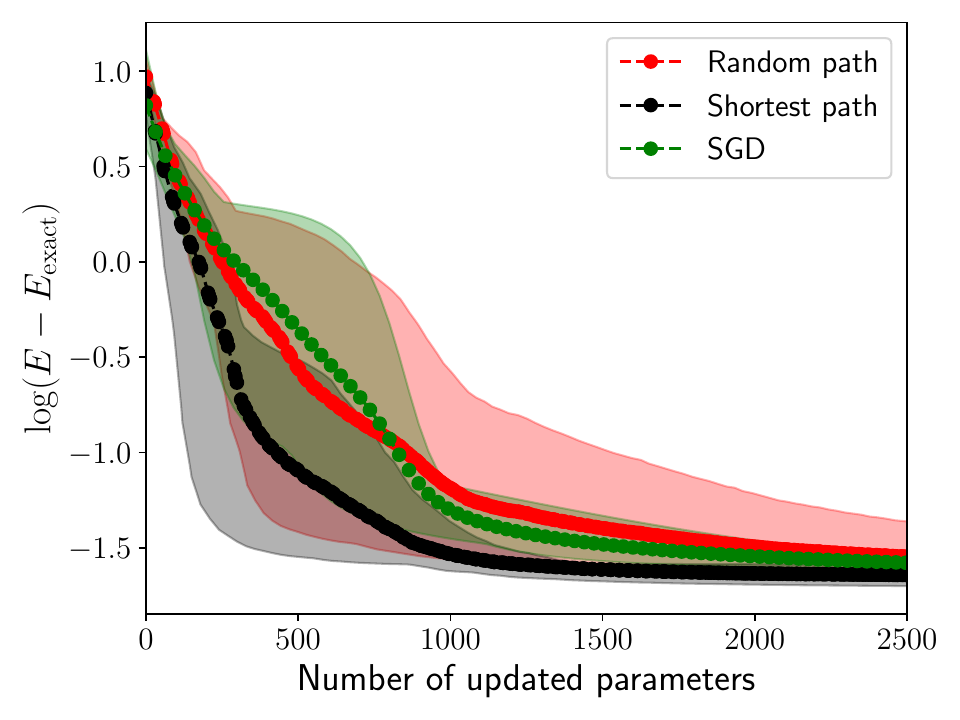} & 
    \includegraphics[width=0.329\textwidth]{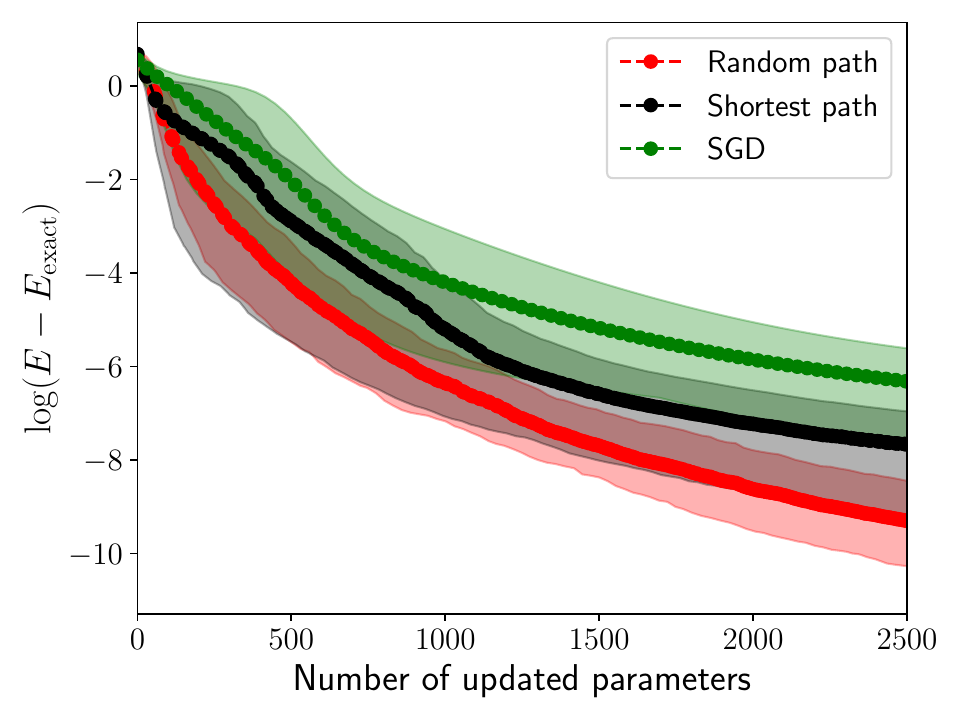} &
    \includegraphics[width=0.329\textwidth]{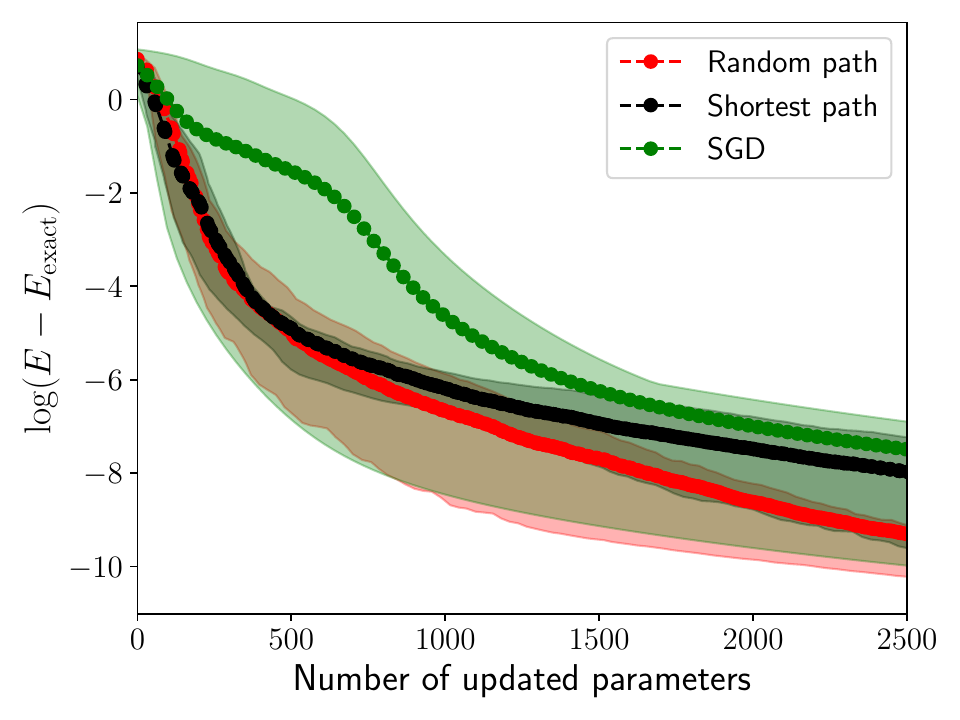}\\
    \end{tabular}
    \caption{\label{fig:vqc_results_updated_params} \color{black} Plot of the XXZ-Heisenberg results from Appendix~\ref{ap:more_heisenberg}, but with the achieved energy plotted as a function of the total number of parameter optimizations rather than the number of optimization iterations. Note the logarithmic energy-axis to better show late-optimization behavior. \color{black}}
\end{figure*}

\color{black}

\subsection{VQC - Binary classification}
\color{black} To investigate the use of our framework for a different use-case with a significantly different cost function, we apply it to the \color{black} VQC framework presented in Section \ref{vqc_intro} for the $n$-bit parity classification problem. The corresponding dataset of such a system consists of $2^n$ distinct binary vectors \color{black}with corresponding labels\color{black}, where each label indicates whether the sum of the $n$ components of the binary vector is odd or even. The Boolean $n$-bit parity function to be modeled is
\begin{align}
    f : \{0,1\}^{\otimes n} \xrightarrow{} \{0,1\}
\end{align}
with the property that $f(x)=1$ if the number of ones in the vector $x\in \{0,1\}^{n}$ is odd. 
Binary classification is an interesting choice for this study because there is only one readout qubit, making the problem \color{black} potentially \color{black} well-suited for our optimization method.

\begin{figure*}[htbp!]
    \centering
    \begin{tabular}{c c c}
    \toprule
    \multicolumn{3}{c}{\textbf{Cost function}}\\
    \midrule
    a) 2 Layers & b) 3 Layers & c) 4 Layers \\
    \midrule
    \includegraphics[width=0.3\textwidth]{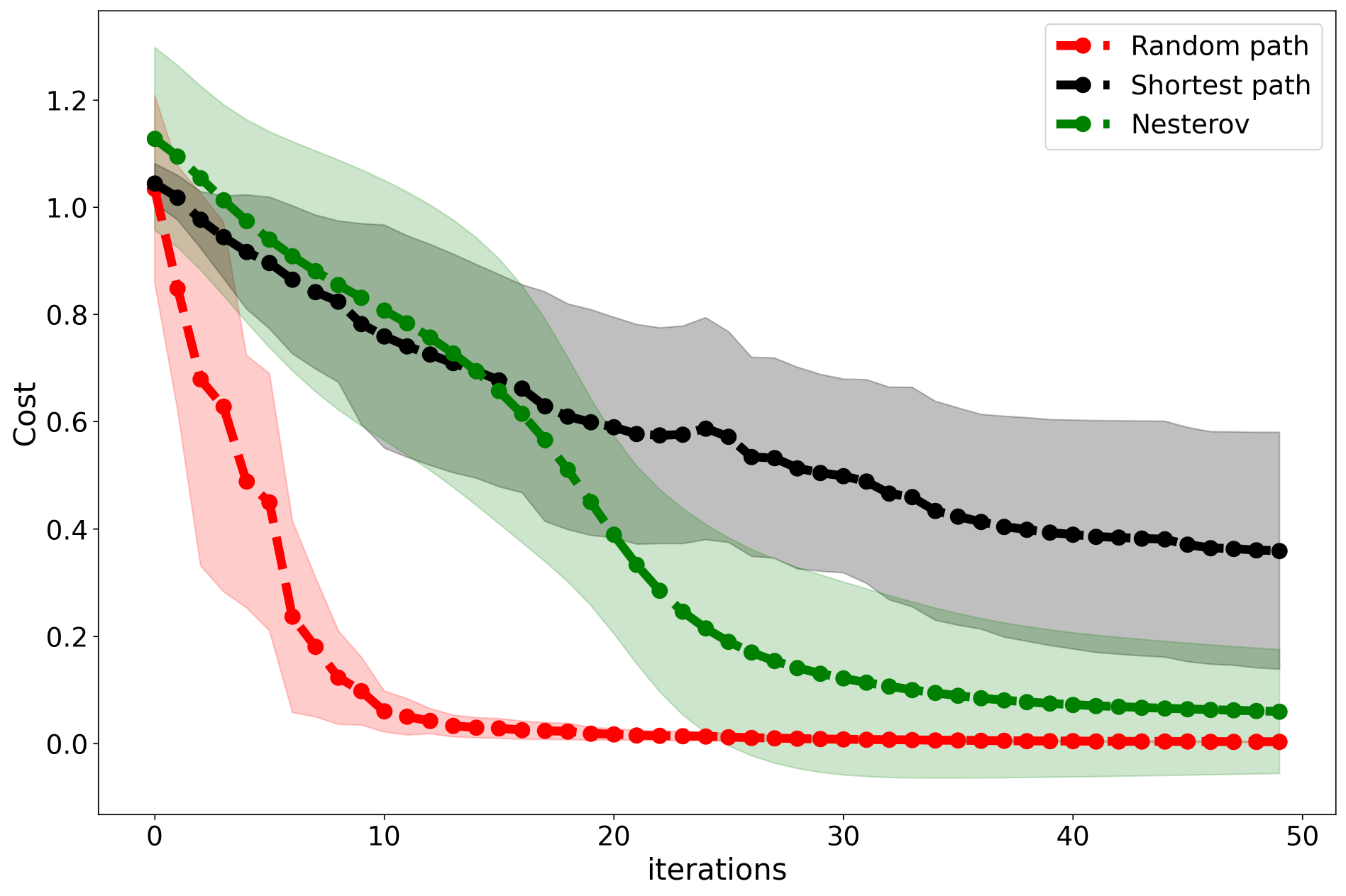} & 
    \includegraphics[width=0.3\textwidth]{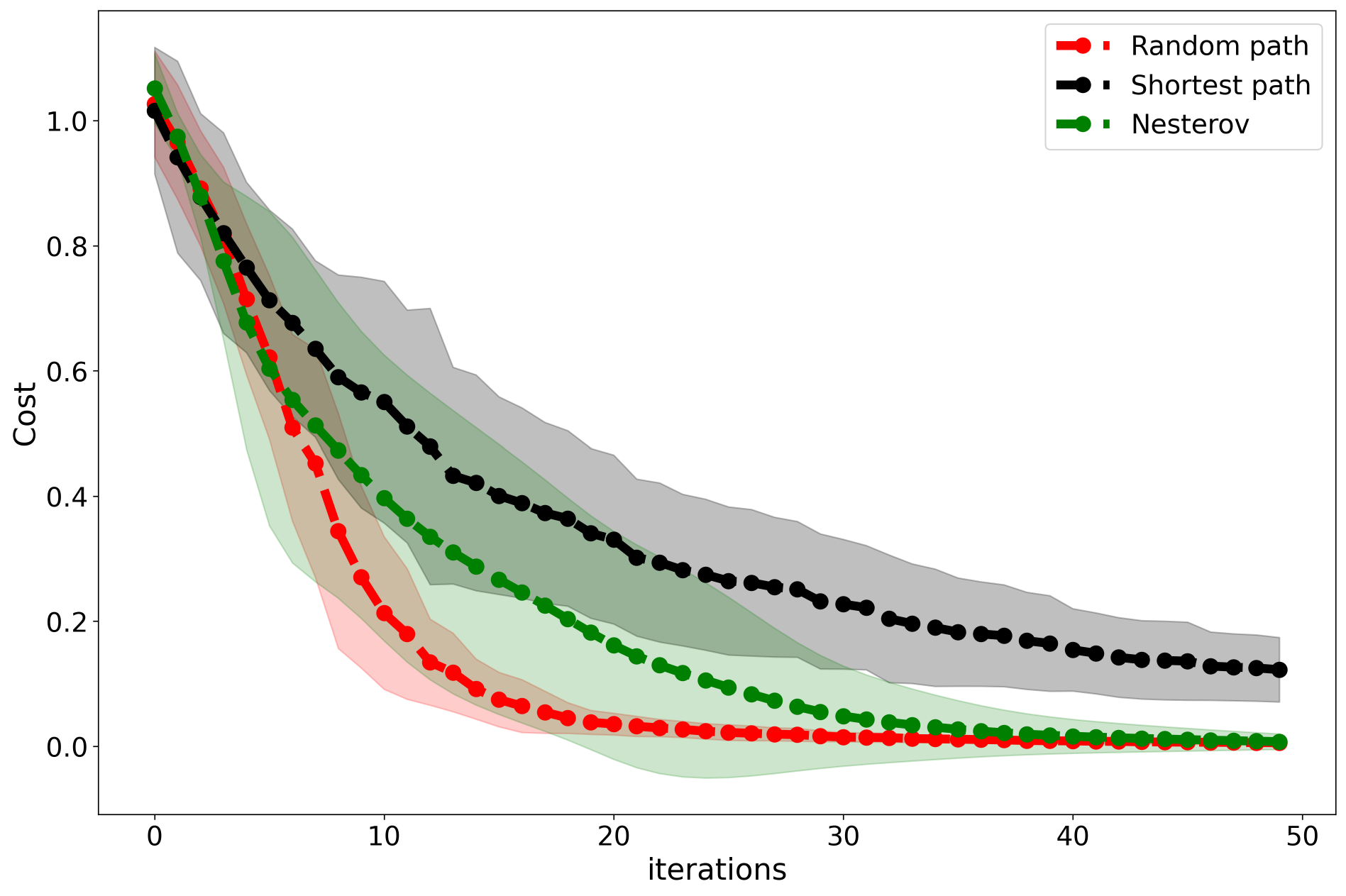} &
    \includegraphics[width=0.3\textwidth]{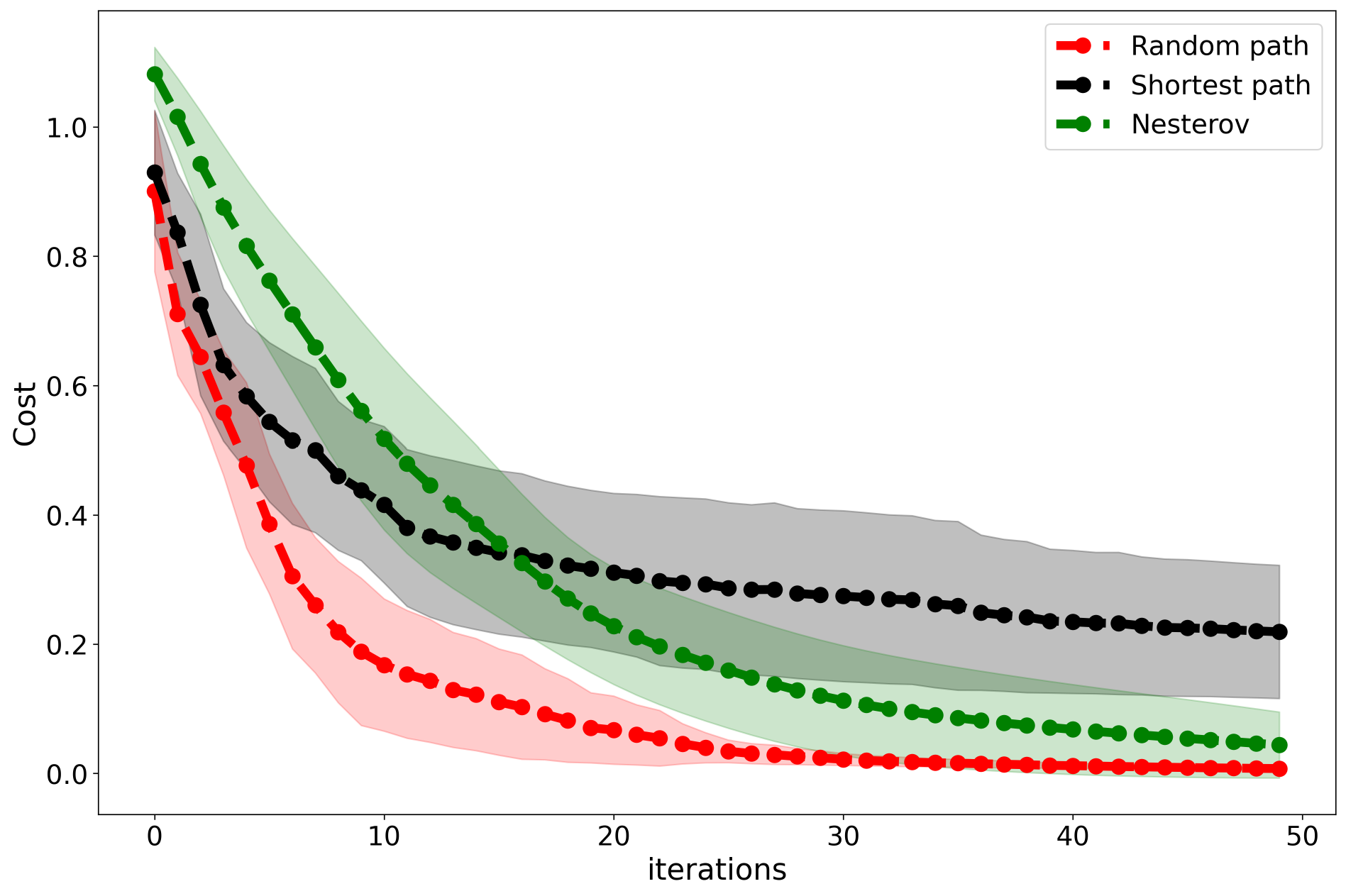}\\
    \midrule
    \multicolumn{3}{c}{\textbf{Accuracy of the model}}\\
    \midrule
     a) 2 Layers & b) 3 Layers & c) 4 Layers \\
    \midrule
    \includegraphics[width=0.3\textwidth]{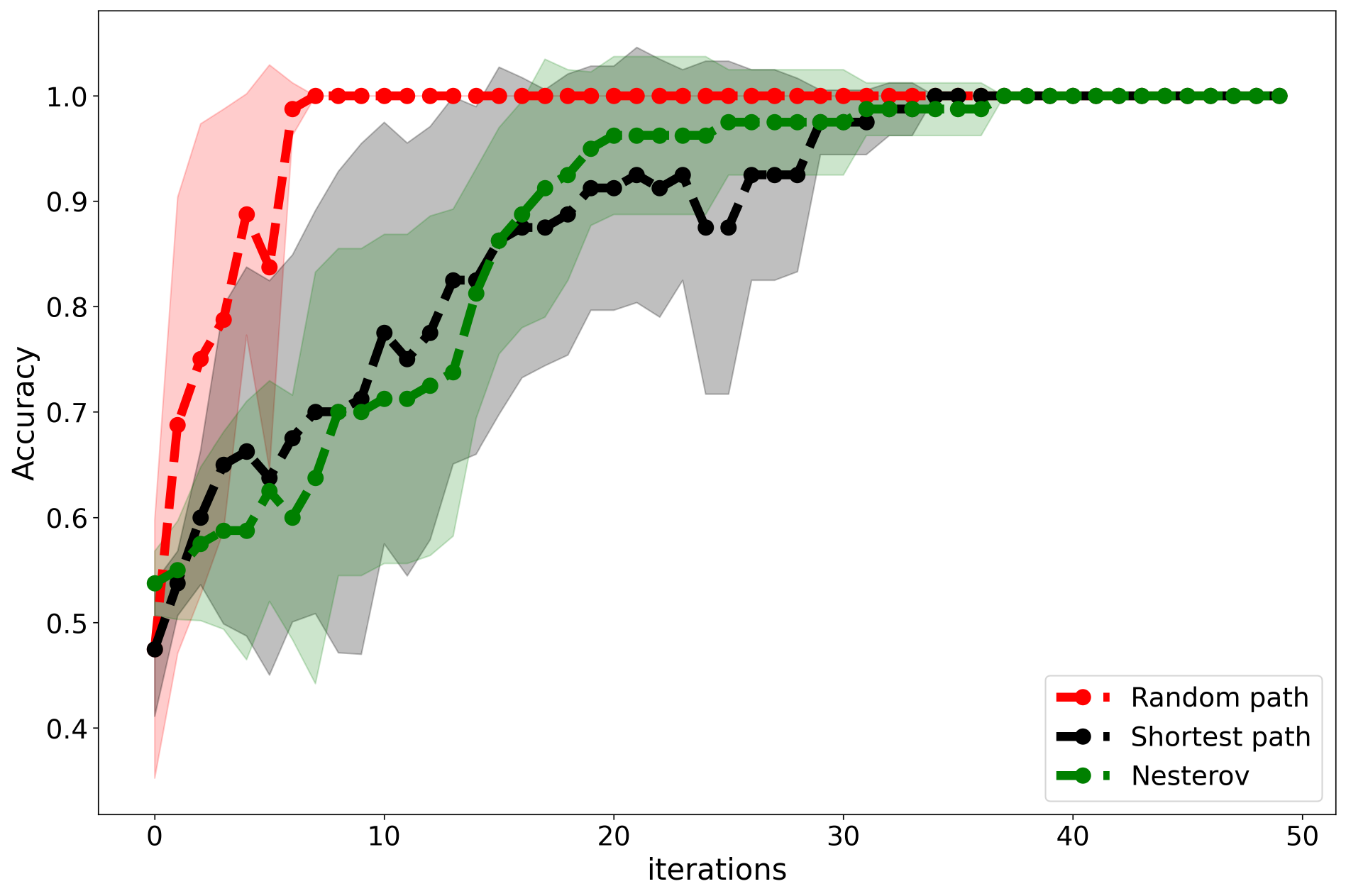} & 
    \includegraphics[width=0.3\textwidth]{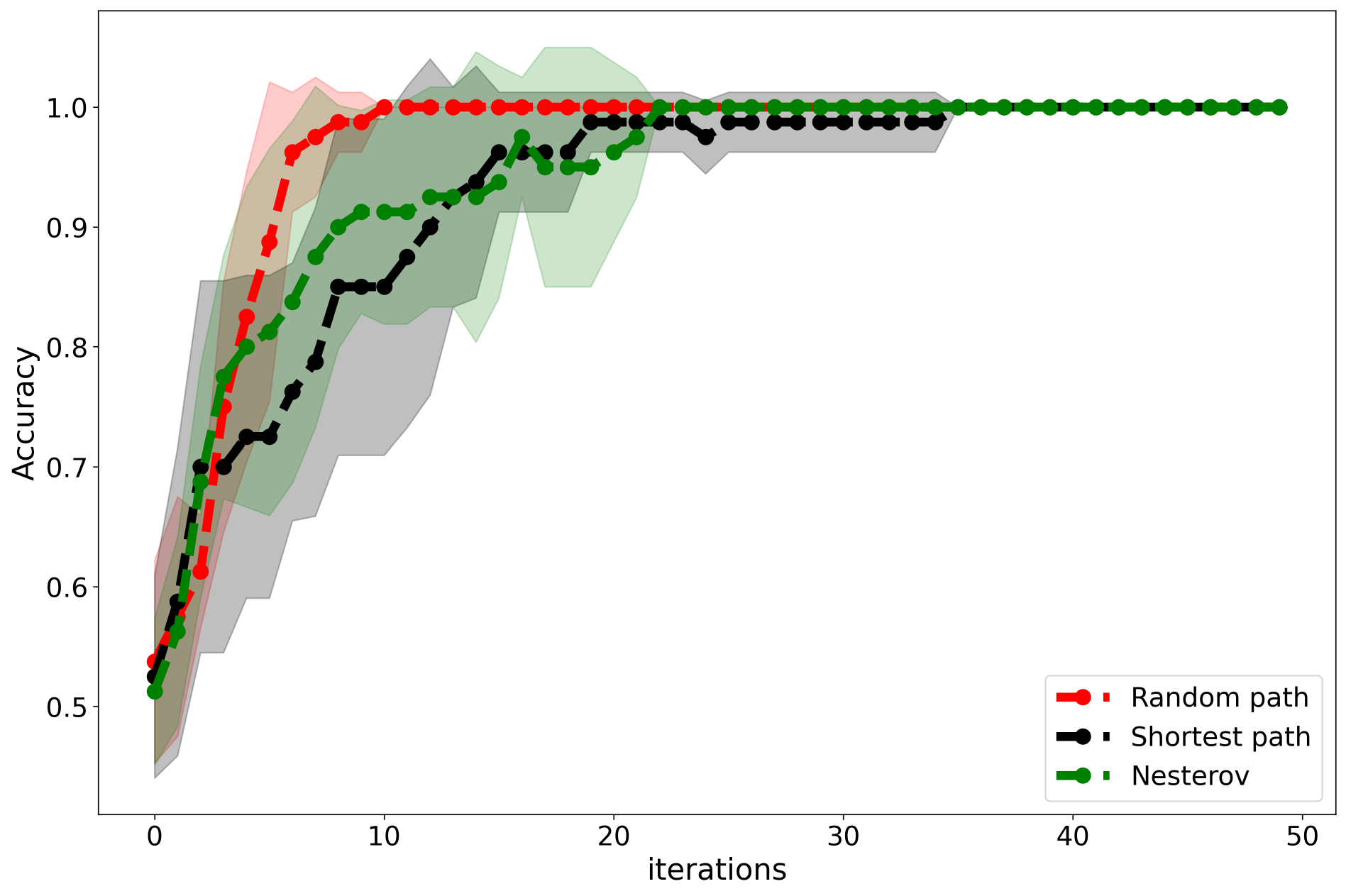} &
    \includegraphics[width=0.3\textwidth]{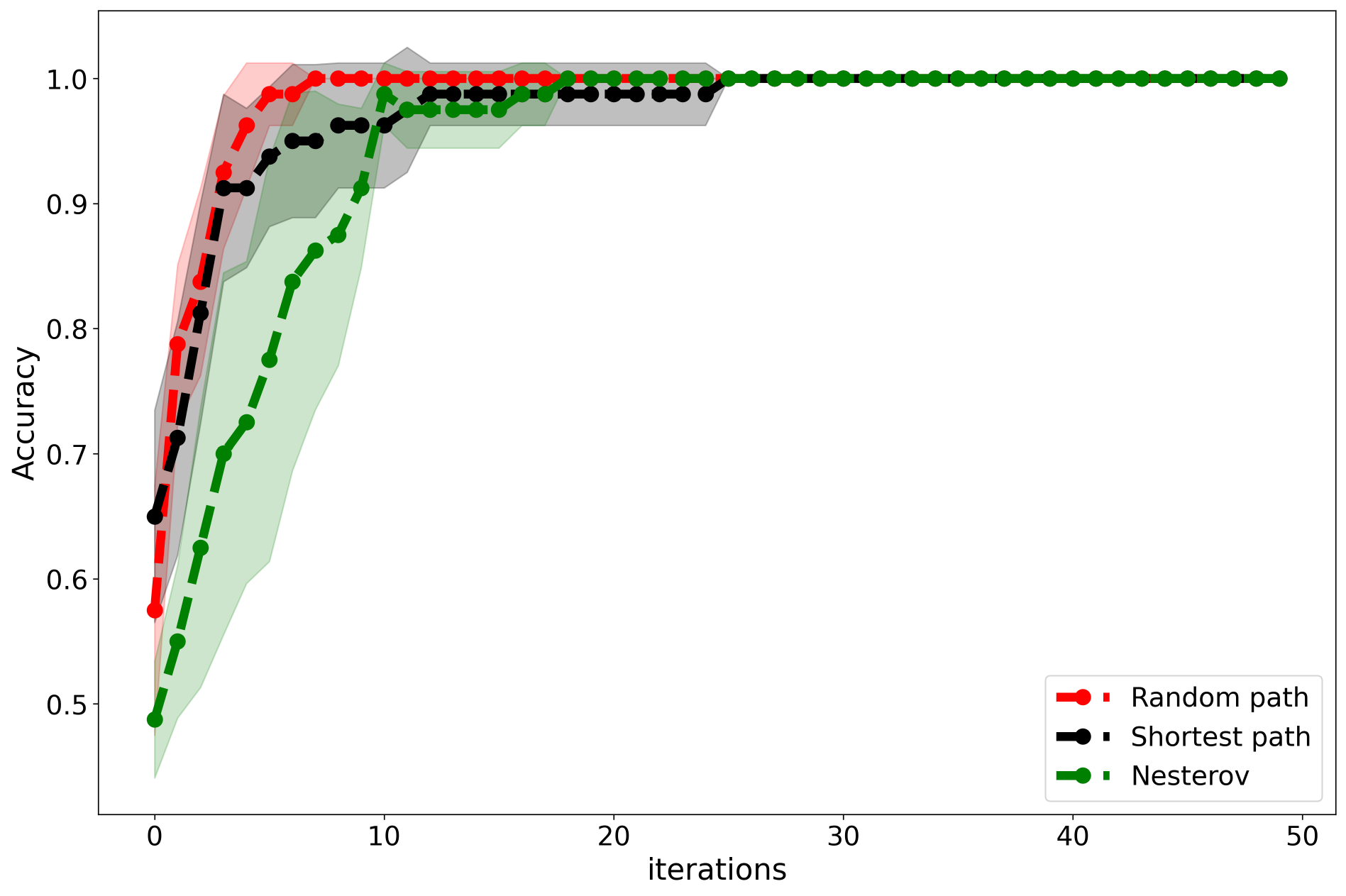}\\
    \end{tabular}
    \caption{\label{fig:vqc_results} Optimization trajectories from different VQC simulations of the n-bit parity problem using a 4-qubit ansatz. The lines correspond to the mean of the trajectories from different runs, and the shadow represents one standard deviation.}
\end{figure*}

Simulations were performed on the 4-qubit parity problem using the ansatz shown in  Fig.~\ref{fig:vqc_1l_ansatz} with $2,3,$ and $4$ layers. 
Again, optimization is performed using the algorithm presented in Algorithm~\ref{al:stoc_op} and Nesterov momentum with a fixed learning rate of 0.1. 
We selected the Nesterov optimizer for comparison, as using Adam failed to improve the model accuracies.
The simulation results are shown in Fig.~\ref{fig:vqc_results}.
We focus on the first 50 training epochs of optimization for all cases and plot the average of 5 instances with random initialization of the gate parameters to collect statistics for comparison. 

\begin{figure*}[htbp!]
    \centering
    \begin{tabular}{c c c}
    \toprule
    \multicolumn{3}{c}{\textbf{Cost function}}\\
    \midrule
    a) 2 Layers & b) 3 Layers & c) 4 Layers \\
    \midrule
    \includegraphics[width=0.3\textwidth]{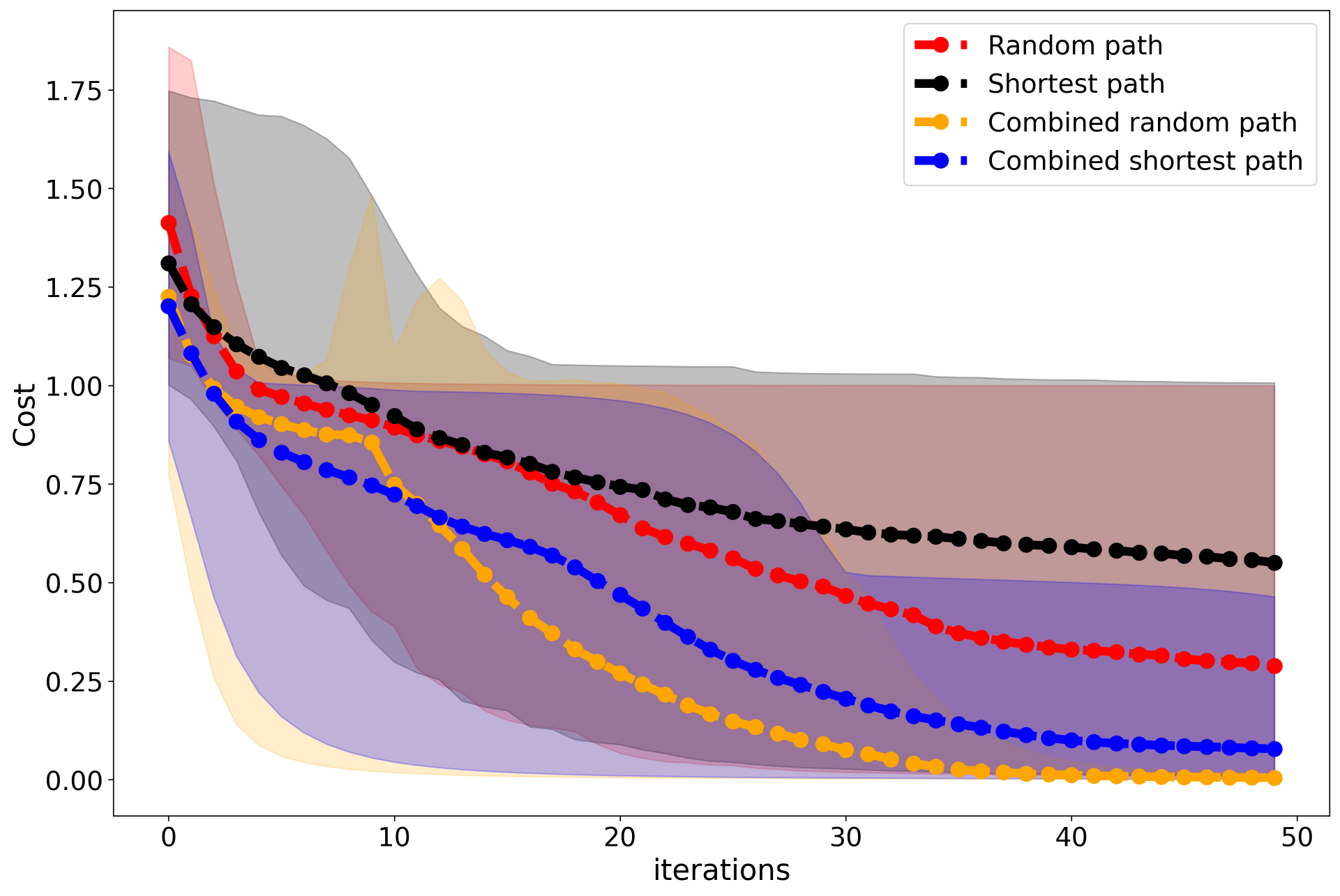} & 
    \includegraphics[width=0.3\textwidth]{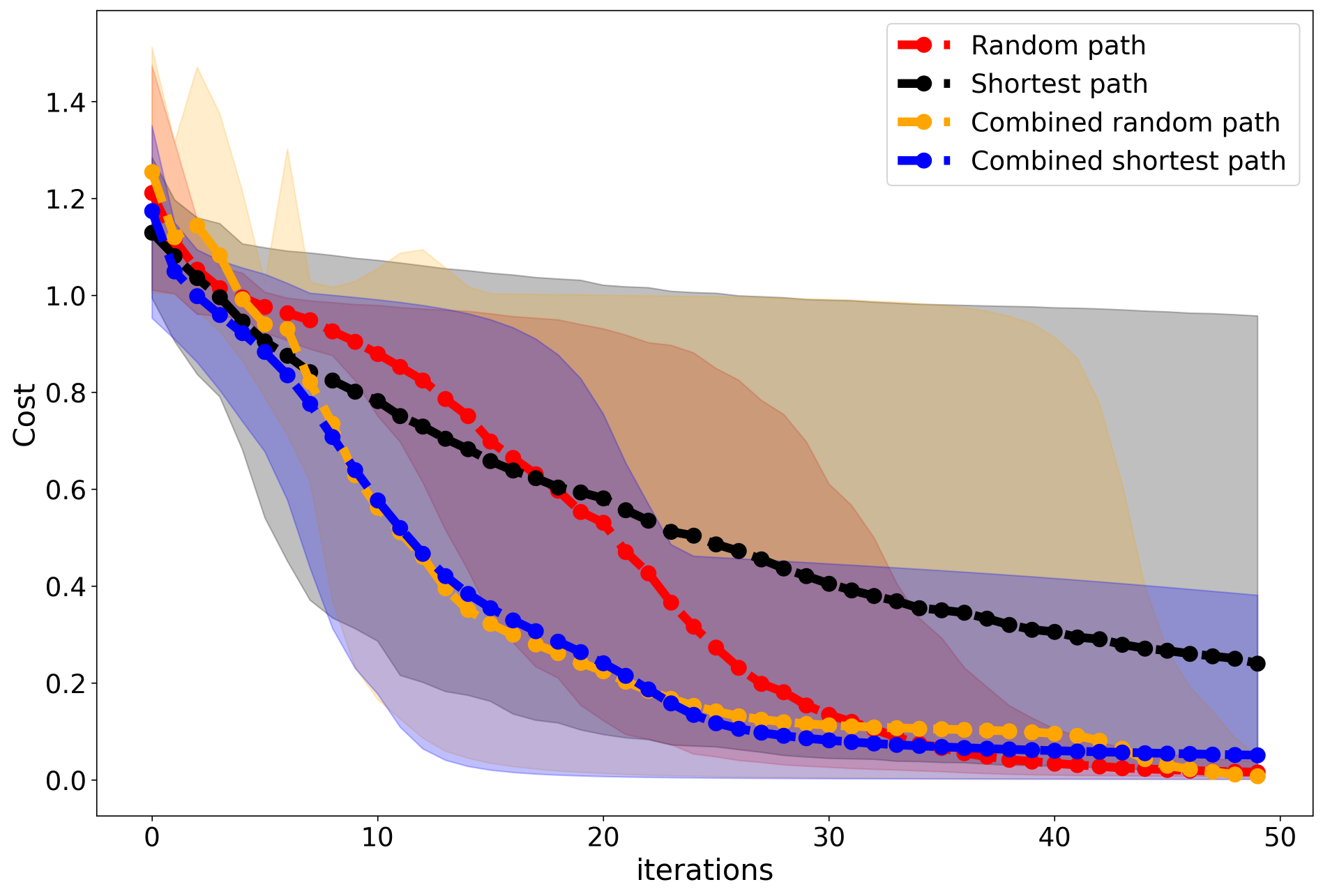} &
    \includegraphics[width=0.3\textwidth]{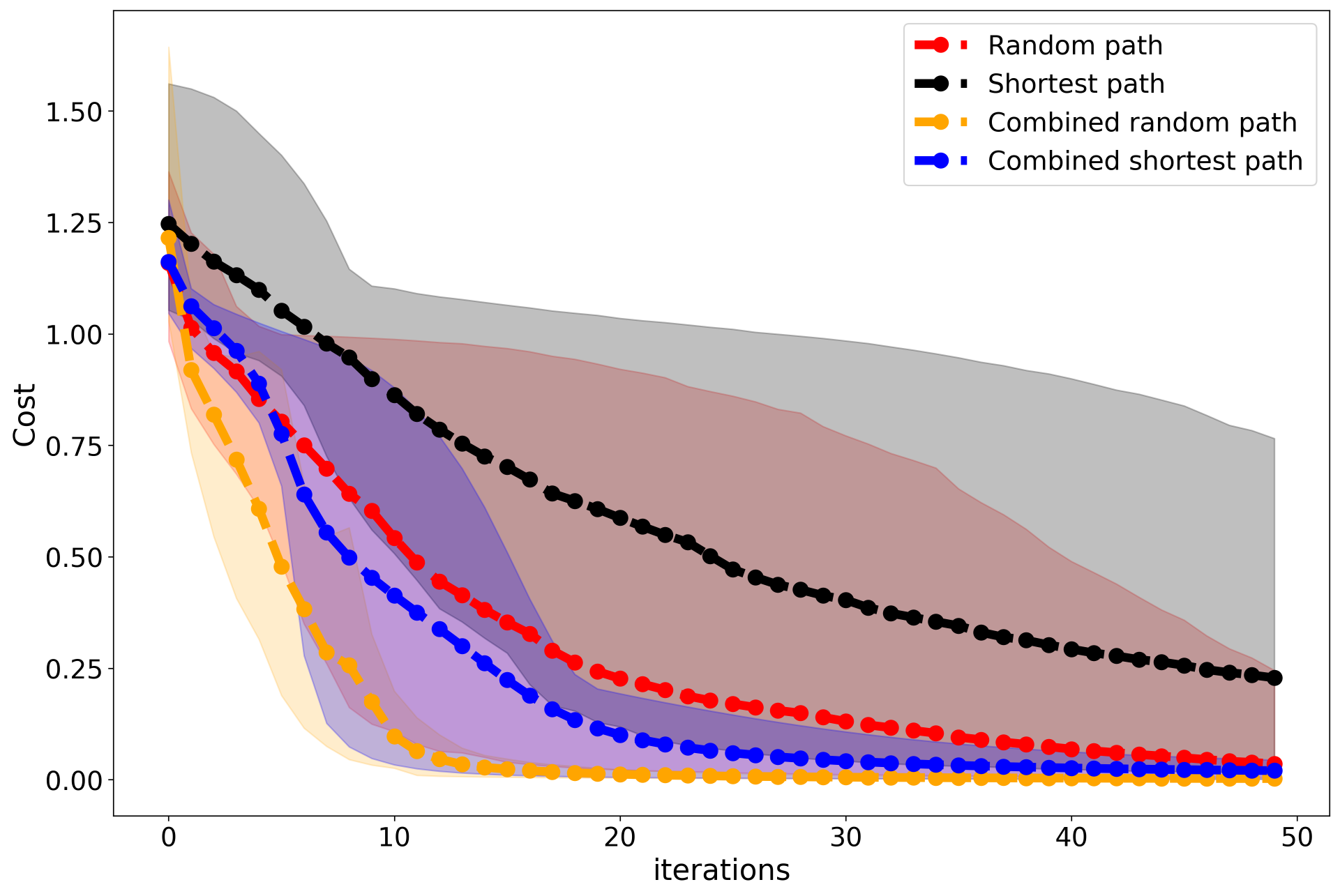}\\
    \midrule
    \multicolumn{3}{c}{\textbf{Accuracy of the model}}\\
    \midrule
     a) 2 Layers & b) 3 Layers & c) 4 Layers \\
    \midrule
    \includegraphics[width=0.3\textwidth]{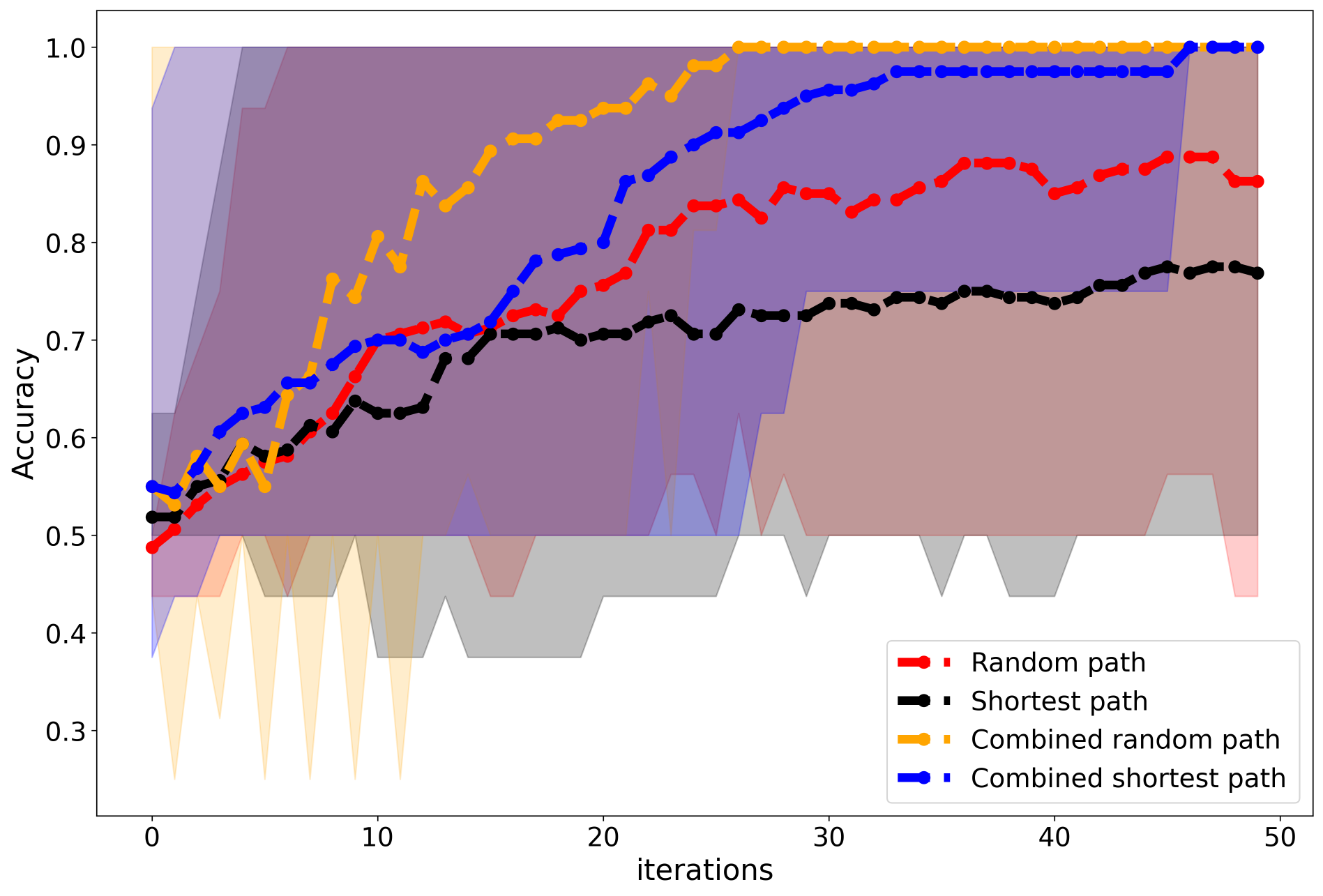} & 
    \includegraphics[width=0.3\textwidth]{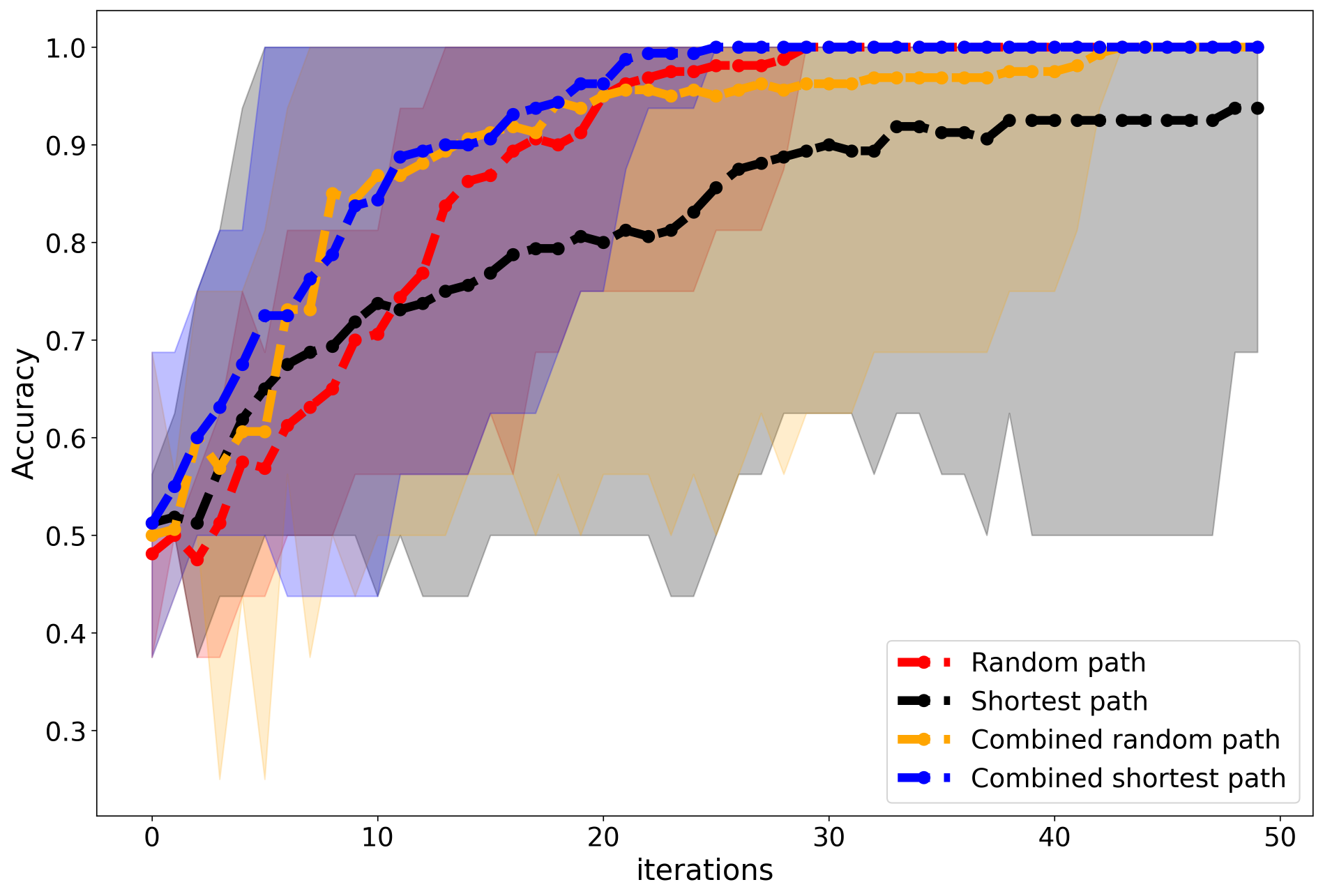} &
    \includegraphics[width=0.3\textwidth]{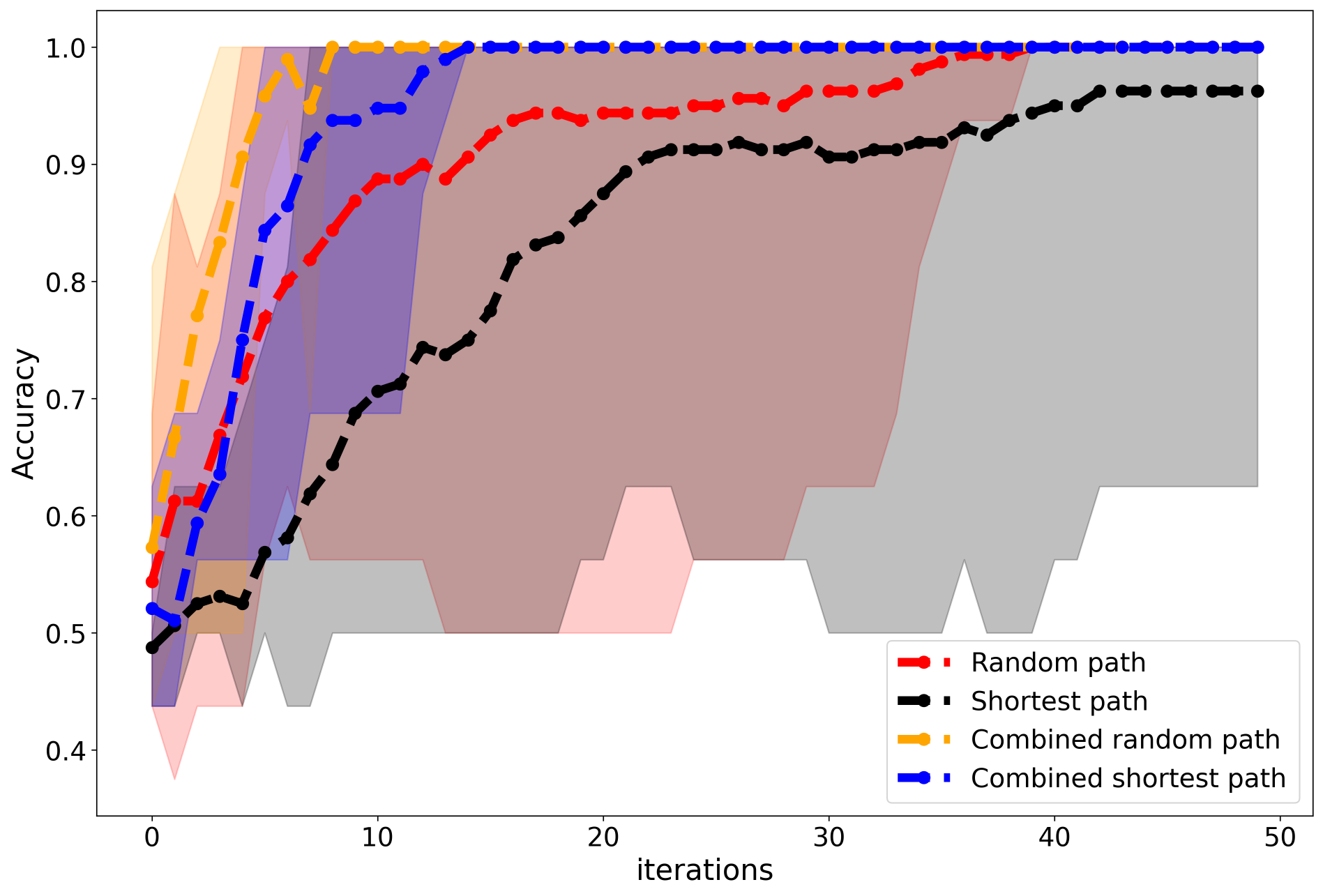}\\
    \end{tabular}
    \caption{\label{fig:vqc_results_1} Optimization trajectories from different VQC simulations of the n-bit parity problem using a 4-qubit ansatz with a reduced learning rate. The lines correspond to the mean of the trajectories from different runs, and the shadow represents the area between the best and worst values from the simulations.} 
\end{figure*}

As can be seen from the optimization trajectories in Fig.~\ref{fig:vqc_results}, the model is able to learn to successfully perform the classification task.
We note that the optimization with random paths consistently outperforms the one with the Nesterov momentum method.
However, we observe that the optimization with the shortest path performs poorly.
 We attribute this to the fact that \color{black} sampling of the shortest path leads to insufficient exploration of the full circuit, and thus to the model having too little knowledge of the full input to compute a global property like the parity at any given step. \color{black}
To test this hypothesis and \color{black} help \color{black} inform the model of the full input state, we perform numerical simulations with paths to all \color{black} input \color{black} qubits instead of a single random path. 
The combined paths in the worst case can correspond to the causal cone of the observable, which has been considered for optimization in previous works.~\cite{benedetti2021hardware}
Furthermore we carry out the simulation with the paths alongside the combined paths with a reduced learning rate of 0.05 and plot the results in Fig.~\ref{fig:vqc_results_1}.
The reason for training with a reduced learning rate is that we observe oscillations in the cost function with the combined paths.

We note that the trajectories from the optimization with the combined paths outperform the optimization with individual paths consistently.
This is the expected behavior as the model has access to the full data for the classification as compared to the case with individual paths.
We further point out that the optimization with the random path on average has a higher convergence rate as compared to the one with the shortest path.
This can be due to the fact that it can have access to a larger number of parameters as compared to the shortest path.
While this suggests that one needs larger number of parameters here, we point that we can still find the optimal solution using single paths which only depend on a subset of the parameters.

Overall, we have presented empirical evidence that our method based on the path (defined using a distance measure) can be used to successfully optimize variational algorithms. \color{black} The observed performances seem to indicate that the current form of the metric works well on problems with a degree of locality (here: the Heisenberg model), while modification in the sampling strategy was needed for problems requiring global information (here: parity classification). We conjecture that this may be a result of a bias towards locality inherent in the metric, i.e., a tendency for the metric-based paths to stay close to the readout qubits from which it originates. Strategies to reduce this local focus might include tuning the magnitude of the two terms in Eq. \ref{eq:metric} or using more complicated metric-based sampling than picking the shortest path. Note that, beyond investigating these modifications, further systematic investigations into similarities and differences across a broader range of optimization strategies based on parameter subsets also seems promising, since it could potentially allow for the development of heuristics to determine which strategy to use for a given problem, as well as the construction of hybrid strategies able to combine the strengths of different approaches. We leave investigations of all of the above questions and strategies as future work. \color{black}

\section{Conclusion}\label{sec:conclusion}
In this work, we have proposed a notion of information flow by defining a path in parameterized quantum circuits.
We have presented a novel measure of distance between two points in the circuit by using mutual information between the quantum states that a local unitary acts on.
The distance can be calculated efficiently as it does not rely on global parameters but only on the local unitary operator.
We also present a strategy for optimizing parameterized circuits using paths for variational quantum algorithms.

We performed numerical experiments to estimate the ground state energy of the XXZ-Heisenberg model as well as do n-bit binary classification, using parameterized circuits of varying size and depth.
The results from the numerical simulations provide empirical evidence that our method can be successfully used for these tasks.

Our work is an initial step toward using path-based information flow for the optimization of quantum circuits.
While we have demonstrated consistent improvement for smaller problem instances, a systematic investigation of the scaling of our method for sufficiently deep circuits could be worth exploring.
Other questions such as if forcing information along paths can help mitigate the observed barren plateau phenomenon or remove redundant parameterization of quantum circuits are left for future research. Furthermore, a future research endeavor is to investigate utilizing the distance metric for defining parameter subsets beyond the shortest path problem, such as the maximum flow problem.
We believe that results from this study can be useful to researchers studying the optimization and design of quantum circuits.

\section*{Acknowledgements}
The authors thank Philipp Schleich and Jakob S. Kottmann for providing valuable feedback regarding the manuscript.
A.A.-G. acknowledges the generous support from Google, Inc.  in the form of a Google Focused Award.
A.A.-G. also acknowledges support from the Canada Industrial Research Chairs  Program and the Canada 150 Research Chairs Program. L.B.K acknowledges support from the Carlsberg Foundation. Computations were performed on the niagara supercomputer at the SciNet HPC Consortium.~\cite{niagara1, niagara2} SciNet is funded by: the Canada Foundation for Innovation; the Government of Ontario; Ontario Research Fund - Research Excellence; and the University of Toronto.

\appendix

\color{black}
\section{Motivation for the Proposed Metric}
\label{ap:Motivation}
\begin{figure}[htbp]
  \centering
   \includegraphics[width=0.85\columnwidth]{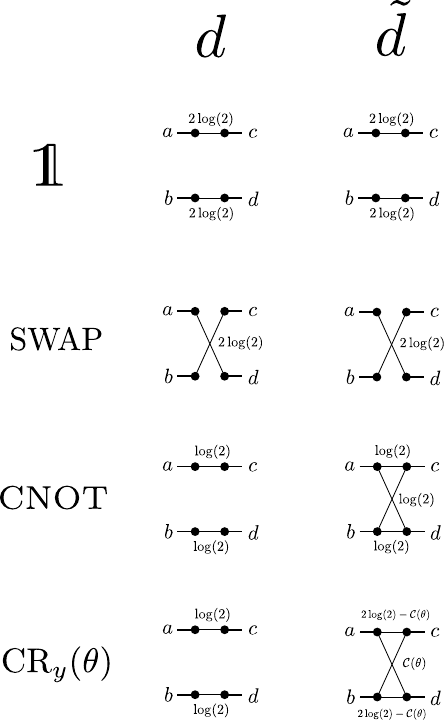}
   \caption{\color{black}Graphical comparison of mutual informations $\{I(i:j)\}$ used to calculate the metric $d$ from \cite{hyatt2017extracting} compared to to the mutual informations $\{I(i:j), \frac{1}{2}I(ac:bd)\}$ used in the calculation of the metric $\tilde{d}$ proposed in this work. Omitted lines correspond to zero mutual information. For a definition of $\mathcal{C}(\theta)$, see Eq.~\ref{eq:C}.\color{black}}
   \label{fig:Examples}
\end{figure}

In this appendix, we provide some motivation for the information metric proposed in Eq.~\ref{eq:metric}. As mentioned in the main text, the metric is based on previous work presented in Ref.~\cite{hyatt2017extracting}. In that paper, a metric was proposed of the form
\begin{align}
    d_{i,j} = - \log\left(\frac{I(i:j)}{2 \log(2)}\right)
    \label{eq:previous_metric}
\end{align}
While this is conceptually simpler than our construction, and was shown to work well for general unitaries, it turns out not to be well-suited for the specific classes of unitaries used in PQCs. Specifically, it does not account correctly for the information transfer facilitated by controlled gates. To see this, consider a gate that applies a unitary $u_{b,d}$ on the lower qubit in Fig.~\ref{fig:Label_Definition} conditioned on the state of the upper qubit. In this case, the unitary takes the form
\begin{align}
    U_{a,b,c,d} = \delta_{a,c} \left( \delta_{a,0} \delta_{b,d} + \delta_{a,1} u_{b,d} \right)  .
\end{align}
An explicit evaluation of $\rho^{a,c}$ in general takes the form
\begin{align*}
    \rho^{a,d} &\propto  \text{Tr}_{b,c} \left( \sum_{\substack{a,b,c,d\\a',b',c',d'}}   U_{a,b,c,d} U_{a',b',c',d'}^* \right.\\
    &\left. \hspace{2cm}\vphantom{\sum_{\substack{a,b,c,d\\a',b',c',d'}}} \times \left| a, b, c, d \right> \left< a', b', c', d' \right| \right) \\
    &= \sum_{\substack{a,b,c,d\\a',d'}}   U_{a,b,c,d} U_{a',b,c,d'}^* \left| a,  d \right> \left< a', d' \right|  .
\end{align*}
Looking closer at the coefficient in this expression and plugging in the expression for the controlled unitary, we see that
\begin{align*}
    &\sum_{a'} U_{a,b,c,d} U_{a',b,c,d'}^* \\
    &= \sum_{a'} \delta_{a,c} \left( \delta_{a,0} \delta_{b,d} + \delta_{a,1} u_{b,d} \right) \delta_{a',c} \left( \delta_{a',0} \delta_{b,d'} + \delta_{a',1} u_{b,d'}^* \right)\\
    &= \left( \delta_{a,0} \delta_{b,d} + \delta_{a,1} u_{b,d} \right)  \left( \delta_{a,0} \delta_{b,d'} + \delta_{a,1} u_{b,d'}^* \right)\\
    &= \delta_{a,0} \delta_{b,d} \delta_{b,d'} + \delta_{a,1} u_{b,d} u_{b,d'}^*   ,
\end{align*}
which means
\begin{align*}
    \rho^{a,d} &\propto \sum_{\substack{a,b,d\\d'}} \left( \delta_{a,0} \delta_{b,d} \delta_{b,d'} + \delta_{a,1} u_{b,d} u_{b,d'}^* \right)  \left| a, d \right> \left< a, d' \right|\\
    &=\sum_{b} \left| 0, b \right> \left< 0,b \right| +\sum_{\substack{a,b,d\\d'}} \delta_{a,1} u_{b,d} u_{b,d'}^*   \left| 1, d \right> \left< 1, d' \right|  .
\end{align*}
However, since $u_{b,d}$ is unitary, we also have that
\begin{align*}
    \sum_{b} u_{b,d} u_{b,d'}^* = \sum_{b} u_{b,d} u^\dagger_{d',b} = \delta_{d,d'}  ,
\end{align*}
from which it follows that
\begin{align*}
    \rho^{a,d} &\propto \sum_{b} \left| 0, b \right> \left< 0,b \right| + \sum_{d} \left| 1, d \right> \left< 1, d \right|\\
    &= \mathds{1}_{4\times4}  ,
\end{align*}
where $\mathds{1}_{4\times4}$ is the 4 by 4 identity matrix. In other words, the density matrix $\rho^{a,d}$ corresponds to the maximally mixed state. Similar arguments applies to $\rho^{b,c}$. Thus, after normalization we have
\begin{align*}
    \rho^{b,c} = \rho^{a,d} = \frac{1}{4} \mathds{1}_{4\times4}.
\end{align*}
Tracing out an additional qubit and evaluating the relevant entropies, we therefore get
\begin{align*}
S\left(\rho^i\right) &= \log(2) \hspace{0.4cm} \forall i \in \{a,b,c,d\}\\
S\left(\rho^{ad} \right) &= S\left(\rho^{bc} \right) = 2\log(2)  ,
\end{align*}
and thus the following mutual informations must evaluate to zero:
\begin{align*}
I(a:d) = I(b:c) = 0  .
\end{align*}
This in turn means that no information transfer between the nodes corresponding to different qubits is predicted by the metric in Eq.~\ref{eq:previous_metric} for controlled gates, a prediction that does not reflect their entanglement-generating capabilities. While controlled rotations represent only a small part of the space of possible 2-qubit unitaries, they play an important role in PQC constructions, making accurate representation of their correlation-creating abilities important in this domain. Since those capabilities seem well-captured by the correlation between what transformation is performed on one qubit (e.g., the $(a,c)$-transformation) and what transformation is performed on the other qubit (e.g., the $(b,d)$-transformation), the quantity $I(ac:bd)$ seems a natural mutual information for quantifying correlations in this setting. As shown in Fig.~\ref{fig:distance} and Fig.~\ref{fig:Examples}, the updated metric indeed seems to work as intended for the classes of gates considered in this work. As an example, explicit calculation for the gate CR${}_y(\theta)$ yields an information transfer characterized by the quantity
\begin{align}
    \mathcal{C} (\theta) &\equiv \frac{1}{2} \left\{ \left(1 + \left|\cos\left(\frac{\theta}{2}\right)\right|\right) \log \left( \frac{2}{1 + \left|\cos\left(\frac{\theta}{2}\right)\right|} \right) \right. \nonumber \\
    &+ \left. \left(1 - \left|\cos\left(\frac{\theta}{2}\right)\right|\right) \log \left( \frac{2}{1 - \left|\cos\left(\frac{\theta}{2}\right)\right|}  \right) \right\} \nonumber\\
    &= \frac{1}{2} I(ac:bd)  ,
    \label{eq:C}
\end{align}
and gives rise to the non-trivial distance metric depicted in Fig.~\ref{fig:distance}.

\section{Properties of the Proposed Metric}
In this appendix, we briefly mention two potentially useful properties of the objects appearing in the definition of the metric. One relates to the entropy of the single-qubit reduced density matrices within our framework, while the second identifies something akin to a local conservation of information.
\subsubsection*{All single-qubit reduced denisty matrices are maximally mixed.}
The first observation is that all single-qubit reduced density matrices resulting from the mapping from unitaries to states defined in Eq.~\ref{eq:Mapping_to_States} are maximally mixed, and thus have entropies independent from the gate considered. To see this, consider a state resulting from this mapping:
\begin{align*}
    \rho \propto \sum_{\substack{a,b,c,d\\a',b',c',d'}} U_{a,b,c,d} \, U_{a',b',c',d'}^* \left| a,b,c,d \right> \left< a',b',c',d'\right|  .
\end{align*}
Tracing out all other degrees of freedom than $a$ yields a reduced density matrix
\begin{align*}
    \rho^a &\propto\sum_{\substack{a,b,c,d\\a'}} U_{a,b,c,d} \, U_{a',b,c,d}^* \left| a \right> \left< a'\right|\\
    &= \sum_{\substack{a,b,c,d\\a'}} U_{a,b,c,d} \, U^\dagger_{c,d,a',b} \left| a \right> \left< a'\right|  .
\end{align*}
However, from unitarity it follows that
\begin{align*}
    \sum_{c,d} U_{a,b,c,d} \, U^\dagger_{c,d,a',b} = \delta_{a,a'}  ,
\end{align*}
meaning
\begin{align*}
    \rho^a &\propto \sum_{\substack{a,b\\a'}} \delta_{a, a'} \left| a \right> \left< a'\right|\\
    & \propto \sum_{a} \left| a \right> \left< a\right|  .
\end{align*}
Thus, $\rho^a = \frac{1}{2} \mathds{1}_{2\times2}$ after taking normalization into account. Similar methods yield the same form for the other single-qubit reduced density matrices, implying the following property:
\begin{align*}
    S(\rho^i) &= \log(2) & \forall i \in \{a,b,c,d\}  .
\end{align*}

\subsubsection*{Conservation law for mutual information}
A simple observation based on the result of the previous section and the definitions of the mutual information used in this work is that the sum of the following re-weighted mutual information across the four links corresponding to a 2-qubit gate fulfills:
\begin{align*}
    I(d:b) &+ I(c:a) + 2\cdot \frac{1}{2} I(ac:bd) \\
    &= S\left(\rho^d \right) + S\left(\rho^b \right) - S\left(\rho^{bd} \right) \\ 
    & \hspace{0.8cm} + S\left(\rho^c \right) + S\left(\rho^a \right) - S\left(\rho^{ac} \right)\\
    &\hspace{0.8cm} + S\left(\rho^{ac} \right) + S\left(\rho^{bd} \right) \\
&= S\left(\rho^d \right) + S\left(\rho^b \right)+  S\left(\rho^c \right) + S\left(\rho^a \right) \\
&= 4 \log(2)  .
\end{align*}
If mutual information is interpreted as a flow of information (as in this work), this represents a conservation-like law stating that the flow of information can be redistributed across different edges, but that the total flow is fixed. While this fact was not explicitly used  in the work here, it may represent an independently useful way of reasoning about information flow in quantum circuits. We leave further investigations of this viewpoint as future work.

\begin{figure*}[htbp!]
    \centering
    \begin{tabular}{c c c}
    \toprule
    \multicolumn{3}{c}{\textbf{6 Qubit Ansatz}}\\
    \midrule
    a) $\Delta = 0.5$ & b) $\Delta = 1.0$ & c) $\Delta = 1.5$ \\
    \midrule
    \includegraphics[width=0.32\textwidth]{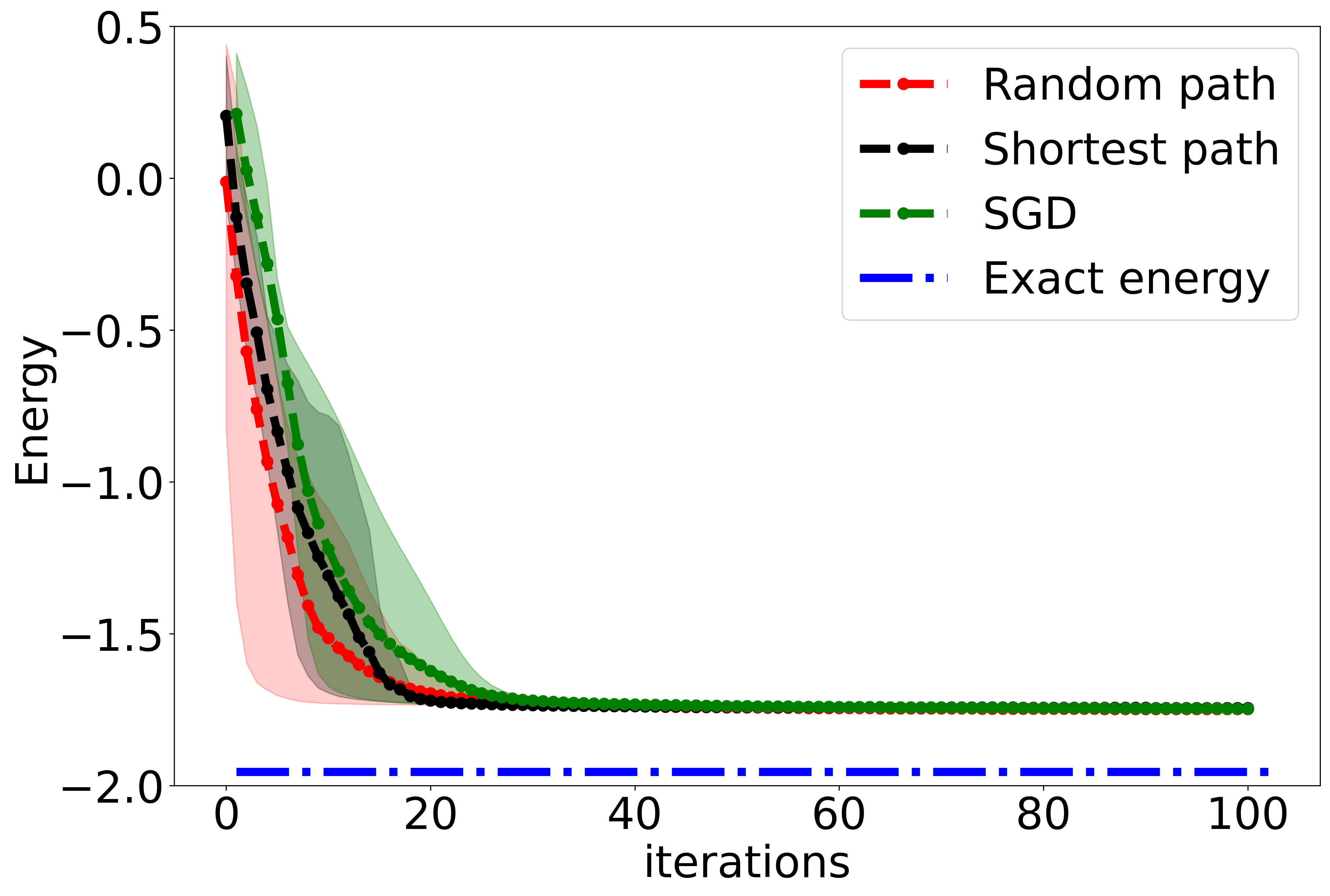} & 
    \includegraphics[width=0.32\textwidth]{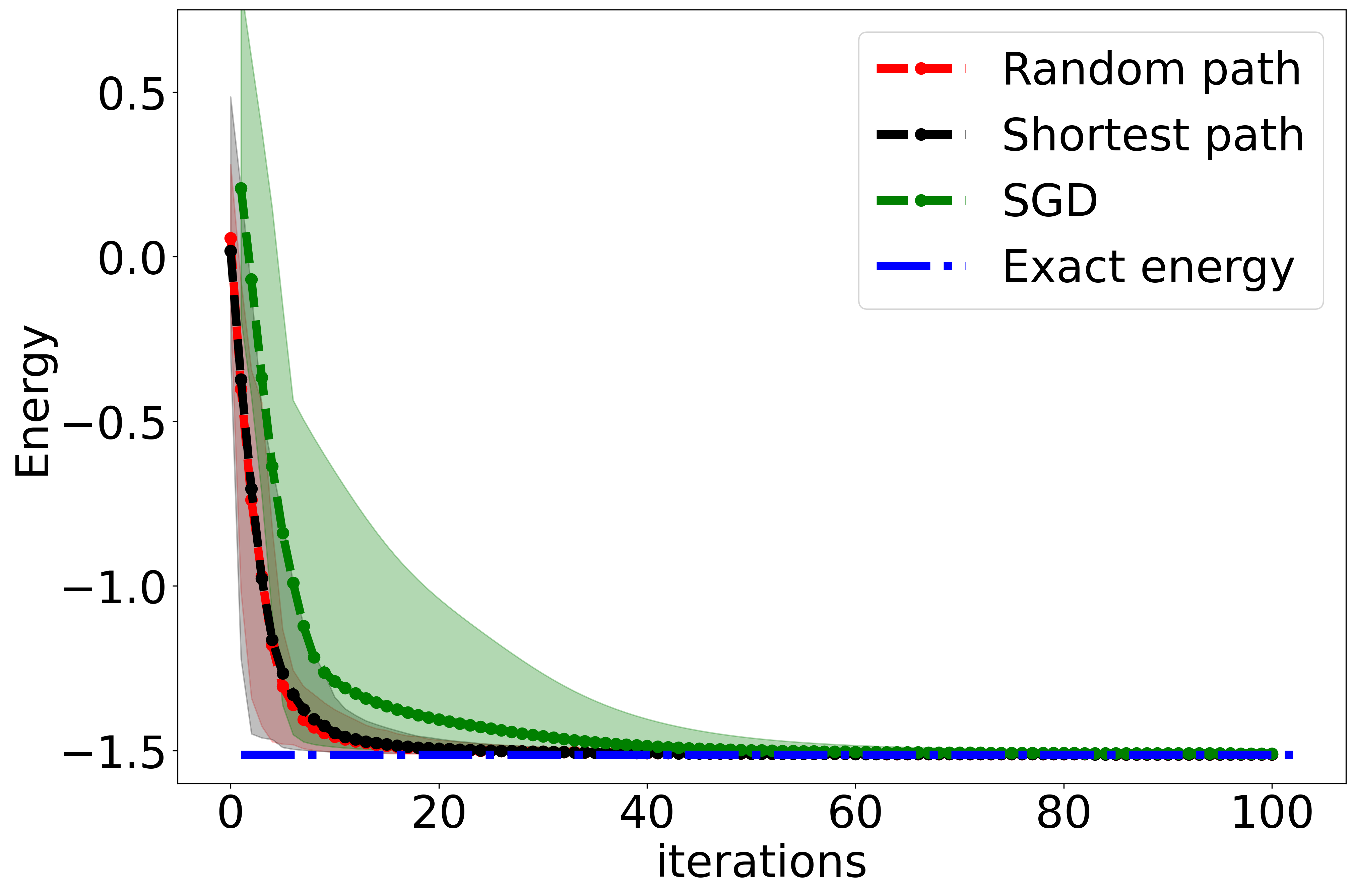} &
    \includegraphics[width=0.32\textwidth]{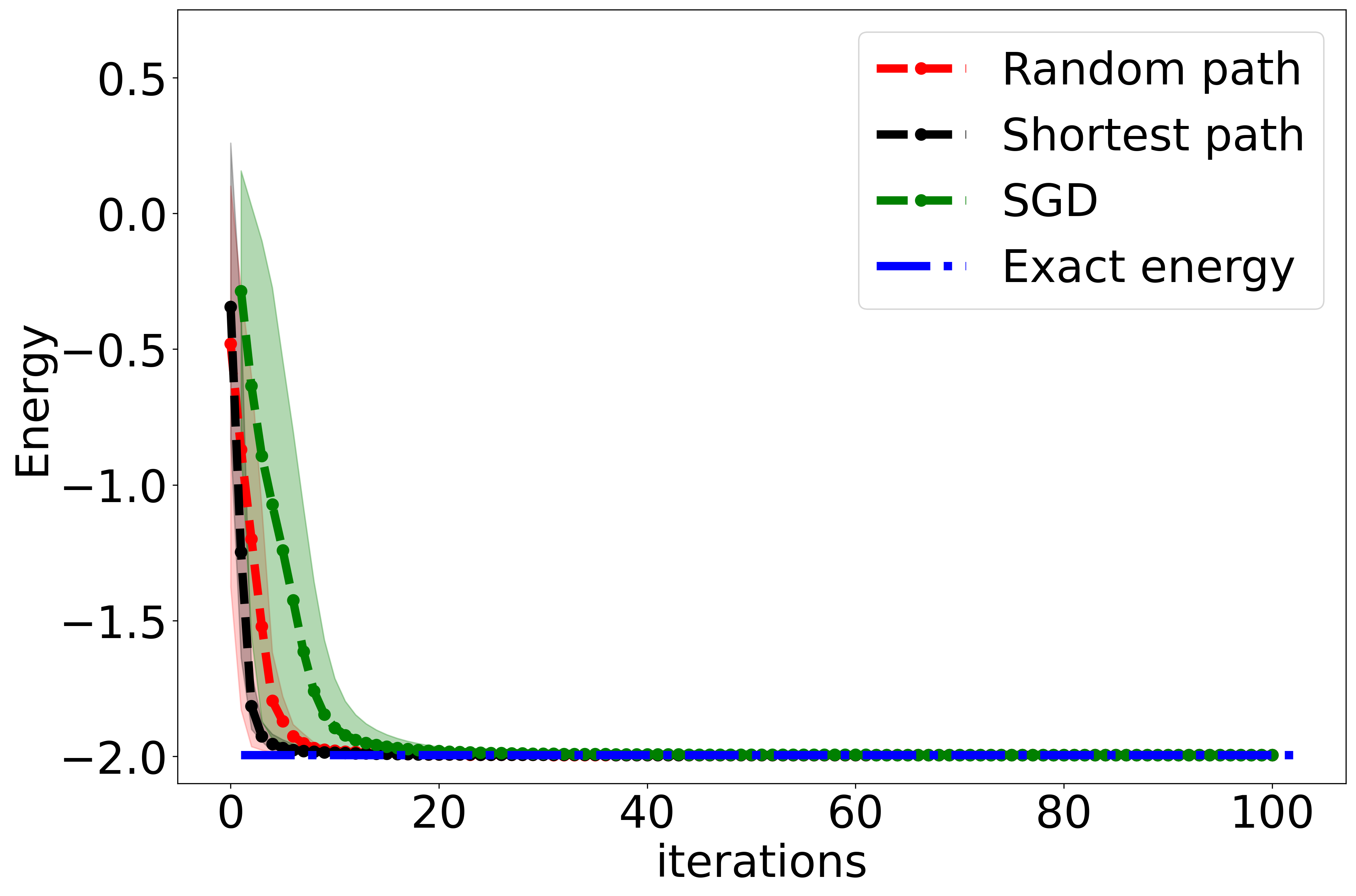}\\
    \midrule
    \multicolumn{3}{c}{\textbf{8 Qubit Ansatz}}\\
    \midrule
    a) $\Delta = 0.5$ & b) $\Delta = 1.0$ & c) $\Delta = 1.5$ \\
    \midrule
    \includegraphics[width=0.329\textwidth]{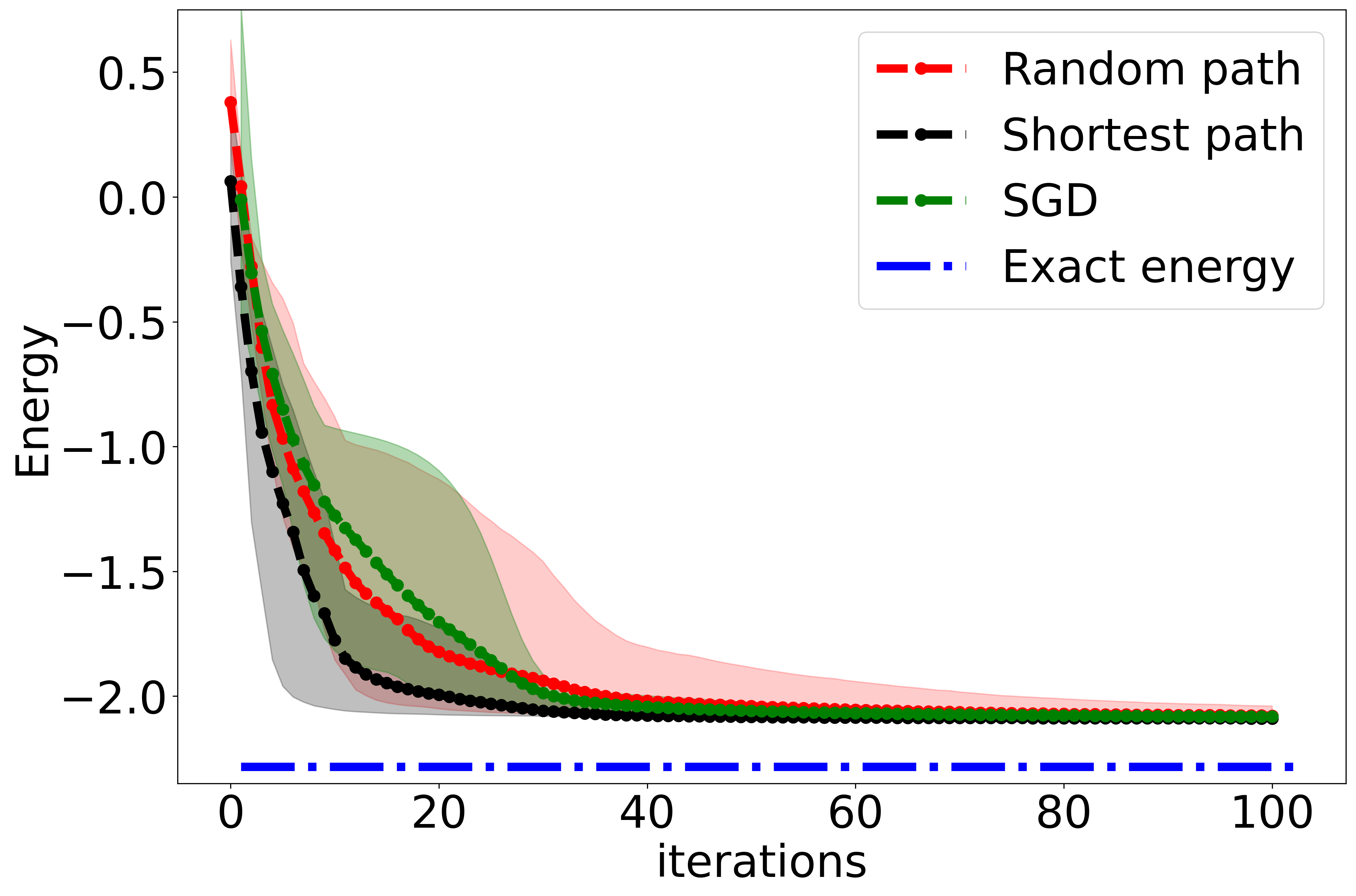} & 
    \includegraphics[width=0.329\textwidth]{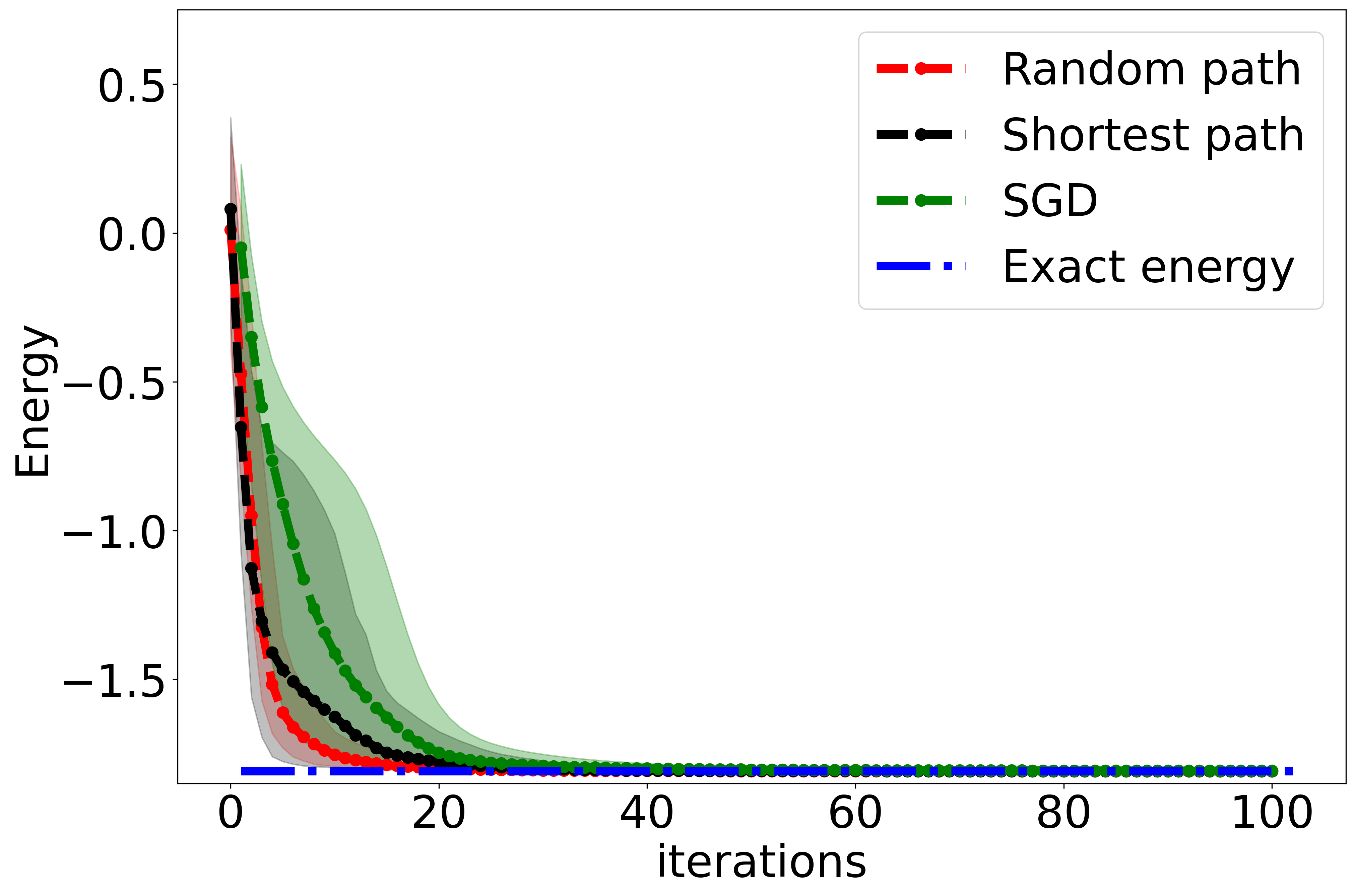} &
    \includegraphics[width=0.329\textwidth]{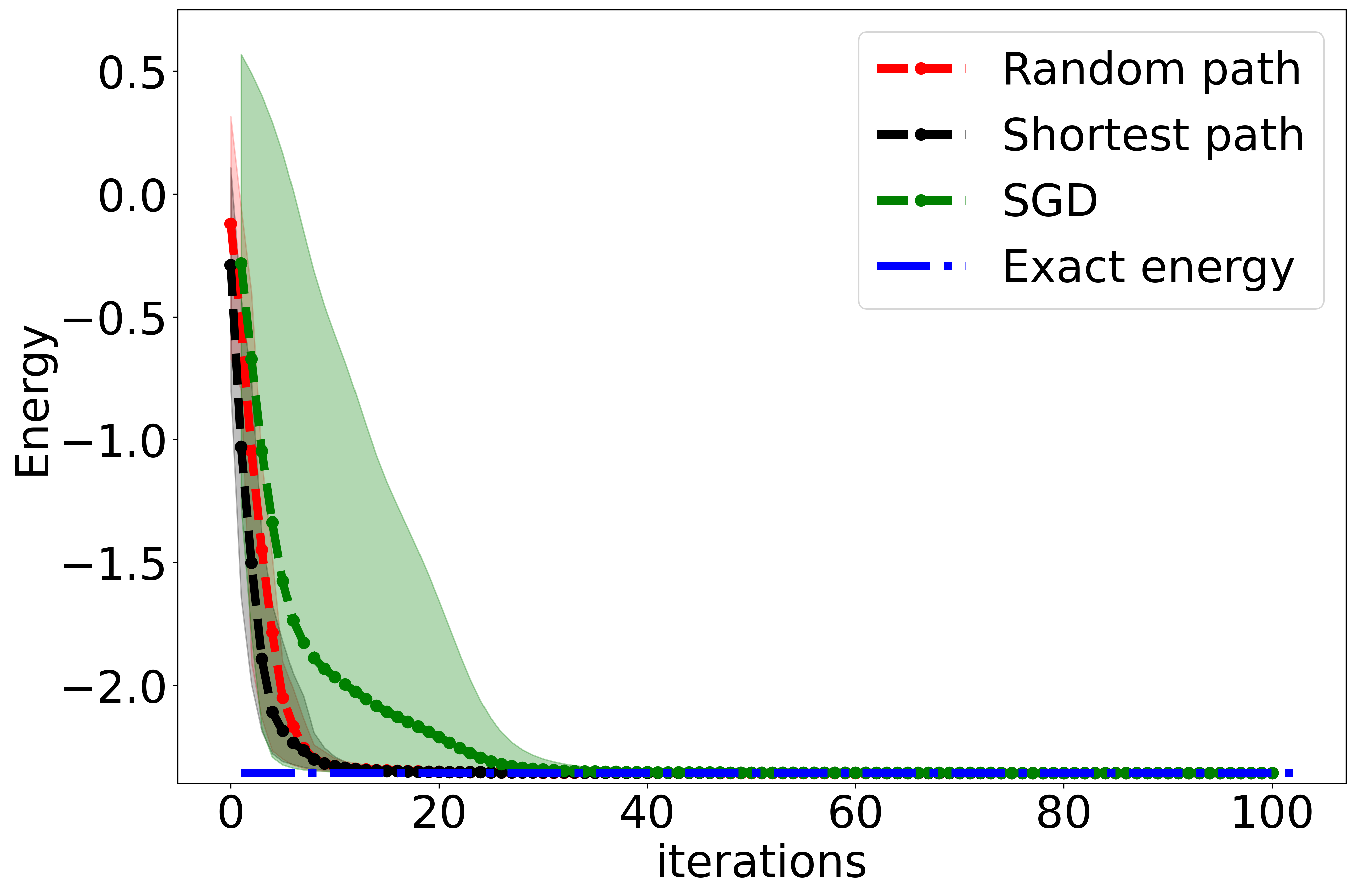}\\
    \end{tabular}
    \caption{\label{fig:vqc_resultsdiff_params} Optimization trajectories from different VQE simulations of the different XXZ-Hamiltonians. The lines correspond to the mean of the trajectories from different runs, and the shadow represent the area between the best and worst values from the simulations.}
\end{figure*}

\section{VQE- XXZ-Hesisenberg model}
\label{ap:more_heisenberg}
In this section we analyze the performance of our method to find the ground state energy of the XXZ model in three different parameter regimes.
We fix the values of the constants to $h=0$ (no external magnetic field) and $J = 1.0$, but vary $\Delta \in \{ 0.5, 1.0 ,1.5\}$ and run experiments for two different lattice sizes, $3\times 2$ qubits, and $4\times 2$ qubits.
For the simulation we use the ansatz shown in Fig.~\ref{fig:cir_graph_conversion}(b) with a single layer.
The optimization is carried out using the algorithm presented in Algorithm~\ref{al:stoc_op} and the stochastic gradient descent algorithm, with a fixed learning rate of $0.5$.
The results from all the simulations are plotted in Fig.~\ref{fig:vqc_resultsdiff_params}.
As can be seen from the optimization trajectories in Fig.~\ref{fig:vqc_resultsdiff_params}, the overall performance of the algorithms remain similar to that observed earlier in section~\ref{sec:simres} A.
However, we do observe that for the case when $\Delta = 0.5$, the final energy of the optimization doesn't converge to the true energy.
This can be attributed to the fact that the ansatz that we use is unable to reach the ground state of that configuration, and one needs to use a better ansatz to get better convergence.

\color{black}

\bibliography{references.bib}

\begin{thebibliography}{49}%
\makeatletter
\providecommand \@ifxundefined [1]{%
 \@ifx{#1\undefined}
}%
\providecommand \@ifnum [1]{%
 \ifnum #1\expandafter \@firstoftwo
 \else \expandafter \@secondoftwo
 \fi
}%
\providecommand \@ifx [1]{%
 \ifx #1\expandafter \@firstoftwo
 \else \expandafter \@secondoftwo
 \fi
}%
\providecommand \natexlab [1]{#1}%
\providecommand \enquote  [1]{``#1''}%
\providecommand \bibnamefont  [1]{#1}%
\providecommand \bibfnamefont [1]{#1}%
\providecommand \citenamefont [1]{#1}%
\providecommand \href@noop [0]{\@secondoftwo}%
\providecommand \href [0]{\begingroup \@sanitize@url \@href}%
\providecommand \@href[1]{\@@startlink{#1}\@@href}%
\providecommand \@@href[1]{\endgroup#1\@@endlink}%
\providecommand \@sanitize@url [0]{\catcode `\\12\catcode `\$12\catcode
  `\&12\catcode `\#12\catcode `\^12\catcode `\_12\catcode `\%12\relax}%
\providecommand \@@startlink[1]{}%
\providecommand \@@endlink[0]{}%
\providecommand \url  [0]{\begingroup\@sanitize@url \@url }%
\providecommand \@url [1]{\endgroup\@href {#1}{\urlprefix }}%
\providecommand \urlprefix  [0]{URL }%
\providecommand \Eprint [0]{\href }%
\providecommand \doibase [0]{https://doi.org/}%
\providecommand \selectlanguage [0]{\@gobble}%
\providecommand \bibinfo  [0]{\@secondoftwo}%
\providecommand \bibfield  [0]{\@secondoftwo}%
\providecommand \translation [1]{[#1]}%
\providecommand \BibitemOpen [0]{}%
\providecommand \bibitemStop [0]{}%
\providecommand \bibitemNoStop [0]{.\EOS\space}%
\providecommand \EOS [0]{\spacefactor3000\relax}%
\providecommand \BibitemShut  [1]{\csname bibitem#1\endcsname}%
\let\auto@bib@innerbib\@empty
\bibitem [{\citenamefont {Bharti}\ \emph {et~al.}(2022)\citenamefont {Bharti},
  \citenamefont {Cervera-Lierta}, \citenamefont {Kyaw}, \citenamefont {Haug},
  \citenamefont {Alperin-Lea}, \citenamefont {Anand}, \citenamefont {Degroote},
  \citenamefont {Heimonen}, \citenamefont {Kottmann}, \citenamefont {Menke}
  \emph {et~al.}}]{bharti2022noisy}%
  \BibitemOpen
  \bibfield  {author} {\bibinfo {author} {\bibfnamefont {K.}~\bibnamefont
  {Bharti}}, \bibinfo {author} {\bibfnamefont {A.}~\bibnamefont
  {Cervera-Lierta}}, \bibinfo {author} {\bibfnamefont {T.~H.}\ \bibnamefont
  {Kyaw}}, \bibinfo {author} {\bibfnamefont {T.}~\bibnamefont {Haug}}, \bibinfo
  {author} {\bibfnamefont {S.}~\bibnamefont {Alperin-Lea}}, \bibinfo {author}
  {\bibfnamefont {A.}~\bibnamefont {Anand}}, \bibinfo {author} {\bibfnamefont
  {M.}~\bibnamefont {Degroote}}, \bibinfo {author} {\bibfnamefont
  {H.}~\bibnamefont {Heimonen}}, \bibinfo {author} {\bibfnamefont {J.~S.}\
  \bibnamefont {Kottmann}}, \bibinfo {author} {\bibfnamefont {T.}~\bibnamefont
  {Menke}}, \emph {et~al.},\ }\href@noop {} {\bibfield  {journal} {\bibinfo
  {journal} {Reviews of Modern Physics}\ }\textbf {\bibinfo {volume} {94}},\
  \bibinfo {pages} {015004} (\bibinfo {year} {2022})}\BibitemShut {NoStop}%
\bibitem [{\citenamefont {Cerezo}\ \emph {et~al.}(2021)\citenamefont {Cerezo},
  \citenamefont {Arrasmith}, \citenamefont {Babbush}, \citenamefont {Benjamin},
  \citenamefont {Endo}, \citenamefont {Fujii}, \citenamefont {McClean},
  \citenamefont {Mitarai}, \citenamefont {Yuan}, \citenamefont {Cincio} \emph
  {et~al.}}]{cerezo2021variational}%
  \BibitemOpen
  \bibfield  {author} {\bibinfo {author} {\bibfnamefont {M.}~\bibnamefont
  {Cerezo}}, \bibinfo {author} {\bibfnamefont {A.}~\bibnamefont {Arrasmith}},
  \bibinfo {author} {\bibfnamefont {R.}~\bibnamefont {Babbush}}, \bibinfo
  {author} {\bibfnamefont {S.~C.}\ \bibnamefont {Benjamin}}, \bibinfo {author}
  {\bibfnamefont {S.}~\bibnamefont {Endo}}, \bibinfo {author} {\bibfnamefont
  {K.}~\bibnamefont {Fujii}}, \bibinfo {author} {\bibfnamefont {J.~R.}\
  \bibnamefont {McClean}}, \bibinfo {author} {\bibfnamefont {K.}~\bibnamefont
  {Mitarai}}, \bibinfo {author} {\bibfnamefont {X.}~\bibnamefont {Yuan}},
  \bibinfo {author} {\bibfnamefont {L.}~\bibnamefont {Cincio}}, \emph
  {et~al.},\ }\href@noop {} {\bibfield  {journal} {\bibinfo  {journal} {Nature
  Reviews Physics}\ }\textbf {\bibinfo {volume} {3}},\ \bibinfo {pages} {625}
  (\bibinfo {year} {2021})}\BibitemShut {NoStop}%
\bibitem [{\citenamefont {Anand}\ \emph
  {et~al.}(2022{\natexlab{a}})\citenamefont {Anand}, \citenamefont {Schleich},
  \citenamefont {Alperin-Lea}, \citenamefont {Jensen}, \citenamefont {Sim},
  \citenamefont {D{\'\i}az-Tinoco}, \citenamefont {Kottmann}, \citenamefont
  {Degroote}, \citenamefont {Izmaylov},\ and\ \citenamefont
  {Aspuru-Guzik}}]{anand2022quantum}%
  \BibitemOpen
  \bibfield  {author} {\bibinfo {author} {\bibfnamefont {A.}~\bibnamefont
  {Anand}}, \bibinfo {author} {\bibfnamefont {P.}~\bibnamefont {Schleich}},
  \bibinfo {author} {\bibfnamefont {S.}~\bibnamefont {Alperin-Lea}}, \bibinfo
  {author} {\bibfnamefont {P.~W.}\ \bibnamefont {Jensen}}, \bibinfo {author}
  {\bibfnamefont {S.}~\bibnamefont {Sim}}, \bibinfo {author} {\bibfnamefont
  {M.}~\bibnamefont {D{\'\i}az-Tinoco}}, \bibinfo {author} {\bibfnamefont
  {J.~S.}\ \bibnamefont {Kottmann}}, \bibinfo {author} {\bibfnamefont
  {M.}~\bibnamefont {Degroote}}, \bibinfo {author} {\bibfnamefont {A.~F.}\
  \bibnamefont {Izmaylov}},\ and\ \bibinfo {author} {\bibfnamefont
  {A.}~\bibnamefont {Aspuru-Guzik}},\ }\href@noop {} {\bibfield  {journal}
  {\bibinfo  {journal} {Chemical Society Reviews}\ } (\bibinfo {year}
  {2022}{\natexlab{a}})}\BibitemShut {NoStop}%
\bibitem [{\citenamefont {Peruzzo}\ \emph {et~al.}(2014)\citenamefont
  {Peruzzo}, \citenamefont {McClean}, \citenamefont {Shadbolt}, \citenamefont
  {Yung}, \citenamefont {Zhou}, \citenamefont {Love}, \citenamefont
  {Aspuru-Guzik},\ and\ \citenamefont {O'Brien}}]{vqe_cite}%
  \BibitemOpen
  \bibfield  {author} {\bibinfo {author} {\bibfnamefont {A.}~\bibnamefont
  {Peruzzo}}, \bibinfo {author} {\bibfnamefont {J.}~\bibnamefont {McClean}},
  \bibinfo {author} {\bibfnamefont {P.}~\bibnamefont {Shadbolt}}, \bibinfo
  {author} {\bibfnamefont {M.-H.}\ \bibnamefont {Yung}}, \bibinfo {author}
  {\bibfnamefont {X.-Q.}\ \bibnamefont {Zhou}}, \bibinfo {author}
  {\bibfnamefont {P.~J.}\ \bibnamefont {Love}}, \bibinfo {author}
  {\bibfnamefont {A.}~\bibnamefont {Aspuru-Guzik}},\ and\ \bibinfo {author}
  {\bibfnamefont {J.~L.}\ \bibnamefont {O'Brien}},\ }\href
  {https://doi.org/10.1038/ncomms5213} {\bibfield  {journal} {\bibinfo
  {journal} {Nature Communications}\ }\textbf {\bibinfo {volume} {5}},\
  \bibinfo {pages} {4213} (\bibinfo {year} {2014})}\BibitemShut {NoStop}%
\bibitem [{\citenamefont {Farhi}\ \emph {et~al.}(2014)\citenamefont {Farhi},
  \citenamefont {Goldstone},\ and\ \citenamefont {Gutmann}}]{qoao_cite}%
  \BibitemOpen
  \bibfield  {author} {\bibinfo {author} {\bibfnamefont {E.}~\bibnamefont
  {Farhi}}, \bibinfo {author} {\bibfnamefont {J.}~\bibnamefont {Goldstone}},\
  and\ \bibinfo {author} {\bibfnamefont {S.}~\bibnamefont {Gutmann}},\ }\href
  {https://doi.org/10.48550/ARXIV.1411.4028} {\bibinfo {title} {A quantum
  approximate optimization algorithm}} (\bibinfo {year} {2014})\BibitemShut
  {NoStop}%
\bibitem [{\citenamefont {Havl{\'i}{\v{c}}ek}\ \emph
  {et~al.}(2019{\natexlab{a}})\citenamefont {Havl{\'i}{\v{c}}ek}, \citenamefont
  {C{\'o}rcoles}, \citenamefont {Temme}, \citenamefont {Harrow}, \citenamefont
  {Kandala}, \citenamefont {Chow},\ and\ \citenamefont {Gambetta}}]{mlc_1}%
  \BibitemOpen
  \bibfield  {author} {\bibinfo {author} {\bibfnamefont {V.}~\bibnamefont
  {Havl{\'i}{\v{c}}ek}}, \bibinfo {author} {\bibfnamefont {A.~D.}\ \bibnamefont
  {C{\'o}rcoles}}, \bibinfo {author} {\bibfnamefont {K.}~\bibnamefont {Temme}},
  \bibinfo {author} {\bibfnamefont {A.~W.}\ \bibnamefont {Harrow}}, \bibinfo
  {author} {\bibfnamefont {A.}~\bibnamefont {Kandala}}, \bibinfo {author}
  {\bibfnamefont {J.~M.}\ \bibnamefont {Chow}},\ and\ \bibinfo {author}
  {\bibfnamefont {J.~M.}\ \bibnamefont {Gambetta}},\ }\href
  {https://doi.org/10.1038/s41586-019-0980-2} {\bibfield  {journal} {\bibinfo
  {journal} {Nature}\ }\textbf {\bibinfo {volume} {567}},\ \bibinfo {pages}
  {209} (\bibinfo {year} {2019}{\natexlab{a}})}\BibitemShut {NoStop}%
\bibitem [{\citenamefont {Farhi}\ and\ \citenamefont
  {Neven}(2018{\natexlab{a}})}]{vqc_ref_1}%
  \BibitemOpen
  \bibfield  {author} {\bibinfo {author} {\bibfnamefont {E.}~\bibnamefont
  {Farhi}}\ and\ \bibinfo {author} {\bibfnamefont {H.}~\bibnamefont {Neven}},\
  }\href {https://doi.org/10.48550/ARXIV.1802.06002} {\bibinfo {title}
  {Classification with quantum neural networks on near term processors}}
  (\bibinfo {year} {2018}{\natexlab{a}})\BibitemShut {NoStop}%
\bibitem [{\citenamefont {Schuld}\ \emph {et~al.}(2020)\citenamefont {Schuld},
  \citenamefont {Bocharov}, \citenamefont {Svore},\ and\ \citenamefont
  {Wiebe}}]{vqc_ref_2}%
  \BibitemOpen
  \bibfield  {author} {\bibinfo {author} {\bibfnamefont {M.}~\bibnamefont
  {Schuld}}, \bibinfo {author} {\bibfnamefont {A.}~\bibnamefont {Bocharov}},
  \bibinfo {author} {\bibfnamefont {K.~M.}\ \bibnamefont {Svore}},\ and\
  \bibinfo {author} {\bibfnamefont {N.}~\bibnamefont {Wiebe}},\ }\bibfield
  {journal} {\bibinfo  {journal} {Physical Review A}\ }\textbf {\bibinfo
  {volume} {101}},\ \href {https://doi.org/10.1103/physreva.101.032308}
  {10.1103/physreva.101.032308} (\bibinfo {year} {2020})\BibitemShut {NoStop}%
\bibitem [{\citenamefont {Romero}\ and\ \citenamefont
  {Aspuru-Guzik}(2019)}]{mlg_1}%
  \BibitemOpen
  \bibfield  {author} {\bibinfo {author} {\bibfnamefont {J.}~\bibnamefont
  {Romero}}\ and\ \bibinfo {author} {\bibfnamefont {A.}~\bibnamefont
  {Aspuru-Guzik}},\ }\href {https://doi.org/10.48550/ARXIV.1901.00848}
  {\bibinfo {title} {Variational quantum generators: Generative adversarial
  quantum machine learning for continuous distributions}} (\bibinfo {year}
  {2019})\BibitemShut {NoStop}%
\bibitem [{\citenamefont {Zhu}\ \emph {et~al.}(2019)\citenamefont {Zhu},
  \citenamefont {Linke}, \citenamefont {Benedetti}, \citenamefont {Landsman},
  \citenamefont {Nguyen}, \citenamefont {Alderete}, \citenamefont
  {Perdomo-Ortiz}, \citenamefont {Korda}, \citenamefont {Garfoot},
  \citenamefont {Brecque}, \citenamefont {Egan}, \citenamefont {Perdomo},\ and\
  \citenamefont {Monroe}}]{mlg_2}%
  \BibitemOpen
  \bibfield  {author} {\bibinfo {author} {\bibfnamefont {D.}~\bibnamefont
  {Zhu}}, \bibinfo {author} {\bibfnamefont {N.~M.}\ \bibnamefont {Linke}},
  \bibinfo {author} {\bibfnamefont {M.}~\bibnamefont {Benedetti}}, \bibinfo
  {author} {\bibfnamefont {K.~A.}\ \bibnamefont {Landsman}}, \bibinfo {author}
  {\bibfnamefont {N.~H.}\ \bibnamefont {Nguyen}}, \bibinfo {author}
  {\bibfnamefont {C.~H.}\ \bibnamefont {Alderete}}, \bibinfo {author}
  {\bibfnamefont {A.}~\bibnamefont {Perdomo-Ortiz}}, \bibinfo {author}
  {\bibfnamefont {N.}~\bibnamefont {Korda}}, \bibinfo {author} {\bibfnamefont
  {A.}~\bibnamefont {Garfoot}}, \bibinfo {author} {\bibfnamefont
  {C.}~\bibnamefont {Brecque}}, \bibinfo {author} {\bibfnamefont
  {L.}~\bibnamefont {Egan}}, \bibinfo {author} {\bibfnamefont {O.}~\bibnamefont
  {Perdomo}},\ and\ \bibinfo {author} {\bibfnamefont {C.}~\bibnamefont
  {Monroe}},\ }\href {https://doi.org/10.1126/sciadv.aaw9918} {\bibfield
  {journal} {\bibinfo  {journal} {Science Advances}\ }\textbf {\bibinfo
  {volume} {5}},\ \bibinfo {pages} {eaaw9918} (\bibinfo {year} {2019})},\
  \Eprint
  {https://arxiv.org/abs/https://www.science.org/doi/pdf/10.1126/sciadv.aaw9918}
  {https://www.science.org/doi/pdf/10.1126/sciadv.aaw9918} \BibitemShut
  {NoStop}%
\bibitem [{\citenamefont {Situ}\ \emph {et~al.}(2020)\citenamefont {Situ},
  \citenamefont {He}, \citenamefont {Wang}, \citenamefont {Li},\ and\
  \citenamefont {Zheng}}]{mlg_3}%
  \BibitemOpen
  \bibfield  {author} {\bibinfo {author} {\bibfnamefont {H.}~\bibnamefont
  {Situ}}, \bibinfo {author} {\bibfnamefont {Z.}~\bibnamefont {He}}, \bibinfo
  {author} {\bibfnamefont {Y.}~\bibnamefont {Wang}}, \bibinfo {author}
  {\bibfnamefont {L.}~\bibnamefont {Li}},\ and\ \bibinfo {author}
  {\bibfnamefont {S.}~\bibnamefont {Zheng}},\ }\href
  {https://doi.org/https://doi.org/10.1016/j.ins.2020.05.127} {\bibfield
  {journal} {\bibinfo  {journal} {Information Sciences}\ }\textbf {\bibinfo
  {volume} {538}},\ \bibinfo {pages} {193} (\bibinfo {year}
  {2020})}\BibitemShut {NoStop}%
\bibitem [{\citenamefont {Zeng}\ \emph {et~al.}(2019)\citenamefont {Zeng},
  \citenamefont {Wu}, \citenamefont {Liu}, \citenamefont {Wang},\ and\
  \citenamefont {Hu}}]{mlg_4}%
  \BibitemOpen
  \bibfield  {author} {\bibinfo {author} {\bibfnamefont {J.}~\bibnamefont
  {Zeng}}, \bibinfo {author} {\bibfnamefont {Y.}~\bibnamefont {Wu}}, \bibinfo
  {author} {\bibfnamefont {J.-G.}\ \bibnamefont {Liu}}, \bibinfo {author}
  {\bibfnamefont {L.}~\bibnamefont {Wang}},\ and\ \bibinfo {author}
  {\bibfnamefont {J.}~\bibnamefont {Hu}},\ }\href
  {https://doi.org/10.1103/PhysRevA.99.052306} {\bibfield  {journal} {\bibinfo
  {journal} {Phys. Rev. A}\ }\textbf {\bibinfo {volume} {99}},\ \bibinfo
  {pages} {052306} (\bibinfo {year} {2019})}\BibitemShut {NoStop}%
\bibitem [{\citenamefont {Lloyd}\ and\ \citenamefont
  {Weedbrook}(2018)}]{mlg_5}%
  \BibitemOpen
  \bibfield  {author} {\bibinfo {author} {\bibfnamefont {S.}~\bibnamefont
  {Lloyd}}\ and\ \bibinfo {author} {\bibfnamefont {C.}~\bibnamefont
  {Weedbrook}},\ }\href {https://doi.org/10.1103/PhysRevLett.121.040502}
  {\bibfield  {journal} {\bibinfo  {journal} {Phys. Rev. Lett.}\ }\textbf
  {\bibinfo {volume} {121}},\ \bibinfo {pages} {040502} (\bibinfo {year}
  {2018})}\BibitemShut {NoStop}%
\bibitem [{\citenamefont {Dallaire-Demers}\ and\ \citenamefont
  {Killoran}(2018)}]{mlg_6}%
  \BibitemOpen
  \bibfield  {author} {\bibinfo {author} {\bibfnamefont {P.-L.}\ \bibnamefont
  {Dallaire-Demers}}\ and\ \bibinfo {author} {\bibfnamefont {N.}~\bibnamefont
  {Killoran}},\ }\href {https://doi.org/10.1103/PhysRevA.98.012324} {\bibfield
  {journal} {\bibinfo  {journal} {Phys. Rev. A}\ }\textbf {\bibinfo {volume}
  {98}},\ \bibinfo {pages} {012324} (\bibinfo {year} {2018})}\BibitemShut
  {NoStop}%
\bibitem [{\citenamefont {Anand}\ \emph {et~al.}(2021)\citenamefont {Anand},
  \citenamefont {Romero}, \citenamefont {Degroote},\ and\ \citenamefont
  {Aspuru-Guzik}}]{anand2021noise}%
  \BibitemOpen
  \bibfield  {author} {\bibinfo {author} {\bibfnamefont {A.}~\bibnamefont
  {Anand}}, \bibinfo {author} {\bibfnamefont {J.}~\bibnamefont {Romero}},
  \bibinfo {author} {\bibfnamefont {M.}~\bibnamefont {Degroote}},\ and\
  \bibinfo {author} {\bibfnamefont {A.}~\bibnamefont {Aspuru-Guzik}},\
  }\href@noop {} {\bibfield  {journal} {\bibinfo  {journal} {Advanced Quantum
  Technologies}\ }\textbf {\bibinfo {volume} {4}},\ \bibinfo {pages} {2000069}
  (\bibinfo {year} {2021})}\BibitemShut {NoStop}%
\bibitem [{\citenamefont {Sim}\ \emph {et~al.}(2019)\citenamefont {Sim},
  \citenamefont {Johnson},\ and\ \citenamefont {Aspuru-Guzik}}]{vqc_ansatz}%
  \BibitemOpen
  \bibfield  {author} {\bibinfo {author} {\bibfnamefont {S.}~\bibnamefont
  {Sim}}, \bibinfo {author} {\bibfnamefont {P.~D.}\ \bibnamefont {Johnson}},\
  and\ \bibinfo {author} {\bibfnamefont {A.}~\bibnamefont {Aspuru-Guzik}},\
  }\href {https://doi.org/10.1002/qute.201900070} {\bibfield  {journal}
  {\bibinfo  {journal} {Advanced Quantum Technologies}\ }\textbf {\bibinfo
  {volume} {2}},\ \bibinfo {pages} {1900070} (\bibinfo {year}
  {2019})}\BibitemShut {NoStop}%
\bibitem [{\citenamefont {Hubregtsen}\ \emph {et~al.}(2020)\citenamefont
  {Hubregtsen}, \citenamefont {Pichlmeier}, \citenamefont {Stecher},\ and\
  \citenamefont {Bertels}}]{exp_metric}%
  \BibitemOpen
  \bibfield  {author} {\bibinfo {author} {\bibfnamefont {T.}~\bibnamefont
  {Hubregtsen}}, \bibinfo {author} {\bibfnamefont {J.}~\bibnamefont
  {Pichlmeier}}, \bibinfo {author} {\bibfnamefont {P.}~\bibnamefont
  {Stecher}},\ and\ \bibinfo {author} {\bibfnamefont {K.}~\bibnamefont
  {Bertels}},\ }\href {https://doi.org/10.48550/ARXIV.2003.09887} {\bibinfo
  {title} {Evaluation of parameterized quantum circuits: on the relation
  between classification accuracy, expressibility and entangling capability}}
  (\bibinfo {year} {2020})\BibitemShut {NoStop}%
\bibitem [{\citenamefont {Anand}\ \emph
  {et~al.}(2022{\natexlab{b}})\citenamefont {Anand}, \citenamefont
  {Alperin-Lea}, \citenamefont {Choquette},\ and\ \citenamefont
  {Aspuru-Guzik}}]{anand2022exploring}%
  \BibitemOpen
  \bibfield  {author} {\bibinfo {author} {\bibfnamefont {A.}~\bibnamefont
  {Anand}}, \bibinfo {author} {\bibfnamefont {S.}~\bibnamefont {Alperin-Lea}},
  \bibinfo {author} {\bibfnamefont {A.}~\bibnamefont {Choquette}},\ and\
  \bibinfo {author} {\bibfnamefont {A.}~\bibnamefont {Aspuru-Guzik}},\
  }\href@noop {} {\bibinfo {title} {Exploring the role of parameters in
  variational quantum algorithms}} (\bibinfo {year} {2022}{\natexlab{b}}),\
  \Eprint {https://arxiv.org/abs/2209.14405} {arXiv:2209.14405 [quant-ph]}
  \BibitemShut {NoStop}%
\bibitem [{\citenamefont {Holmes}\ \emph {et~al.}(2022)\citenamefont {Holmes},
  \citenamefont {Sharma}, \citenamefont {Cerezo},\ and\ \citenamefont
  {Coles}}]{exp_barren}%
  \BibitemOpen
  \bibfield  {author} {\bibinfo {author} {\bibfnamefont {Z.}~\bibnamefont
  {Holmes}}, \bibinfo {author} {\bibfnamefont {K.}~\bibnamefont {Sharma}},
  \bibinfo {author} {\bibfnamefont {M.}~\bibnamefont {Cerezo}},\ and\ \bibinfo
  {author} {\bibfnamefont {P.~J.}\ \bibnamefont {Coles}},\ }\bibfield
  {journal} {\bibinfo  {journal} {{PRX} Quantum}\ }\textbf {\bibinfo {volume}
  {3}},\ \href {https://doi.org/10.1103/prxquantum.3.010313}
  {10.1103/prxquantum.3.010313} (\bibinfo {year} {2022})\BibitemShut {NoStop}%
\bibitem [{\citenamefont {Nakanishi}\ \emph {et~al.}(2020)\citenamefont
  {Nakanishi}, \citenamefont {Fujii},\ and\ \citenamefont
  {Todo}}]{nakanishi2020sequential}%
  \BibitemOpen
  \bibfield  {author} {\bibinfo {author} {\bibfnamefont {K.~M.}\ \bibnamefont
  {Nakanishi}}, \bibinfo {author} {\bibfnamefont {K.}~\bibnamefont {Fujii}},\
  and\ \bibinfo {author} {\bibfnamefont {S.}~\bibnamefont {Todo}},\ }\href@noop
  {} {\bibfield  {journal} {\bibinfo  {journal} {Physical Review Research}\
  }\textbf {\bibinfo {volume} {2}},\ \bibinfo {pages} {043158} (\bibinfo {year}
  {2020})}\BibitemShut {NoStop}%
\bibitem [{\citenamefont {Skolik}\ \emph {et~al.}(2021)\citenamefont {Skolik},
  \citenamefont {McClean}, \citenamefont {Mohseni}, \citenamefont {van~der
  Smagt},\ and\ \citenamefont {Leib}}]{skolik2021layerwise}%
  \BibitemOpen
  \bibfield  {author} {\bibinfo {author} {\bibfnamefont {A.}~\bibnamefont
  {Skolik}}, \bibinfo {author} {\bibfnamefont {J.~R.}\ \bibnamefont {McClean}},
  \bibinfo {author} {\bibfnamefont {M.}~\bibnamefont {Mohseni}}, \bibinfo
  {author} {\bibfnamefont {P.}~\bibnamefont {van~der Smagt}},\ and\ \bibinfo
  {author} {\bibfnamefont {M.}~\bibnamefont {Leib}},\ }\href@noop {} {\bibfield
   {journal} {\bibinfo  {journal} {Quantum Machine Intelligence}\ }\textbf
  {\bibinfo {volume} {3}},\ \bibinfo {pages} {1} (\bibinfo {year}
  {2021})}\BibitemShut {NoStop}%
\bibitem [{\citenamefont {Xiao}\ \emph {et~al.}(2022)\citenamefont {Xiao},
  \citenamefont {Wen}, \citenamefont {Wei},\ and\ \citenamefont
  {Long}}]{xiao2022reconstructing}%
  \BibitemOpen
  \bibfield  {author} {\bibinfo {author} {\bibfnamefont {J.}~\bibnamefont
  {Xiao}}, \bibinfo {author} {\bibfnamefont {J.}~\bibnamefont {Wen}}, \bibinfo
  {author} {\bibfnamefont {S.}~\bibnamefont {Wei}},\ and\ \bibinfo {author}
  {\bibfnamefont {G.}~\bibnamefont {Long}},\ }\href@noop {} {\bibfield
  {journal} {\bibinfo  {journal} {Frontiers of Physics}\ }\textbf {\bibinfo
  {volume} {17}},\ \bibinfo {pages} {51501} (\bibinfo {year}
  {2022})}\BibitemShut {NoStop}%
\bibitem [{\citenamefont {Jattana}\ \emph {et~al.}(2023)\citenamefont
  {Jattana}, \citenamefont {Jin}, \citenamefont {De~Raedt},\ and\ \citenamefont
  {Michielsen}}]{jattana2023improved}%
  \BibitemOpen
  \bibfield  {author} {\bibinfo {author} {\bibfnamefont {M.~S.}\ \bibnamefont
  {Jattana}}, \bibinfo {author} {\bibfnamefont {F.}~\bibnamefont {Jin}},
  \bibinfo {author} {\bibfnamefont {H.}~\bibnamefont {De~Raedt}},\ and\
  \bibinfo {author} {\bibfnamefont {K.}~\bibnamefont {Michielsen}},\
  }\href@noop {} {\bibfield  {journal} {\bibinfo  {journal} {Physical Review
  Applied}\ }\textbf {\bibinfo {volume} {19}},\ \bibinfo {pages} {024047}
  (\bibinfo {year} {2023})}\BibitemShut {NoStop}%
\bibitem [{\citenamefont {Campos}\ \emph {et~al.}(2021)\citenamefont {Campos},
  \citenamefont {Rabinovich}, \citenamefont {Akshay},\ and\ \citenamefont
  {Biamonte}}]{campos2021training}%
  \BibitemOpen
  \bibfield  {author} {\bibinfo {author} {\bibfnamefont {E.}~\bibnamefont
  {Campos}}, \bibinfo {author} {\bibfnamefont {D.}~\bibnamefont {Rabinovich}},
  \bibinfo {author} {\bibfnamefont {V.}~\bibnamefont {Akshay}},\ and\ \bibinfo
  {author} {\bibfnamefont {J.}~\bibnamefont {Biamonte}},\ }\href@noop {}
  {\bibfield  {journal} {\bibinfo  {journal} {Physical Review A}\ }\textbf
  {\bibinfo {volume} {104}},\ \bibinfo {pages} {L030401} (\bibinfo {year}
  {2021})}\BibitemShut {NoStop}%
\bibitem [{\citenamefont {Halder}\ \emph {et~al.}(2023)\citenamefont {Halder},
  \citenamefont {Patra}, \citenamefont {Mondal},\ and\ \citenamefont
  {Maitra}}]{halder2023machine}%
  \BibitemOpen
  \bibfield  {author} {\bibinfo {author} {\bibfnamefont {S.}~\bibnamefont
  {Halder}}, \bibinfo {author} {\bibfnamefont {C.}~\bibnamefont {Patra}},
  \bibinfo {author} {\bibfnamefont {D.}~\bibnamefont {Mondal}},\ and\ \bibinfo
  {author} {\bibfnamefont {R.}~\bibnamefont {Maitra}},\ }\href@noop {}
  {\bibfield  {journal} {\bibinfo  {journal} {The Journal of Chemical Physics}\
  }\textbf {\bibinfo {volume} {158}} (\bibinfo {year} {2023})}\BibitemShut
  {NoStop}%
\bibitem [{\citenamefont {Liu}\ \emph {et~al.}(2023)\citenamefont {Liu},
  \citenamefont {Zhang}, \citenamefont {Jian},\ and\ \citenamefont
  {Yao}}]{liu2023training}%
  \BibitemOpen
  \bibfield  {author} {\bibinfo {author} {\bibfnamefont {S.}~\bibnamefont
  {Liu}}, \bibinfo {author} {\bibfnamefont {S.-X.}\ \bibnamefont {Zhang}},
  \bibinfo {author} {\bibfnamefont {S.-K.}\ \bibnamefont {Jian}},\ and\
  \bibinfo {author} {\bibfnamefont {H.}~\bibnamefont {Yao}},\ }\bibfield
  {journal} {\bibinfo  {journal} {Physical Review Research}\ }\textbf {\bibinfo
  {volume} {5}},\ \href {https://doi.org/10.1103/physrevresearch.5.l032040}
  {10.1103/physrevresearch.5.l032040} (\bibinfo {year} {2023})\BibitemShut
  {NoStop}%
\bibitem [{\citenamefont {Zhang}\ \emph
  {et~al.}(2021{\natexlab{a}})\citenamefont {Zhang}, \citenamefont {Kyaw},
  \citenamefont {Kottmann}, \citenamefont {Degroote},\ and\ \citenamefont
  {Aspuru-Guzik}}]{zhang2021mutual}%
  \BibitemOpen
  \bibfield  {author} {\bibinfo {author} {\bibfnamefont {Z.-J.}\ \bibnamefont
  {Zhang}}, \bibinfo {author} {\bibfnamefont {T.~H.}\ \bibnamefont {Kyaw}},
  \bibinfo {author} {\bibfnamefont {J.~S.}\ \bibnamefont {Kottmann}}, \bibinfo
  {author} {\bibfnamefont {M.}~\bibnamefont {Degroote}},\ and\ \bibinfo
  {author} {\bibfnamefont {A.}~\bibnamefont {Aspuru-Guzik}},\ }\href@noop {}
  {\bibfield  {journal} {\bibinfo  {journal} {Quantum Science and Technology}\
  }\textbf {\bibinfo {volume} {6}},\ \bibinfo {pages} {035001} (\bibinfo {year}
  {2021}{\natexlab{a}})}\BibitemShut {NoStop}%
\bibitem [{\citenamefont {Zhang}\ \emph
  {et~al.}(2021{\natexlab{b}})\citenamefont {Zhang}, \citenamefont {Cincio},
  \citenamefont {Negre}, \citenamefont {Czarnik}, \citenamefont {Coles},
  \citenamefont {Anisimov}, \citenamefont {Mniszewski}, \citenamefont
  {Tretiak},\ and\ \citenamefont {Dub}}]{zhang2021variational}%
  \BibitemOpen
  \bibfield  {author} {\bibinfo {author} {\bibfnamefont {Y.}~\bibnamefont
  {Zhang}}, \bibinfo {author} {\bibfnamefont {L.}~\bibnamefont {Cincio}},
  \bibinfo {author} {\bibfnamefont {C.~F.}\ \bibnamefont {Negre}}, \bibinfo
  {author} {\bibfnamefont {P.}~\bibnamefont {Czarnik}}, \bibinfo {author}
  {\bibfnamefont {P.}~\bibnamefont {Coles}}, \bibinfo {author} {\bibfnamefont
  {P.~M.}\ \bibnamefont {Anisimov}}, \bibinfo {author} {\bibfnamefont {S.~M.}\
  \bibnamefont {Mniszewski}}, \bibinfo {author} {\bibfnamefont
  {S.}~\bibnamefont {Tretiak}},\ and\ \bibinfo {author} {\bibfnamefont {P.~A.}\
  \bibnamefont {Dub}},\ }\href@noop {} {\bibfield  {journal} {\bibinfo
  {journal} {arXiv preprint arXiv:2106.07619}\ } (\bibinfo {year}
  {2021}{\natexlab{b}})}\BibitemShut {NoStop}%
\bibitem [{\citenamefont {Schuld}\ \emph {et~al.}(2019)\citenamefont {Schuld},
  \citenamefont {Bergholm}, \citenamefont {Gogolin}, \citenamefont {Izaac},\
  and\ \citenamefont {Killoran}}]{schuld2019evaluating}%
  \BibitemOpen
  \bibfield  {author} {\bibinfo {author} {\bibfnamefont {M.}~\bibnamefont
  {Schuld}}, \bibinfo {author} {\bibfnamefont {V.}~\bibnamefont {Bergholm}},
  \bibinfo {author} {\bibfnamefont {C.}~\bibnamefont {Gogolin}}, \bibinfo
  {author} {\bibfnamefont {J.}~\bibnamefont {Izaac}},\ and\ \bibinfo {author}
  {\bibfnamefont {N.}~\bibnamefont {Killoran}},\ }\href@noop {} {\bibfield
  {journal} {\bibinfo  {journal} {Physical Review A}\ }\textbf {\bibinfo
  {volume} {99}},\ \bibinfo {pages} {032331} (\bibinfo {year}
  {2019})}\BibitemShut {NoStop}%
\bibitem [{\citenamefont {Mari}\ \emph {et~al.}(2021)\citenamefont {Mari},
  \citenamefont {Bromley},\ and\ \citenamefont
  {Killoran}}]{mari2021estimating}%
  \BibitemOpen
  \bibfield  {author} {\bibinfo {author} {\bibfnamefont {A.}~\bibnamefont
  {Mari}}, \bibinfo {author} {\bibfnamefont {T.~R.}\ \bibnamefont {Bromley}},\
  and\ \bibinfo {author} {\bibfnamefont {N.}~\bibnamefont {Killoran}},\
  }\href@noop {} {\bibfield  {journal} {\bibinfo  {journal} {Physical Review
  A}\ }\textbf {\bibinfo {volume} {103}},\ \bibinfo {pages} {012405} (\bibinfo
  {year} {2021})}\BibitemShut {NoStop}%
\bibitem [{\citenamefont {Kottmann}\ \emph
  {et~al.}(2021{\natexlab{a}})\citenamefont {Kottmann}, \citenamefont {Anand},\
  and\ \citenamefont {Aspuru-Guzik}}]{kottmann2021feasible}%
  \BibitemOpen
  \bibfield  {author} {\bibinfo {author} {\bibfnamefont {J.~S.}\ \bibnamefont
  {Kottmann}}, \bibinfo {author} {\bibfnamefont {A.}~\bibnamefont {Anand}},\
  and\ \bibinfo {author} {\bibfnamefont {A.}~\bibnamefont {Aspuru-Guzik}},\
  }\href@noop {} {\bibfield  {journal} {\bibinfo  {journal} {Chemical science}\
  }\textbf {\bibinfo {volume} {12}},\ \bibinfo {pages} {3497} (\bibinfo {year}
  {2021}{\natexlab{a}})}\BibitemShut {NoStop}%
\bibitem [{\citenamefont {Wierichs}\ \emph {et~al.}(2022)\citenamefont
  {Wierichs}, \citenamefont {Izaac}, \citenamefont {Wang},\ and\ \citenamefont
  {Lin}}]{wierichs2022general}%
  \BibitemOpen
  \bibfield  {author} {\bibinfo {author} {\bibfnamefont {D.}~\bibnamefont
  {Wierichs}}, \bibinfo {author} {\bibfnamefont {J.}~\bibnamefont {Izaac}},
  \bibinfo {author} {\bibfnamefont {C.}~\bibnamefont {Wang}},\ and\ \bibinfo
  {author} {\bibfnamefont {C.~Y.-Y.}\ \bibnamefont {Lin}},\ }\href@noop {}
  {\bibfield  {journal} {\bibinfo  {journal} {Quantum}\ }\textbf {\bibinfo
  {volume} {6}},\ \bibinfo {pages} {677} (\bibinfo {year} {2022})}\BibitemShut
  {NoStop}%
\bibitem [{\citenamefont {Harrow}\ and\ \citenamefont
  {Napp}(2021)}]{harrow2021low}%
  \BibitemOpen
  \bibfield  {author} {\bibinfo {author} {\bibfnamefont {A.~W.}\ \bibnamefont
  {Harrow}}\ and\ \bibinfo {author} {\bibfnamefont {J.~C.}\ \bibnamefont
  {Napp}},\ }\href@noop {} {\bibfield  {journal} {\bibinfo  {journal} {Physical
  Review Letters}\ }\textbf {\bibinfo {volume} {126}},\ \bibinfo {pages}
  {140502} (\bibinfo {year} {2021})}\BibitemShut {NoStop}%
\bibitem [{\citenamefont {Sweke}\ \emph {et~al.}(2020)\citenamefont {Sweke},
  \citenamefont {Wilde}, \citenamefont {Meyer}, \citenamefont {Schuld},
  \citenamefont {F{\"a}hrmann}, \citenamefont {Meynard-Piganeau},\ and\
  \citenamefont {Eisert}}]{sweke2020stochastic}%
  \BibitemOpen
  \bibfield  {author} {\bibinfo {author} {\bibfnamefont {R.}~\bibnamefont
  {Sweke}}, \bibinfo {author} {\bibfnamefont {F.}~\bibnamefont {Wilde}},
  \bibinfo {author} {\bibfnamefont {J.}~\bibnamefont {Meyer}}, \bibinfo
  {author} {\bibfnamefont {M.}~\bibnamefont {Schuld}}, \bibinfo {author}
  {\bibfnamefont {P.~K.}\ \bibnamefont {F{\"a}hrmann}}, \bibinfo {author}
  {\bibfnamefont {B.}~\bibnamefont {Meynard-Piganeau}},\ and\ \bibinfo {author}
  {\bibfnamefont {J.}~\bibnamefont {Eisert}},\ }\href@noop {} {\bibfield
  {journal} {\bibinfo  {journal} {Quantum}\ }\textbf {\bibinfo {volume} {4}},\
  \bibinfo {pages} {314} (\bibinfo {year} {2020})}\BibitemShut {NoStop}%
\bibitem [{\citenamefont {McClean}\ \emph {et~al.}(2018)\citenamefont
  {McClean}, \citenamefont {Boixo}, \citenamefont {Smelyanskiy}, \citenamefont
  {Babbush},\ and\ \citenamefont {Neven}}]{nn_1}%
  \BibitemOpen
  \bibfield  {author} {\bibinfo {author} {\bibfnamefont {J.~R.}\ \bibnamefont
  {McClean}}, \bibinfo {author} {\bibfnamefont {S.}~\bibnamefont {Boixo}},
  \bibinfo {author} {\bibfnamefont {V.~N.}\ \bibnamefont {Smelyanskiy}},
  \bibinfo {author} {\bibfnamefont {R.}~\bibnamefont {Babbush}},\ and\ \bibinfo
  {author} {\bibfnamefont {H.}~\bibnamefont {Neven}},\ }\href
  {https://doi.org/10.1038/s41467-018-07090-4} {\bibfield  {journal} {\bibinfo
  {journal} {Nature Communications}\ }\textbf {\bibinfo {volume} {9}},\
  \bibinfo {pages} {4812} (\bibinfo {year} {2018})}\BibitemShut {NoStop}%
\bibitem [{\citenamefont {Farhi}\ and\ \citenamefont
  {Neven}(2018{\natexlab{b}})}]{nn_2}%
  \BibitemOpen
  \bibfield  {author} {\bibinfo {author} {\bibfnamefont {E.}~\bibnamefont
  {Farhi}}\ and\ \bibinfo {author} {\bibfnamefont {H.}~\bibnamefont {Neven}},\
  }\href {https://doi.org/10.48550/ARXIV.1802.06002} {\bibinfo {title}
  {Classification with quantum neural networks on near term processors}}
  (\bibinfo {year} {2018}{\natexlab{b}})\BibitemShut {NoStop}%
\bibitem [{\citenamefont {Schuld}\ and\ \citenamefont
  {Killoran}(2019)}]{kernel_1}%
  \BibitemOpen
  \bibfield  {author} {\bibinfo {author} {\bibfnamefont {M.}~\bibnamefont
  {Schuld}}\ and\ \bibinfo {author} {\bibfnamefont {N.}~\bibnamefont
  {Killoran}},\ }\href {https://doi.org/10.1103/PhysRevLett.122.040504}
  {\bibfield  {journal} {\bibinfo  {journal} {Phys. Rev. Lett.}\ }\textbf
  {\bibinfo {volume} {122}},\ \bibinfo {pages} {040504} (\bibinfo {year}
  {2019})}\BibitemShut {NoStop}%
\bibitem [{\citenamefont {Havl{\'i}{\v{c}}ek}\ \emph
  {et~al.}(2019{\natexlab{b}})\citenamefont {Havl{\'i}{\v{c}}ek}, \citenamefont
  {C{\'o}rcoles}, \citenamefont {Temme}, \citenamefont {Harrow}, \citenamefont
  {Kandala}, \citenamefont {Chow},\ and\ \citenamefont {Gambetta}}]{kernel_2}%
  \BibitemOpen
  \bibfield  {author} {\bibinfo {author} {\bibfnamefont {V.}~\bibnamefont
  {Havl{\'i}{\v{c}}ek}}, \bibinfo {author} {\bibfnamefont {A.~D.}\ \bibnamefont
  {C{\'o}rcoles}}, \bibinfo {author} {\bibfnamefont {K.}~\bibnamefont {Temme}},
  \bibinfo {author} {\bibfnamefont {A.~W.}\ \bibnamefont {Harrow}}, \bibinfo
  {author} {\bibfnamefont {A.}~\bibnamefont {Kandala}}, \bibinfo {author}
  {\bibfnamefont {J.~M.}\ \bibnamefont {Chow}},\ and\ \bibinfo {author}
  {\bibfnamefont {J.~M.}\ \bibnamefont {Gambetta}},\ }\href
  {https://doi.org/10.1038/s41586-019-0980-2} {\bibfield  {journal} {\bibinfo
  {journal} {Nature}\ }\textbf {\bibinfo {volume} {567}},\ \bibinfo {pages}
  {209} (\bibinfo {year} {2019}{\natexlab{b}})}\BibitemShut {NoStop}%
\bibitem [{\citenamefont {Hyatt}\ \emph {et~al.}(2017)\citenamefont {Hyatt},
  \citenamefont {Garrison},\ and\ \citenamefont {Bauer}}]{hyatt2017extracting}%
  \BibitemOpen
  \bibfield  {author} {\bibinfo {author} {\bibfnamefont {K.}~\bibnamefont
  {Hyatt}}, \bibinfo {author} {\bibfnamefont {J.~R.}\ \bibnamefont
  {Garrison}},\ and\ \bibinfo {author} {\bibfnamefont {B.}~\bibnamefont
  {Bauer}},\ }\href@noop {} {\bibfield  {journal} {\bibinfo  {journal}
  {Physical review letters}\ }\textbf {\bibinfo {volume} {119}},\ \bibinfo
  {pages} {140502} (\bibinfo {year} {2017})}\BibitemShut {NoStop}%
\bibitem [{\citenamefont {Benedetti}\ \emph {et~al.}(2021)\citenamefont
  {Benedetti}, \citenamefont {Fiorentini},\ and\ \citenamefont
  {Lubasch}}]{benedetti2021hardware}%
  \BibitemOpen
  \bibfield  {author} {\bibinfo {author} {\bibfnamefont {M.}~\bibnamefont
  {Benedetti}}, \bibinfo {author} {\bibfnamefont {M.}~\bibnamefont
  {Fiorentini}},\ and\ \bibinfo {author} {\bibfnamefont {M.}~\bibnamefont
  {Lubasch}},\ }\href@noop {} {\bibfield  {journal} {\bibinfo  {journal}
  {Physical Review Research}\ }\textbf {\bibinfo {volume} {3}},\ \bibinfo
  {pages} {033083} (\bibinfo {year} {2021})}\BibitemShut {NoStop}%
\bibitem [{\citenamefont {Hagberg}\ \emph {et~al.}(2008)\citenamefont
  {Hagberg}, \citenamefont {Swart},\ and\ \citenamefont
  {S~Chult}}]{hagberg2008exploring}%
  \BibitemOpen
  \bibfield  {author} {\bibinfo {author} {\bibfnamefont {A.}~\bibnamefont
  {Hagberg}}, \bibinfo {author} {\bibfnamefont {P.}~\bibnamefont {Swart}},\
  and\ \bibinfo {author} {\bibfnamefont {D.}~\bibnamefont {S~Chult}},\
  }\href@noop {} {\emph {\bibinfo {title} {Exploring network structure,
  dynamics, and function using NetworkX}}},\ \bibinfo {type} {Tech. Rep.}\
  (\bibinfo  {institution} {Los Alamos National Lab.(LANL), Los Alamos, NM
  (United States)},\ \bibinfo {year} {2008})\BibitemShut {NoStop}%
\bibitem [{\citenamefont {Kivlichan}\ \emph {et~al.}(2019)\citenamefont
  {Kivlichan}, \citenamefont {Granade},\ and\ \citenamefont
  {Wiebe}}]{kivlichan2019phase}%
  \BibitemOpen
  \bibfield  {author} {\bibinfo {author} {\bibfnamefont {I.~D.}\ \bibnamefont
  {Kivlichan}}, \bibinfo {author} {\bibfnamefont {C.~E.}\ \bibnamefont
  {Granade}},\ and\ \bibinfo {author} {\bibfnamefont {N.}~\bibnamefont
  {Wiebe}},\ }\href@noop {} {\bibfield  {journal} {\bibinfo  {journal} {arXiv
  preprint arXiv:1907.10070}\ } (\bibinfo {year} {2019})}\BibitemShut {NoStop}%
\bibitem [{\citenamefont {Kottmann}\ \emph
  {et~al.}(2021{\natexlab{b}})\citenamefont {Kottmann}, \citenamefont
  {Alperin-Lea}, \citenamefont {Tamayo-Mendoza}, \citenamefont
  {Cervera-Lierta}, \citenamefont {Lavigne}, \citenamefont {Yen}, \citenamefont
  {Verteletskyi}, \citenamefont {Schleich}, \citenamefont {Anand},
  \citenamefont {Degroote}, \citenamefont {Chaney}, \citenamefont {Kesibi},
  \citenamefont {Curnow}, \citenamefont {Solo}, \citenamefont
  {Tsilimigkounakis}, \citenamefont {Zendejas-Morales}, \citenamefont
  {Izmaylov},\ and\ \citenamefont {Aspuru-Guzik}}]{kottmann2020tequila}%
  \BibitemOpen
  \bibfield  {author} {\bibinfo {author} {\bibfnamefont {J.}~\bibnamefont
  {Kottmann}}, \bibinfo {author} {\bibfnamefont {S.}~\bibnamefont
  {Alperin-Lea}}, \bibinfo {author} {\bibfnamefont {T.}~\bibnamefont
  {Tamayo-Mendoza}}, \bibinfo {author} {\bibfnamefont {A.}~\bibnamefont
  {Cervera-Lierta}}, \bibinfo {author} {\bibfnamefont {C.}~\bibnamefont
  {Lavigne}}, \bibinfo {author} {\bibfnamefont {T.-C.}\ \bibnamefont {Yen}},
  \bibinfo {author} {\bibfnamefont {V.}~\bibnamefont {Verteletskyi}}, \bibinfo
  {author} {\bibfnamefont {P.}~\bibnamefont {Schleich}}, \bibinfo {author}
  {\bibfnamefont {A.}~\bibnamefont {Anand}}, \bibinfo {author} {\bibfnamefont
  {M.}~\bibnamefont {Degroote}}, \bibinfo {author} {\bibfnamefont
  {S.}~\bibnamefont {Chaney}}, \bibinfo {author} {\bibfnamefont
  {M.}~\bibnamefont {Kesibi}}, \bibinfo {author} {\bibfnamefont {N.~G.}\
  \bibnamefont {Curnow}}, \bibinfo {author} {\bibfnamefont {B.}~\bibnamefont
  {Solo}}, \bibinfo {author} {\bibfnamefont {G.}~\bibnamefont
  {Tsilimigkounakis}}, \bibinfo {author} {\bibfnamefont {C.}~\bibnamefont
  {Zendejas-Morales}}, \bibinfo {author} {\bibfnamefont {A.~F.}\ \bibnamefont
  {Izmaylov}},\ and\ \bibinfo {author} {\bibfnamefont {A.}~\bibnamefont
  {Aspuru-Guzik}},\ }\href
  {http://iopscience.iop.org/article/10.1088/2058-9565/abe567} {\bibfield
  {journal} {\bibinfo  {journal} {Quantum Science and Technology}\ } (\bibinfo
  {year} {2021}{\natexlab{b}})}\BibitemShut {NoStop}%
\bibitem [{\citenamefont {Suzuki}\ \emph {et~al.}(2021)\citenamefont {Suzuki},
  \citenamefont {Kawase}, \citenamefont {Masumura}, \citenamefont {Hiraga},
  \citenamefont {Nakadai}, \citenamefont {Chen}, \citenamefont {Nakanishi},
  \citenamefont {Mitarai}, \citenamefont {Imai}, \citenamefont {Tamiya} \emph
  {et~al.}}]{suzuki2021qulacs}%
  \BibitemOpen
  \bibfield  {author} {\bibinfo {author} {\bibfnamefont {Y.}~\bibnamefont
  {Suzuki}}, \bibinfo {author} {\bibfnamefont {Y.}~\bibnamefont {Kawase}},
  \bibinfo {author} {\bibfnamefont {Y.}~\bibnamefont {Masumura}}, \bibinfo
  {author} {\bibfnamefont {Y.}~\bibnamefont {Hiraga}}, \bibinfo {author}
  {\bibfnamefont {M.}~\bibnamefont {Nakadai}}, \bibinfo {author} {\bibfnamefont
  {J.}~\bibnamefont {Chen}}, \bibinfo {author} {\bibfnamefont {K.~M.}\
  \bibnamefont {Nakanishi}}, \bibinfo {author} {\bibfnamefont {K.}~\bibnamefont
  {Mitarai}}, \bibinfo {author} {\bibfnamefont {R.}~\bibnamefont {Imai}},
  \bibinfo {author} {\bibfnamefont {S.}~\bibnamefont {Tamiya}}, \emph
  {et~al.},\ }\href@noop {} {\bibfield  {journal} {\bibinfo  {journal}
  {Quantum}\ }\textbf {\bibinfo {volume} {5}},\ \bibinfo {pages} {559}
  (\bibinfo {year} {2021})}\BibitemShut {NoStop}%
\bibitem [{\citenamefont {Bergholm}\ \emph {et~al.}(2018)\citenamefont
  {Bergholm}, \citenamefont {Izaac}, \citenamefont {Schuld}, \citenamefont
  {Gogolin}, \citenamefont {Alam}, \citenamefont {Ahmed}, \citenamefont
  {Arrazola}, \citenamefont {Blank}, \citenamefont {Delgado}, \citenamefont
  {Jahangiri} \emph {et~al.}}]{bergholm2018pennylane}%
  \BibitemOpen
  \bibfield  {author} {\bibinfo {author} {\bibfnamefont {V.}~\bibnamefont
  {Bergholm}}, \bibinfo {author} {\bibfnamefont {J.}~\bibnamefont {Izaac}},
  \bibinfo {author} {\bibfnamefont {M.}~\bibnamefont {Schuld}}, \bibinfo
  {author} {\bibfnamefont {C.}~\bibnamefont {Gogolin}}, \bibinfo {author}
  {\bibfnamefont {M.~S.}\ \bibnamefont {Alam}}, \bibinfo {author}
  {\bibfnamefont {S.}~\bibnamefont {Ahmed}}, \bibinfo {author} {\bibfnamefont
  {J.~M.}\ \bibnamefont {Arrazola}}, \bibinfo {author} {\bibfnamefont
  {C.}~\bibnamefont {Blank}}, \bibinfo {author} {\bibfnamefont
  {A.}~\bibnamefont {Delgado}}, \bibinfo {author} {\bibfnamefont
  {S.}~\bibnamefont {Jahangiri}}, \emph {et~al.},\ }\href@noop {} {\bibfield
  {journal} {\bibinfo  {journal} {arXiv preprint arXiv:1811.04968}\ } (\bibinfo
  {year} {2018})}\BibitemShut {NoStop}%
\bibitem [{\citenamefont {Marrero}\ \emph {et~al.}(2021)\citenamefont
  {Marrero}, \citenamefont {Kieferov{\'a}},\ and\ \citenamefont
  {Wiebe}}]{marrero2021entanglement}%
  \BibitemOpen
  \bibfield  {author} {\bibinfo {author} {\bibfnamefont {C.~O.}\ \bibnamefont
  {Marrero}}, \bibinfo {author} {\bibfnamefont {M.}~\bibnamefont
  {Kieferov{\'a}}},\ and\ \bibinfo {author} {\bibfnamefont {N.}~\bibnamefont
  {Wiebe}},\ }\href@noop {} {\bibfield  {journal} {\bibinfo  {journal} {PRX
  Quantum}\ }\textbf {\bibinfo {volume} {2}},\ \bibinfo {pages} {040316}
  (\bibinfo {year} {2021})}\BibitemShut {NoStop}%
\bibitem [{\citenamefont {Swenson}\ \emph {et~al.}(2020)\citenamefont
  {Swenson}, \citenamefont {Murray}, \citenamefont {Kar},\ and\ \citenamefont
  {Poor}}]{swenson2020distributed}%
  \BibitemOpen
  \bibfield  {author} {\bibinfo {author} {\bibfnamefont {B.}~\bibnamefont
  {Swenson}}, \bibinfo {author} {\bibfnamefont {R.}~\bibnamefont {Murray}},
  \bibinfo {author} {\bibfnamefont {S.}~\bibnamefont {Kar}},\ and\ \bibinfo
  {author} {\bibfnamefont {H.~V.}\ \bibnamefont {Poor}},\ }\href@noop {}
  {\bibfield  {journal} {\bibinfo  {journal} {arXiv preprint arXiv:2003.02818}\
  } (\bibinfo {year} {2020})}\BibitemShut {NoStop}%
\bibitem [{\citenamefont {Ponce}\ \emph {et~al.}(2019)\citenamefont {Ponce},
  \citenamefont {van Zon}, \citenamefont {Northrup}, \citenamefont {Gruner},
  \citenamefont {Chen}, \citenamefont {Ertinaz}, \citenamefont {Fedoseev},
  \citenamefont {Groer}, \citenamefont {Mao}, \citenamefont {Mundim} \emph
  {et~al.}}]{niagara1}%
  \BibitemOpen
  \bibfield  {author} {\bibinfo {author} {\bibfnamefont {M.}~\bibnamefont
  {Ponce}}, \bibinfo {author} {\bibfnamefont {R.}~\bibnamefont {van Zon}},
  \bibinfo {author} {\bibfnamefont {S.}~\bibnamefont {Northrup}}, \bibinfo
  {author} {\bibfnamefont {D.}~\bibnamefont {Gruner}}, \bibinfo {author}
  {\bibfnamefont {J.}~\bibnamefont {Chen}}, \bibinfo {author} {\bibfnamefont
  {F.}~\bibnamefont {Ertinaz}}, \bibinfo {author} {\bibfnamefont
  {A.}~\bibnamefont {Fedoseev}}, \bibinfo {author} {\bibfnamefont
  {L.}~\bibnamefont {Groer}}, \bibinfo {author} {\bibfnamefont
  {F.}~\bibnamefont {Mao}}, \bibinfo {author} {\bibfnamefont {B.~C.}\
  \bibnamefont {Mundim}}, \emph {et~al.},\ }in\ \href@noop {} {\emph {\bibinfo
  {booktitle} {Proceedings of the Practice and Experience in Advanced Research
  Computing on Rise of the Machines (learning)}}}\ (\bibinfo {year} {2019})\
  pp.\ \bibinfo {pages} {1--8}\BibitemShut {NoStop}%
\bibitem [{\citenamefont {Loken}\ \emph {et~al.}(2010)\citenamefont {Loken},
  \citenamefont {Gruner}, \citenamefont {Groer}, \citenamefont {Peltier},
  \citenamefont {Bunn}, \citenamefont {Craig}, \citenamefont {Henriques},
  \citenamefont {Dempsey}, \citenamefont {Yu}, \citenamefont {Chen} \emph
  {et~al.}}]{niagara2}%
  \BibitemOpen
  \bibfield  {author} {\bibinfo {author} {\bibfnamefont {C.}~\bibnamefont
  {Loken}}, \bibinfo {author} {\bibfnamefont {D.}~\bibnamefont {Gruner}},
  \bibinfo {author} {\bibfnamefont {L.}~\bibnamefont {Groer}}, \bibinfo
  {author} {\bibfnamefont {R.}~\bibnamefont {Peltier}}, \bibinfo {author}
  {\bibfnamefont {N.}~\bibnamefont {Bunn}}, \bibinfo {author} {\bibfnamefont
  {M.}~\bibnamefont {Craig}}, \bibinfo {author} {\bibfnamefont
  {T.}~\bibnamefont {Henriques}}, \bibinfo {author} {\bibfnamefont
  {J.}~\bibnamefont {Dempsey}}, \bibinfo {author} {\bibfnamefont {C.-H.}\
  \bibnamefont {Yu}}, \bibinfo {author} {\bibfnamefont {J.}~\bibnamefont
  {Chen}}, \emph {et~al.},\ }in\ \href@noop {} {\emph {\bibinfo {booktitle}
  {Journal of Physics-Conference Series}}},\ Vol.\ \bibinfo {volume} {256}\
  (\bibinfo {year} {2010})\ p.\ \bibinfo {pages} {012026}\BibitemShut {NoStop}%
\end{thebibliography}%

\end{document}